\documentclass[11pt,letter]{article}
\usepackage{jheppub}
\usepackage{csquotes}
\usepackage{amsmath,amssymb,amsfonts}
\usepackage{booktabs}
\usepackage{epsfig}
\usepackage{graphicx}
\usepackage{color}
\usepackage{shuffle}
\usepackage{float}
\usepackage{hyperref}
\usepackage[dvipsnames]{xcolor}
\usepackage{colordvi}
\usepackage{colortbl}
\usepackage{float}
\usepackage{floatflt}
\usepackage{graphicx}
\usepackage{hhline}
\usepackage{contour}
\usepackage{tikz}
\usepackage[tableposition=above]{caption}
\usepackage[normalem]{ulem}

\usepackage{wrapfig}
\usepackage{cases}
\usepackage[utf8]{inputenc}
\usepackage{mathtools,tabu}
\usepackage{slashed}
\usepackage{caption}
\usepackage{subcaption}
\usepackage{array}
\usepackage[english]{babel}

\newcounter{nomfigcpt} 
\newcommand{\numberedfig}[1][]{%
 \includegraphics[#1]{images/ckFermionBootstrap-figure\arabic{nomfigcpt}.pdf} %
  \stepcounter{nomfigcpt}%
}

\definecolor{NUpurple}{RGB}{078,042,132}

\newcommand{\nn}[0]{\nonumber}
\newcommand{\be}[0]{\begin{equation}}
\newcommand{\ee}[0]{\end{equation}}

\newcommand{\eqn}[1]{Eqn.~\ref{#1}}
\newcommand{\Fig}[1]{Fig.~\ref{#1}}
\newcommand{\fig}[1]{Fig.~\ref{#1}}
\def\sect#1{Section~{\ref{#1}}}

\def\Rev{\mathcal{R}}

\title{\boldmath Loop-level double-copy for massive fermions in the fundamental.}
\author[a]{John Joseph Carrasco,}
\author[a]{Aslan Seifi}

\affiliation[a]{Department of Physics and Astronomy
	Northwestern University, Evanston, IL 60208, USA}

\emailAdd{carrasco@northwestern.edu}
\emailAdd{aslan.seifi@gmail.com}

\abstract{We find that unitarity cuts and the duality between color and kinematics are sufficient constraints to bootstrap $D$-dimensional QCD scattering amplitudes starting from three-particle tree-level.  Specifically, we calculate tree level amplitudes through six-points, as well as the four-point one-loop correction for massive fermions in the fundamental representation of the gauge group -- constructing a color-dual representation of the latter for the first time. To do so we clarify a prescription for functional kinematic ansatze involving fermionic matter. The advantages of color-dual calculation, familiar from particles in the adjoint, also apply here: only a small number of basis topologies must be constrained via physical information of the theory, and algebraic relations propagate this to a full solution. As all the QCD amplitudes we construct here are color-dual, they trivially generate $D$-dimensional amplitudes in gravitational theories via double-copy construction. }

\begin{document} 

\maketitle
\flushbottom

\section{Introduction}\label{sec:Introduction}
Traditionally, scattering amplitudes for relativistic quantum field theories are calculated by deriving Feynman rules from a given Lagrangian. There are, however,  inconveniences with this approach. First, Feynman rules propagate a tremendous redundancy of off-shell information, especially for external gluons and gravitons, that must cancel in physical observables.  This means that individual diagrams, themselves unphysical, tend to be large and unwieldy, even if the final amplitude can be written in a very compact form.  A second issue is that the number of individual diagrams grows factorially as multiplicity and loop level increases. Unitarity methods deal with the first problem by restricting to on-shell physical expressions when writing down amplitudes. The duality between color and kinematics  begins to alleviate the second because we only need to feed physical information to a smaller set of basis diagrams, with algebraic relations propagating the information to the entire amplitude. We take advantage of both of these complementary ideas to bootstrap  QCD amplitudes with massive fermions. 

Unitarity demands that amplitudes factorize appropriately to sums over products of lower multiplicity amplitudes when virtual (internal) particles become physical -- i.e. when the momenta of internal particles are taken on-shell~\cite{Bern:1994zx} -- described as cut. These sums are over all physical states that could cross such cuts. This imposes strict constraints on loop integrands -- i.e. fusing~\cite{Bern:1994cg,Bern:1995db} tree-level data into multiloop integrands. In color-ordered unitarity cuts of loop amplitudes, only an exponential subset of  graphs contribute rather than the factorial number contributing to the full amplitude.

The duality between color and kinematics is manifest in gauge theories when one can satisfy the same identities between kinematics factors as for colors. This duality was first proposed by Bern, Johansson and one of the current authors~(BCJ) , and is intimately related to the double-copy construction, whereby one can construct gravity amplitudes from gauge theory expressions~\cite{Bern:2008qj, Bern:2010ue}.  In the last decade, these ideas have been applied to a web of theories both at tree and loop-integrand level.  For recent reviews, see refs.~\cite{Bern:2019prr,Bern:2022wqg,Adamo:2022dcm}.

At tree-level, QCD amplitudes with massive quarks in the fundamental are known to respect the duality between color and kinematics~\cite{Johansson:2014zca,Johansson:2015oia,delaCruz:2015dpa}.  At loop level, super-QCD amplitudes with massless matter have been shown to be color-dual through two-loops with external gluons~\cite{Johansson:2017bfl,Kalin:2018thp,Duhr:2019ywc,Kalin:2019vjc,Mogull:2020sak}.  In this paper we consider massive fermionic matter through four-point  one-loop.  We do not use Feynman rules, except to verify our results, but instead take advantage of both on-shell $D$-dimensional methods and the color-kinematics duality to construct scattering amplitudes for QCD with fundamental fermions at both tree and one-loop level. This necessitates clarifying appropriate rules for describing kinematic weights of fermionic matter functionally. This is intimately related to the work of ref.~\cite{Carrasco:2020ywq} where it was found that scalar QCD amplitudes respect color-kinematics duality at tree and one-loop level and can be similarly bootstrapped via simply imposing mass-dimension, color-kinematics duality, and factorization. 

One of our key results is the construction of {BCJ-dual numerators} which can be used in the double-copy procedure to straightforwardly generate gravitational amplitudes involving massive matter \cite{delaCruz:2016wbr,Johansson:2019dnu,Bautista:2019evw}.  For example, the integrands we generate can be double-copied with that of ~\cite{Carrasco:2020ywq} to generate the one-loop correction to massive fermion scattering mediated by axion-dilaton gravity, or with itself to generate the one-loop correction to two-to-two scattering of massive vectors mediated by axion-dilaton gravity.  Results in Einstein-Hilbert gravitation can be straightforwardly constructed via projective double-copy~\cite{Carrasco:2021bmu}.  We note that this is complementary to recent work in the effective worldline formalism that has seen the consideration of color-dual massive spinning particles with an eye towards double-copy and relevance to gravitational wave astrophysics in the classical limit~\cite{Shi:2021qsb,Plefka:2018dpa,Goldberger:2016iau,Shen:2018ebu,Goldberger:2017ogt,Li:2018qap,Goldberger:2019xef,Comberiati:2022ldk,Gonzo:2021drq}, including massive color-dual fermionic amplitudes \cite{Comberiati:2022cpm}.

We briefly review the duality between color and kinematics  and unitarity methods in \sect{sec:review}.  A famous quote in the amplitudes community\footnote{e.g.~Lance Dixon's discussion at \href{https://www.preposterousuniverse.com/blog/2013/10/03/guest-post-lance-dixon-on-calculating-amplitudes/}{https://www.preposterousuniverse.com/blog/2013/10/03/guest-post-lance-dixon-on-calculating-amplitudes/}.}:  frequently ascribed to Lev Landau goes as follows, 
\begin{displayquote}
A method is more important than a discovery, since the right method will lead to new and even more important discoveries.
\end{displayquote}
The method  we develop here is establishing color-dual functional ansatze for external $D$-dimensional massive fermions which we apply largely here to Dirac fermions in the fundamental. We present our rules for functional ansatze involving fermionic matter in \sect{sec:FermionicAnsatze}.  The following three sections clarify the use of this approrach in developing specific color-dual results presented as $D$-dimensional color-dual numerators. In \sect{sec:MasslessTreeQCD}, we walk through examples of bootstrapping tree-level QCD amplitudes in the massless limit, and we see that unitarity cuts and color-kinematics duality are sufficient to fix any ansatze for numerators.  We emphasize that the construction is $D$-dimensional for Dirac fermions in the fundamental.  If the duality between color-and-kinematics were to place dimensional constraints then the duality would only be satisfied in particular dimensions.  We present an example what of such a constraint would look like by considering same-flavor fermions in the adjoint at four-points in \sect{adjointExample} and see that the adjoint color-dual relations are only satisfied under four-dimensional fiertz identities -- consistent with dimensional constraints due to supersymmetry.

We generalize to massive matter at tree-level in \sect{sec:MassiveTreeQCD}.  We use these results to construct various four-point one-loop amplitudes in QCD in \sect{sec:one-loopQCD}, where  we verify bootstrapped gauge theory results against the known results. Specfically we present the basis numerators for four-external fermions at 1-loop in \eqn{eq:Numerator1FourFermions}, \eqn{eq:Numerator3FourFermions}, and \eqn{eq:Numerator5FourFermions}, as well as the color-dual basis for external glue with $N_f$ fermions running around the loop.  The latter expressions, depending as they do upon formal polarizations, are too lengthy to conveniently typeset, so we relegate them to ancilary machine-readable files. Instead in \sect{sec:one-loopQCDFourExternalGlueOneFermionicLoop} we describe in detail both the steps involved in performing the calculation, and its verification involving {Passarino-Veltman (PV)}~\cite{Passarino:1978jh} reduction to the known basis of one-loop integrals. Furthermore we specialize to external four-dimensions to recover the simple rational result for all-plus helicity gluons which is compact enough to present in \eqn{eq:OneFermionicLoopBoxCoefficient}. Recovering this known result is a critical verification that we correctly handled internal $D$-dimensional cut-construction as four-dimensional cut-construction will famously miss such rational terms.  We summarize and outline intriguing next steps in \sect{sec:conclusion}.

\section{Review}\label{sec:review}
\subsection{Color-kinematics duality at tree and loop level}\label{sec:CKreview}
In general, one can write~\cite{Melia:2013bta,Melia:2013epa,Johansson:2015oia}  QCD  scattering amplitudes at tree level for $m$-particles interacting with $k$ distinct fermion pairs in terms of cubic graphs as:
\begin{equation}\label{eq:genericTreeQCDAmplitude}
\mathcal{A}^{\text{QCD-tree}}_{m,k} = g^{m-2} \sum^{\nu(k,m)}_{i} \frac{c_{(i)}\, n_{(i)}}{d_{(i)}}\,.
\end{equation}
The sum runs over the $\nu(k,m)$ distinct cubic graphs, with $c_{(i)}$,$n_{(i)}$ and $d_{(i)}$ denoting their color factors, kinematic numerator factors, and denominators respectively. We use $g$ for the coupling constant.  The number of cubic graphs is given,
\begin{equation}\label{eq:numberQubicGraphs}
\nu(k,m)=\frac{(2m-5)!!}{(2k-1)!!}\,.
\end{equation}  
The maximal number of graphs occurs for the cases where either all external particles are gluons or there is a single pair of fermions.  Additional quark-antiquark pairs mean fewer distinct graphs.  Contact terms are absorbed into such a cubic representation by multiplying the contact term  by appropriate factors of unity represented by propagator over propagator terms.   Generally, color factors will be a product of adjoint and fundamental color structures. 

\begin{figure}[H]
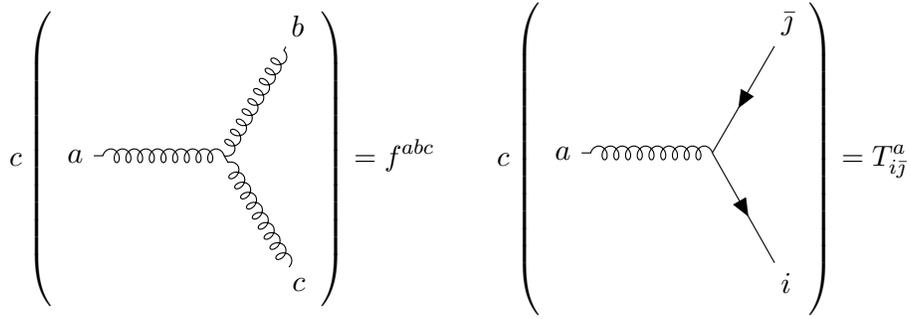

\begin{center}
\begin{equation*}
\begin{split}
c\left( {\begin{array}{c}
{\numberedfig[width=1.35in]}
\end{array}
} \right)	&=f^{abc}\,
\qquad
c\left( {\begin{array}{c}
\numberedfig[width=1.35in]
\end{array}
} \right) =T^{a}_{i\bar{\jmath}}
\end{split}
\end{equation*}
\end{center}
\caption{We depict the diagrammatic building blocks of the cubic graphs associated with QCD amplitudes. The pure gluonic vertex (left) is dressed in the adjoint representation, while quark-antiquark pair with a gluon (right) is dressed with a generator in the fundamental.  Contact terms associated with higher vertices are encoded in cubic graphs by appropriate factors unity in the form of propagators divided by propagators. }
\label{fig:QCDbuildingBlocks}
\end{figure}

There are two different three-point  building blocks to construct QCD color-weights. One of them consists of dressing vertices with pure gluons, which are dressed in with color-structure constants $f^{abc}$, and the other is a  quark-antiquark pair with a gluon, which is dressed in the fundamental, $T^a_{i \bar{\jmath}}$,  as depicted in Fig.~\ref{fig:QCDbuildingBlocks}.  There is a certain freedom in assigning complementary phases to color-factors and kinematic weights --- as long as their product remains unchanged in physical amplitudes. We say more about the power of exploiting this freedom to accommodate fermionic signs shortly.

Color-factors of graphs obey similar Jacobi and commutation identities which can be understood as three-term identities between graphs:
\begin{align}\label{eq:JacobiRelations}
f^{dae}f^{ebc}-f^{dbe}f^{eac} &= f^{abe}f^{ecd} {~~~} \mbox{(pure adjoint)} \nonumber \\
T^{a}_{i \bar\jmath}T^{b}_{j \bar k}-T^{b}_{i \bar \jmath}T^{a}_{j \bar k} &= f^{abc}T^{c}_{i \bar k} {~~~} \mbox{(mixed adjoint-fundamental)} 
\end{align}
with sums over repeated indices.  We refer to both relations as Jacobi-like.

Once color-weights for an amplitude are expressed in a minimal basis via antisymmetry and Jacobi-like relations, any coefficient of the remaining color-weights must be gauge-invariant, as it can not cancel against any other contribution to the amplitude.  Such coefficients are called color-ordered or color-stripped amplitudes and generically consist of sums over graphs contributing to that color-order.  In the most severe case, such as all-gluon or one fermion pair, an exponential number of graphs can contribute to a color-ordered amplitude. This stands in contrast to the factorial number of graphs that can contribute to the full color-dressed amplitude. 

\begin{figure}[H]
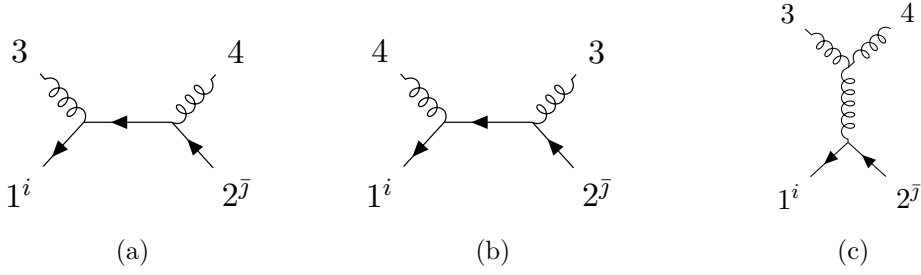

	\centering
\begin{subfigure}{0.3\textwidth}
	\centering \setcounter{nomfigcpt}{4}
\numberedfig[width=1.4in]
\caption{}
\end{subfigure}
\begin{subfigure}{0.3\textwidth}
\centering
\numberedfig[width=1.4in]

\caption{}
\end{subfigure}
\begin{subfigure}{0.3\textwidth}
	\centering
\numberedfig[width=.85in]
	\caption{}
\end{subfigure}
\caption{Feynman diagrams (fixed) for one fermionic pair and two gluons. }
\label{fig:TreeOnePairTwoGlue}
\end{figure}

In recent years there has been a renewed interest in finding independent color basis for all multiplicity fundamental representations at tree and loop level~\cite{Melia:2013bta,Melia:2013epa,Johansson:2015oia,Kalin:2017oqr,Ochirov:2019mtf}.
Here, on a case by case basis, we will find a relevant minimal basis by simply reducing the linear equations relating redundant color factors as they arise.  
For example, in scattering of one distinct fermion pair with two gluons, three cubic graphs contribute, as depicted in Fig.~\ref{fig:TreeOnePairTwoGlue}.

The color factors for these graphs are:
\begin{equation}\label{eq:ExplicitColorFactorTwoGlueTwoFermion}
\begin{split}
&c(1_f,3_A,4_A,2_{\bar{f}})=T^{a}_{i\bar{x}}T^{b}_{x\bar{\jmath}}\,,\\
&c(1_f,4_A,3_A,2_{\bar{f}}) =T^{b}_{i\bar{x}}T^{a}_{x\bar{\jmath}}\,,\\
&c(2_{\bar{f}},1_f,3_A,4_A) =f^{abc}T^{c}_{i\bar{\jmath}}\,.
\end{split}
\end{equation}
The total amplitude is given by:
\begin{equation}\label{eq:TotalAmplitudesOnePairTwoGlue}
\begin{split}
-i\mathcal{A} &= \frac{c[1_f,3_A,4_A,2_{\bar{f}}] n(1_f,3_A,4_A,2_{\bar{f}})}{(k_1 + k_3)^2-m^2} + \frac{c[1_f,4_A,3_A,2_{\bar{f}}] n(1_f,4_A,3_A,2_{\bar{f}})}{(k_1 + k_4)^2-m^2}\\
&+\frac{c[2_{\bar{f}},1_f,3_A,4_A] n(2_{\bar{f}},1_f,3_A,4_A)}{(k_1 + k_2)^2}\,.
\end{split}
\end{equation}
Using the Jacobi identity, 
\begin{equation}
    \label{toyColorJacobiEqn}
    c(2_{\bar{f}},1_f,3_A,4_A) = c(1_f,3_A,4_A,2_{\bar{f}})  - c(1_f,4_A,3_A,2_{\bar{f}}),
\end{equation}
we can rewrite  the amplitude in terms of a minimal color basis. This yields the full amplitude written as,
\begin{equation}
	\begin{split}
		-i\mathcal{A} &= c(1_f,3_A,4_A,2_{\bar{f}}) \left(\frac{n(1_f,3_A,4_A,2_{\bar{f}})}{(k_1 + k_3)^2-m^2}+\frac {n(2_{\bar{f}},1_f,3_A,4_A)}{(k_1 + k_2)^2} \right)\\
	&+c(1_f,4_A,3_A,2_{\bar{f}}) \left(\frac{n(1_f,4_A,3_A,2_{\bar{f}})}{(k_1 + k_4)^2-m^2}-\frac {n(2_{\bar{f}},1_f,3_A,4_A)}{(k_1 + k_2)^2} \right)\,.
	\end{split}
\end{equation}
The coefficient of each independent color factor must be gauge-invariant and correspond to ordered amplitudes:
\begin{equation}\label{eq:OrderedAmplitudesOnePairTwoGlue}
	\begin{split}
		&A[1_f,3_A,4_A,2_{\bar{f}}] =\frac{n(1_f,3_A,4_A,2_{\bar{f}})}{(k_1 + k_3)^2-m^2}+\frac {n(2_{\bar{f}},1_f,3_A,4_A)}{(k_1 + k_2)^2}\,,\\ 
		&A[1_f,4_A,3_A,2_{\bar{f}}] = \frac{n(1_f,4_A,3_A,2_{\bar{f}})}{(k_1 + k_4)^2-m^2}-\frac {n(2_{\bar{f}},1_f,3_A,4_A)}{(k_1 + k_2)^2}\,.
	\end{split}
\end{equation}

The duality between color and kinematics is satisfied if we can find a representation of kinematic weights which satisfy the same Jacobi-like relations as the color weights, 
\begin{equation}
    n_i +n_j +n_k=0 \text{ for graphs } \{i,j,k\} \text{ such that } c_i + c_j + c_k = 0 \,.
\end{equation}
 With such color-dual kinematic representations we can replace the color factors in \eqn{eq:genericTreeQCDAmplitude} with concordant kinematics factors to build a gravitational amplitude:
\begin{equation}\label{eq:genericGRtreeAmplitudes}
\mathcal{A}^{\text{GR-tree}}_{m,k} = i\left(\frac{\kappa}{2}\right)^{m-2} \sum^{\nu(k,m)}_{i} \frac{\tilde{n}_{(i)} n_{(i)}}{d_{(i)}}\,.
\end{equation}
This procedure is called the double-copy construction of gravitational amplitudes.

Similarly, there is a double-copy procedure for loop amplitudes at the integrand level~\cite{Bern:2010ue}. The general form of $m$-point $L$-loop scattering amplitudes with $k$ distinguished quark-antiquark pairs in D-space-time dimensions in QCD is given,
\begin{equation}\label{eq:genericLoopQCDAmplitudes}
\mathcal{A}^{QCD}_{L,m,k} = i^{L}g^{m-2+2L}\sum_{i} \int \prod^{L}_{i=l} \frac{d^D p_l}{(2\pi)^L}\frac{1}{S_i}\frac{c_{(i)} n_{(i)}}{d_{(i)}}\,,
\end{equation}
where the sum is over all cubic $m$-point $L$-loop graphs and the integral is over the independent loop momentum $p_l$.  The (multi)-loop case requires $S_i$ for the automorphic symmetric factor of the diagram under redundant relabelings of either external or loop momenta. When kinematic weights are found that satisfy the duality between color and kinematics on all cuts, we can build the gravitational amplitudes at loop level by replacing the color factors in the above equation by their dual kinematics factors:
\begin{equation}\label{eq:genericLoopGRAmplitudes}
\mathcal{A}^{GR}_{L,m,k} = i^{L+1}\left(\frac{k}{2}\right)^{m-2+2L}\sum_{i} \int \prod^{L}_{i=l} \frac{d^D p_l}{(2\pi)^L}\frac{1}{S_i}\frac{\tilde{n}_{(i)} n_{(i)}}{d_{(i)}}
\end{equation}
In this paper, we focus on finding the color-dual representation of numerators at tree and one-loop level when pairs of fermions are present.  We find representations that satisfy the duality between color and kinematics for all off-shell values of internal legs, and therefore satisfy all physical cuts.

\subsection{Unitarity cuts}\label{sec:UnitarityReview}
We include here a lightning review of relevant on-shell methods, but for pedagogic lecture notes on graph-organized cut verification and construction see e.g. ref~\cite{Carrasco:2015iwa} as well as a more specialized review with detailed discussion of unitarity methods in for perturbative QCD~\cite{Bern:2007dw}.  

Consider the verification problem for loop-level amplitudes at the integrand level.  In other words, how can one verify that an integrand is correct for an intended theory?  It is necessary that any integrand must integrate to the same expression that one would have arrived at from Feynman rules for a given theory.  A sufficient condition\footnote{This condition is not strictly necessary as one can imagine including terms that violate individual cuts but vanish upon integration -- so still yielding the correct amplitude.} is that all on-shell unitarity cuts are satisfied.   Unitarity cuts place on-shell the internal lines of sets of graphs contributing to an integrand.  The resulting cut integrand must match the sum over states across cut legs of the product of corresponding tree-amplitudes, where only physical states are allowed to cross the cut.   One can invert sufficient verifications to constructively build integrand representations by constraining a spanning ansatz to satisfy all unitarity cuts.  One can think of this as a way to systematically guarantee that no relevant off-shell information contained in Feynman rules has been discarded from an integrand, while only evaluating more compact on-shell quantities. Any free parameters after the constraints of all physical cuts must vanish for any physical observable. This constructive approach is efficiently organized in terms of graphs.  When organized and applied in terms of operations on graphs, one finds the same methodology is applied towards considering factorization consistency at tree-level. So here we do not distinguish between tree-level factorization and unitarity methods, describing the former in terms of the latter.

\captionsetup[subfigure]{labelformat=empty}
\begin{figure}[H]
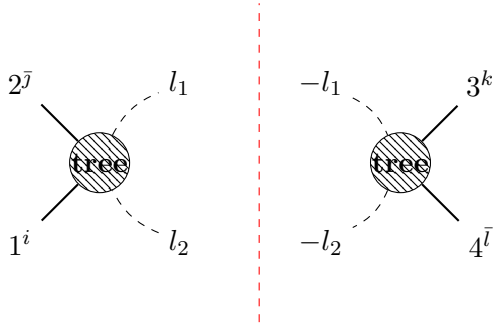

	\centering
          \numberedfig
	\caption{In the two-particle unitarity cut of the four-point  one-loop amplitude, legs $l_1$ and $l_2$ are taken on-shell.  The resulting dressed expression (summed over all graphs that can contribute) should be equal to the product of tree-amplitudes summed over all physical states that can cross the cut. The vertical dashed line simply indicates cutting the internal legs -- i.e. placing them on shell.}
	\label{fig:UnitarityCuts4pointOneLoop}
\end{figure}

\begin{figure}[H]
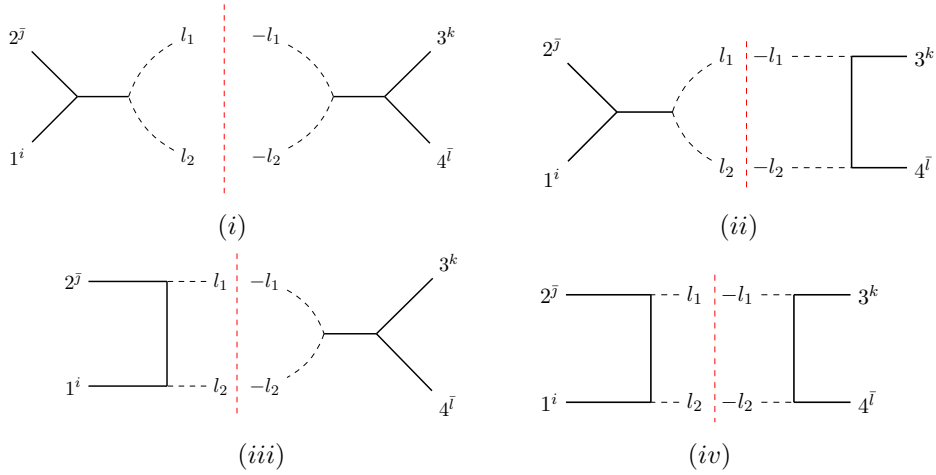

	\centering
			\begin{subfigure}{0.45\textwidth}
		\centering
	\resizebox {.9\textwidth} {!} {	
 \numberedfig  	
}
	\caption{\((i)\)}
	\end{subfigure}
	\begin{subfigure}{0.4\textwidth}
		\centering
	\resizebox {.9\textwidth} {!} {	
	\numberedfig
}
	\caption{\((ii)\)}
	\end{subfigure}
\begin{subfigure}{0.4\textwidth}
		\centering
	\resizebox {.9\textwidth} {!} {	
 \numberedfig
}
	\caption{\((iii)\)}
	\end{subfigure}
\begin{subfigure}{0.35\textwidth}
		\centering
	\resizebox {.9\textwidth} {!} {	
   \numberedfig	
}
	\caption{\((iv)\)}
	\end{subfigure}

		\caption{Sewing different cubic graphs which contribute to the tree level ordered amplitudes, $A^{\text{tree}}(a,b,l_1,l_2)$ and $A^{\text{tree}}(-l_2,-l_1,c,d)$}
	\label{fig:SewingTopologies}
\end{figure}

There are technicalities  that can simplify both verification and construction with gauge theory amplitudes.  A primary one is the advantage of considering color-ordered cuts.  As color-ordered tree-amplitudes depend on at most an exponential number of graphs (compared to a combinatorial number), a potentially vastly smaller number of graphs contribute to color-ordered cuts rather than color-dressed cuts. As a concrete example we present here a two-particle cut of four-point  one-loop amplitude. This cut can be described as the product between two four-point  tree level amplitudes, as per Fig.~\ref{fig:UnitarityCuts4pointOneLoop}.  If this were a color-dressed cut we would need to consider the contribution of nine graphs to the integrand, but here we need only consider four-graphs.  Why? Unitarity implies that the one-loop ordered integrand evaluated with on-shell conditions for cut-legs be equal to the product of two ordered tree level amplitudes summed over all physical states of the theory:
\begin{equation}\label{eq:UnitarityCuts}
A^{\text{one-loop}}(a,b,c,d) \mid_{\text{2-particle cut}} = \sum_{s_i \in \text{states}} A^{\text{tree}}(a,b,l_1^{s_1},l_2^{s_2})A^{\text{tree}}(-l_2^{\bar{s}_2},-l_1^{\bar{s}_1},c,d)
\end{equation}
Since we are considering representations with only cubic topologies, we need only write down the trivalent  graphs that contribute to each of the ordered amplitudes,
\begin{equation}\label{eq:fusingTreeToLoops}
    \Gamma^{\text{one-loop}} = \Gamma^{\text{tree}}_{L} \otimes  \Gamma^{\text{tree}}_{R} \, ,
\end{equation}
where $\Gamma$ are cubic graphs. If all of the channels at tree-level contribute, there are two graphs for each of the ordered trees, and so their outer product generates four distinct one-loop graphs dressed as per the integrand on the left hand side of \eqn{eq:UnitarityCuts}. We show these graphs in Fig.~\ref{fig:SewingTopologies}. 

It is necesary that any kinematic weights of the integrand this must satisfy the constraints of this cut up to terms that must vanish upon integration. Here we insist on the sufficient condition that the integrand simply satisfies cut constraints, which informs the parameters of our ansatze.
Once a spanning set of constraints have been consistently imposed on an integrand ansatz, any unconstrained parameters can not contribute to any physical observables.  \\

Regarding the sum over physical states that can cross a cut, we now review the identities and completeness relations used to simplify the right hand side of \eqn{eq:UnitarityCuts}. In this paper, we impose on-shell cuts on both fermionic and gluonic propagators.  We exploit the following completeness relations for gluons,
\begin{equation}\label{eq:StateSumGluon}
\sum_{s\in \rm{pols}} \epsilon^{\mu,s}(k)  \epsilon^{\nu,\bar{s}}(-k) = \eta^{\mu \nu} - \frac{k^{\mu} q^{\nu} + q^{\mu} p^{\nu}}{k\cdot q}\,,
\end{equation}
and fermionic state sums:
\begin{equation}\label{eq:StateSumFermion}
\sum_{s\in\rm{states}} v(-l,s) \bar{u}(l,\bar{s}) = \slashed{l} +m\,.
\end{equation}
where $q$ is an arbitrary reference null-vector. The freedom in $q$ corresponds to gauge-choice, and as such any gauge-independent observable like an ordered amplitude must ultimately be independent of $q$. When the sewing of trees closes a fermion loop this will result in a trace over the fermionic indices associated with that loop. While much of the literature for unitarity methods goes hand in hand with spinor-helicity methods for compact expressions in fixed dimensions, $D$-dimensional unitarity methods for massive particles has a relatively venerable legacy, e.g.~\cite{Bern:1995db}.

In any calculations with fermions contributing via a closed loop we can employ trace identities to canonicalize the resulting expressions.  These critically depend on dimensions and the nature of the fermions. In even dimensions, for Dirac fermions, the trace of odd number of gamma matrices is zero, and for an even number of gamma matrices we have the following recursive relation:
\begin{equation}\label{eq:RecursiveRelationTraceGammaMatrices}
Tr (\gamma^{\mu_1},\gamma^{\mu_2},...,\gamma^{\mu_n}) = \sum_{k=2} ^{n} (-1)^k \eta^{\mu_1 \mu_k} Tr (\gamma^{\mu_1},...,\gamma^{\mu_{k-1}},\gamma^{\mu_{k+1}},...,\gamma^{\mu_n})\,.
\end{equation}
Some familiar examples are for $n=2$ and $n=4$: 
\begin{equation}\label{eq:ExamplesTraceGamma}
\begin{split}
    Tr (\gamma^{\mu} \gamma^{\nu}) &= 2^{D/2} \times \eta^{\mu \nu} \\
Tr (\gamma^{\mu} \gamma^{\nu} \gamma^{\rho} \gamma^{\sigma}) &=2^{D/2} \times (\eta^{\mu \nu} \eta^{\rho \sigma} -\eta^{\mu \rho}\eta^{\nu \sigma} + \eta^{\mu \sigma}\eta^{\nu \rho})
\end{split}
\end{equation} 
One can use analogous relations for $n=6$ and $n=8$ to reduce expressions in our one-loop calculations. As fermion traces dressing graphs are only matched against fermion traces in cuts, the contribution of fermion traces can be left formal but for ease of comparison with standard results we evaluate in even dimensions using the identities above.

A known difficulty for unitarity methods is to access any information that is only contained in subtle or difficult to reach cuts such as forward limits -- e.g.~a tree sewn with itself which will ofen need to be regulated. This is relevant at one-loop for tadpole diagrams or snail-like corrections to external legs. We will not concern ourself with such contributions in this paper, but point the interested reader to ref.~\cite{Bern:1995db} which describes how such diagrams seperate from $D$-dimensional constructable contributions, and consideration of known UV and IR behavior can be used to constrain any such resulting ambiguity. 

\section{Functional fermionic ansatze and conventions}\label{sec:FermionicAnsatze}
In bosonic theories such as pure gluonic amplitudes or scalar non-linear sigma models, one can construct an ansatz solely from Lorentz invariants depending only on momenta and any required polarizations required for little-group scaling. The inclusion of fermions necessitates the use of  spinor bilinears such as $\bar{u}_{1}\gamma^{\mu}v_{2}$ in our ansatze as well. 

Beside conservation of energy and transverse property of polarization vectors (i.e $k_i . \epsilon_{(i)} = 0$) we require some identities between Gamma matrices, in $D$ dimensions, to reduce to a linearly minimal basis of dimensionally appropriate terms.
\begin{align}\label{eq:GammaMatriceIdentities}
\gamma^{\mu} \gamma_{\mu} &= D \,,\\ 
\gamma^{\mu} \gamma^{\nu} \gamma_{\mu} &= -(D-2)\gamma^{\nu} \,,\\ 
\gamma^{\mu} \gamma^{\nu} \gamma^{\rho} \gamma_{\mu} &=4\eta^{\nu \rho} -(4-D)\gamma^{\nu} \gamma^{\rho} \,,\\ 
\gamma^{\mu} \gamma^{\nu} \gamma^{\rho} \gamma^{\sigma} \gamma_{\mu} &= -2\gamma^{\sigma} \gamma^{\rho} \gamma^{\nu} +(4-D) \gamma^{\nu} \gamma^{\rho} \gamma^{\sigma} \,,\\
\gamma^{\mu} \gamma^{\nu} + \gamma^{\nu} \gamma^{\mu} &= 2\eta^{\mu \nu}\,.
\end{align}
In addition, we can exploit the Dirac equation to simplify some combinations of spinor and momenta:
\begin{equation}
\bar{u}_1 (\slashed{k}_1 -m) = 0 \,, \qquad (\slashed{k}_2 +m)v_2 = 0 \,.
\end{equation}

In general, our amplitudes will consist of quark-antiquark pairs as well as gluons. For little group scaling, we require each term in our ansatz to have a $\bar{u}_i$ for every external fermion, a ${v}_j$ for every external anti-fermion, and a polarization vector for every external gluon. The maximum number of gamma matrices we need to include  is equal to the sum of the number of fermionic vertices plus the number of fermionic propagators. Then, using independent momenta, we construct an ansatz with the correct mass dimension. If we have expressions for maximally distinct number of fermion flavors, we can always easily build expressions where some of our quark-anti-quark pairs share the same flavor.  As such, we consider each pair as having a different flavor in our amplitudes.  In this paper, we construct a general ansatz for a minimal basis of graphs (under the duality between color and kinematics), then we fix coefficients via any remaining color-dual constraints and cut conditions.

The kinematic numerators for graphs with pairs of quark-antiquark are constructed from spinor bilinears. In general, these take the form:
\begin{equation}\label{eq:GeneralSpinorBilinears}
\bar{u}_1 \gamma^{\mu_1} ...  \gamma^{\mu_n}  v_2
\end{equation}
where we run from anti-particle $2^{\bar{\jmath}}$ to particle $1^i$ along the fermionic line.

We use a convention where we assume the momenta of all particles are outgoing, $\sum_i k_i=0$. Any arrows on graphs are used to label external quarks and antiquarks and the fermionic line connecting them, not momentum flow. As such, we dress the outgoing fermion $1^{i}$ with the $\bar{u}_1$ spinor solution to the Dirac equation, and the outgoing anti-fermion, $2^{\bar{\jmath}}$, with the $v_2$ spinor solution.   We leave fermionic color-indices unbarred, and bar the anti-fermionic color-indices as a notational convenience.

\subsection{Swapping fermions and anti-fermions within a pair}\label{sec:ReorderingFermions}

\begin{figure}[H]
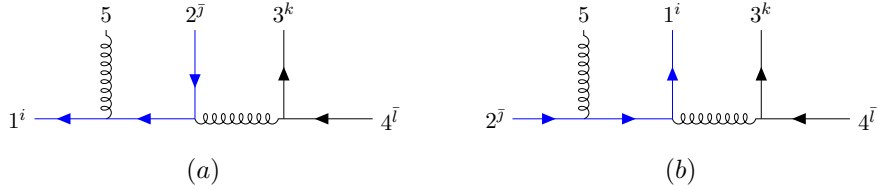

	\centering
	\begin{subfigure}{0.4\textwidth}
		\centering
	\resizebox {.9\textwidth} {!} {	
        \numberedfig	
}
	\caption{\((a)\)}
	
	\end{subfigure}
	\begin{subfigure}{0.4\textwidth}
		\centering
		\resizebox {.9\textwidth} {!} {	
            \numberedfig

		}
		\caption{\((b)\)}
		
	\end{subfigure}
	\caption{Graphs with topologies differing only by the direction of one fermion line. Their kinematics weights not simply related by a trivial relabeling.  Rather we track such book keeping by applying an $\Rev_{1\bar{2}}$ operation to the kinematic weight (see \eqn{eq:RelationReorderedGraphs}).}
	\label{fig:ExamplesForReorderingRelation}
\end{figure}

Consider the two graphs in Fig.~\ref{fig:ExamplesForReorderingRelation}.  These contribute to scattering involving two distinct quark-anti-quark pairs with an external gluon. They have the same topologies except quark-antiquark particles $1^{i}$ and $2^{\bar{\jmath}}$ have been swapped.  Fermion kinematics work differently for particles and anti-particles so in kinematic dressings we can not swap particles with antiparticles by simply exchanging labels.  But of course, assuming underlying Feynman rules (as there are for QCD), the dressings of the two graphs must be functionally related, and we can exploit these relations to give the overall topology a canonical dressing for a canonical ordering of particle-antiparticle in each fermionic line, and then appropriately manipulate that dressing to relabel particles and anti-particles as necessary.

Schematically, the difference in dressing the two diagrams of \Fig{fig:ExamplesForReorderingRelation} is that the path between pair $\{1^i,2^{\bar{\jmath}}\}$ is reversed. So, any relabeling from one to the other will want to reverse the order of Gamma matrices in the bilinear dressing. A subtle difference in this reversing procedure is flipping the direction of propagators. This implies that we need to flip the sign of the mass of the particle anywhere it explicitly appears in the expression. Functionally, the momenta of particle $1^i$ and antiparticle $2^{\bar{\jmath}}$ should be swapped in such dressings.

In general, if two graphs, $(a)$ and $(b)$, only differ by reversing the fermion path between quark $1$ and it's antiquark  $\bar{2}$, as per the graphs in \Fig{fig:ExamplesForReorderingRelation}, we impose the following functional relation between their kinematic weights:
\begin{equation}\label{eq:RelationReorderedGraphs}
n_{(b)} =\Rev \left(n_{(a)} \right)\,.
\end{equation}
The reordering operator $\Rev_{1\overline{2}}$ acts as,
\begin{equation}\label{eq:ReorderingRules}
\Rev_{1\bar{2}}  \left(\ldots\bar{u}_1 \gamma^{\mu_1} \ldots  \gamma^{\mu_n}  v_2\ldots \right) = (-1) \left(\ldots \bar{u}_1 \gamma^{\mu_n} ...  \gamma^{\mu_1}  v_2 \ldots \right)|_{k_1\leftrightarrow k_2, m\rightarrow-m}\,.
\end{equation}
Here $k_1$ ($k_2$) are the momenta of the particle (antiparticle), and $m$ is their mass. 

The most familiar setting to see how this rule plays out may be $e^-e^+\to\gamma\gamma$.  There are two diagrams that contribute.  In our convention, with incoming electron with momenta $-k_1$, incoming positron with momenta $-k_2$, and outgoing photons with momenta $k_3$ and $k_4$,
the $t=(k_1+k_4)^2$ channel would be dressed with standard Feynman rules to have kinematic weight given,
\begin{equation}\label{eq:tChannelNumeratorOnePairTwoGlueFeynman}
n\left( \begin{array}{c}
 \numberedfig
 \end{array}
\right)
\equiv n_t= \bar{u}_1\slashed{\epsilon}_4(\slashed{k}_1 +\slashed{k}_4 +m)\slashed{\epsilon}_3 v_2 \,.
\end{equation}
Traditionally one might read off the $u=(k_1+k_3)^2$-channel then by simply exchanging labels $k_3 \leftrightarrow k_4$,
\begin{equation}\label{eq:uChannelNumeratorOnePairTwoGlueFeynman}
n\left( \begin{array}{c}
 \numberedfig
 \end{array}
\right)
\equiv n_u = \bar{u}_1\slashed{\epsilon}_3(\slashed{k}_1 +\slashed{k}_3 +m)\slashed{\epsilon}_4 v_2
\end{equation}
It is not difficult to see \eqn{eq:uChannelNumeratorOnePairTwoGlueFeynman} and \eqn{eq:tChannelNumeratorOnePairTwoGlueFeynman} are equivalently related by the reordering rule of \eqn{eq:ReorderingRules}:
\begin{align}\label{eq:ReorderingRuleExample}
    n_t&= \Rev_{1\overline{2}}n_u \nonumber \\
       &= - \bar{u}_1\slashed{\epsilon}_4(\slashed{k}_2 +\slashed{k}_3 -m)\slashed{\epsilon}_3 v_2 \nonumber \\
       &=  \bar{u}_1\slashed{\epsilon}_4(\slashed{k}_1 +\slashed{k}_4 +m)\slashed{\epsilon}_3 v_2\,.
\end{align}
In the final line we have simply used conservation of momentum to replace $k_2=-k_1-k_3-k_4$.

When tracking edge order around vertices, as one is required to do for adjoint representations, a quark-antiquark vertex that has order: $\{$anti-fermion, fermion, gluon$\}$ is considered distinct from one that has order $\{$fermion, anti-fermion, gluon$\}$ -- they are related only by a non-cyclic permutation.  As we are free to introduce complementary phases between color and kinematics, we find it convenient to introduce adjunct fundamental generators to allow an antisymmetric dressing of such distinct quark-antiquark vertices in concordance with the antisymmetry of dressing acyclically related adjoint vertices:
\begin{equation}
\label{eq:anti-symmetricColor}
 T^{a}_{\bar\jmath i} \equiv - T^{a}_{i\bar{\jmath}} ~~~\iff~~~f^{cab}=-f^{bac}.
\end{equation}
This means that the kinematics for quark-antiquark-glue, $n(g,f,\bar{f})$ should be dressed with a minus sign relative to the kinematics of an antiquark-quark-glue $n(g,\bar{f},f)$, so that either graph could be used in writing down the full color-dressed amplitude:
\begin{equation}
\mathcal{A}(f^i,\bar{f}^{\bar \jmath},g^a)=T^a_{\bar{\jmath} i} n(g,f,\bar{f}) = T^{a}_{i \bar \jmath} n(g,\bar{f},f)\,.
\end{equation}
So, somewhat trivially, we can see how the sign in our definition of $\Rev$ allows this convenience:
\begin{align}
  \bar{u} \slashed{\epsilon} v &= n(g,\bar{f},f)   \nn \\
                      &=- n(g,f,\bar{f}) \nn \\
                      &= - \Rev_{f \bar{f}} \left(n(g,\bar{f},f) \right) \nn \\
                      &= - (- \bar{u} \slashed{\epsilon} v) \nn \,. 
\end{align}             

A nice example of how functional reordering can be useful involves minimizing the necessary basis graphs at loop-level. This is familiar from adjoint representations of bosons. For example, the following equation depicts a Jacobi relation between box and triangle diagrams,
\begin{equation}\label{eq:GenericJacobiBoxToTriangle}
\vcenter{\hbox{\resizebox {.27\textwidth} {!} {
\numberedfig}}}~-~
\vcenter{\hbox{\resizebox {.27\textwidth} {!} {
\numberedfig}}}~=~
\vcenter{\hbox{\resizebox {.27\textwidth} {!} {
\numberedfig}}}
\end{equation}
 The numerators for the two box diagrams can be related by simply swapping the labels of leg $3$ and $4$. 

 However, for fermions in the fundamental representation, simple relabeling is not an option.  However we can use the reordering operator of \eqn{eq:RelationReorderedGraphs} to relate the box numerators in the following analogous Jacobi-like equation,
\begin{equation}\label{eq:FermionicJacobiBoxToTriangle}
\vcenter{\hbox{\resizebox {.27\textwidth} {!} {
\numberedfig}}}~-~
\vcenter{\hbox{\resizebox {.27\textwidth} {!} {
\numberedfig}}}~=~
\vcenter{\hbox{\resizebox {.27\textwidth} {!} {
\numberedfig}}}
\end{equation}

We can write down the relation between kinematic weights in graphs in \eqn{eq:GenericJacobiBoxToTriangle} and \ref{eq:FermionicJacobiBoxToTriangle} as:
\begin{equation}\label{eq:RelabelingVsReversingJacobies}
\begin{split}
N_{\square} - N_{\square} \mid_{3\leftrightarrow 4} &= N_{\triangle} ~~~(\mbox{for graph relation \ref{eq:GenericJacobiBoxToTriangle}})\\
N_{\square} -\Rev_{3\bar{4}} (N_{\square} )&= N_{\triangle} ~~~(\mbox{for graph relation \ref{eq:FermionicJacobiBoxToTriangle}})\\
\end{split}
\end{equation}

\subsection{Swapping (anti) fermion labels between pairs of the same-flavor}\label{sec:SwappingSameFlaveredFermions}
\label{sect:swappingSameFlavorFermions}
Consider the two diagrams which contribute to Bhabha scattering: 
\be
\mathcal{A}/Q^2=\frac{ n\left(
 \begin{array}{c}
 \numberedfig
 \end{array}
\right)}{(k_1+k_2)^2} + 
\frac{
n\left(
 \begin{array}{c}
 \numberedfig
 \end{array}
\right)}{(k_1+k_4)^2}
\ee
with $Q$ standing in for the normalized charge of the scattering particles.
The only difference between the two graphs is the exchange of external outgoing momenta $k_2\leftrightarrow k_4$ between the same type of external particles. Famously, for Dirac-Fermi statistics we must be able to account for a relative minus sign between contributions to the exchange.  We can do so with the traditional signature around argument labels involving identical particle (same flavor) pairs:
\be
\label{eq:exampleSignature}
n(1_f,2_{\bar{f}},3_{f},4_{\bar{f}}) = \text{sig}(1_f,2_{\bar{f}},3_f,4_{\bar{f}}) \bar{u}_1 \gamma^\mu v_{2} \bar{u}_3 \gamma_\mu v_{4}
\ee
To avoid cluttering expressions we will keep such signatures implicit for the most part, but wish to emphasize for readers the importance of tracking such exchanges.
 
\section{Tree level QCD amplitudes, massless fermions}\label{sec:MasslessTreeQCD}
We begin with massless fermions for simplicity and to better emphasize what small changes occur when considering fermions of generic masses per flavor.
	 
\subsection{Three-point  amplitudes} \label{sec:MasslessTreeQCD.1}
The three-point  two-fermion amplitude is the simplest case.  Since there is no propagator in this case the numerator should have mass dimension one.  The only ansatz we can write down consistent with our rules is as follows,
\begin{equation}\label{eq:ThreePointQCDwithFermions}
	n\left( \begin{array}{c}
 \numberedfig
 \end{array}
	\right)\equiv{~} n(1_f,2_{\bar{f}},3_{A})=\bar{u}_1\slashed{\epsilon_3} v_2\,.
\end{equation}
Note that if we replace the polarization vector with the corresponding momentum, the numerator will vanish via conservation of momentum. This is not surprising because there is only one independent diagram contributing to the amplitude so the kinematic weight must be gauge-invariant by itself.

The on-shell three-point  amplitude for external kinematics $k_1,k_2,k_3$ is thus given by:
\be
\mathcal{A}(1_f^i,2_{\bar{f}}^{\bar{\jmath}},3_A^a)=g \,T^{a}_{\bar{\jmath}i}\,  n(1_f,2_{\bar{f}}, 3_A)
\ee
where we have chosen to set the coupling to $g$.

Bootstrapping the three-point  gluon amplitude via the duality between color and kinematics has been discussed previously, see e.g.~\cite{Bern:2019prr,Carrasco:2020ywq}, so we simply quote here for completeness and easy reference that imposing antisymmetry on a minimal basis of the appropriate mass dimension reproduces exactly what imposing gauge-invariance does on the same ansatz:
\begin{equation}\label{eq:ThreePointQCDwithGlue}
	n\left(
	 \begin{array}{c}
 \numberedfig
 \end{array}
	\right)\equiv{~} n(1_A,2_A,3_{A}) \propto (\epsilon_1 \cdot \epsilon_2) \left( (k_1-k_2)\cdot \epsilon_3 \right) + \text{cyclic}\,,
\end{equation}
with the associated three-point  gluon amplitude simply given as:
\be
\mathcal{A}(1_A^a,2_A^b,3_A^c)= g f^{abc} n(1_A,2_A,3_{A})\,.
\ee
Strictly speaking at this stage the gluon coupling could be allowed to float from the fermion coupling but consistency of factorization of higher point amplitudes fixes them to be identical. 

\subsection{Four-point  amplitudes with two distinct massless fermionic pairs}\label{sec:MasslessTreeQCD.2}	
\begin{figure}[H]
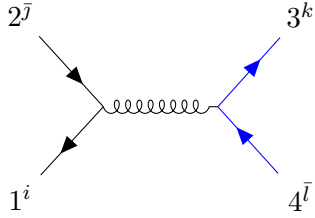


\begin{center}
 \numberedfig
\end{center}
\caption{The only topology that can contribute with two distinct-flavor fermion pairs scattering at four-point  tree-level.}
\label{fig:TreeTwoDistinctFermions}
\end{figure}

In this case, the amplitude should be dimensionless and the numerator must have mass dimension two. We have two distinct fermionic pairs, say $\{\{1_f,2_{\bar{f}}\},\{3_{f},4_{\bar{f}}\}\}$ and thus only one non-vanishing topology, \fig{fig:TreeTwoDistinctFermions}.

Like the three-point  quark-antiquark example, we can only write down one term for the kinematic numerator of this graph consistent with our rules and this mass dimension:
\begin{equation}\label{eq:TwoPairsNumerator}
n(1_f,2_{\bar{f}},3_f, 4_{\bar{f}}) = (\bar{u}_1\gamma_{\mu} v_2)(\bar{u}_3\gamma^{\mu}v_4)
\end{equation}
The ordered and color-dressed amplitudes are simply,
\begin{align}
\label{eq:TwoPairsOrderedAmp}
 A(1_f,2_{\bar{f}},3_f,4_{\bar{f}}) &= \frac{n(1_f,2_{\bar{f}},3_f,4_{\bar{f}})}{s_{12}} \\
\label{eq:TwoPairsColorDressedAmp}
\mathcal{A}(1_f^i,2_{\bar{f}}^{\bar{\jmath}}, 3_{f}^k,4_{\bar{f}}^{\bar{l}})&=g^2\, T^a_{i\bar{\jmath}} T^a_{k\bar{l}}  \, A(1_f,2_{\bar{f}},3_f,4_{\bar{f}})
\end{align}
with $s_{ij}=(k_i+k_j)^2$.

This amplitude respects the only non-trivial ordered cut to three-point  amplitudes:
\begin{equation}\label{eq:CutConditionTwoPairs}
	\sum_{s\in \text{states}}^{}A(1_f,2_{\bar{f}},l_{A,s})A(-l_{A,\bar{s}},3_f,4_{\bar{f}})=\lim_{s_{12}\rightarrow 0}\,s_{12}\,A(1_f,2_{\bar{f}},3_f,4_{\bar{f}})=n(1_f,2_{\bar{f}},3_f,4_{\bar{f}}) \mid_{s_{12} \rightarrow 0}
\end{equation}
The sum over physical gluonic states (physical polarizations) is carried out by use of the gluonic completeness relation mentioned in \eqn{eq:StateSumGluon}. 

It will be useful to see how this works in detail as related cuts will be used continuously throughout this paper.  As ordered amplitudes at three-points are simply the kinematic numerators for the correctly oriented graphs (no propagators), we begin by substituting numerators for ordered amplitudes on the LHS using \eqn{eq:ThreePointQCDwithFermions}
\begin{align}
 \sum_{s\in \text{states}}^{}A(1_f,2_{\bar{f}},l_{A,s})A(-l_{A,\bar{s}},3_f,4_{\bar{f}}) &=  \sum_{s} n(1_f,2_{\bar{f}}, l_{A,s})n(- l_{A,s},3_f,4_{\bar{f}})\\
 &=  \sum_{s}  (\bar{u}_1 \gamma_\mu v_2)  \epsilon_s(l)^\mu \epsilon_{\bar{s}}(-l)^\nu (\bar{u}_3 \gamma_\nu v_4) \\
 &=    (\bar{u}_1 \gamma_\mu v_2)  (\bar{u}_3 \gamma_\nu v_4) \left( \eta^{\mu\nu} - \frac{l^\mu q^\nu+l^\nu q^\mu}{l \cdot q} \right )\\
 &=   (\bar{u}_1 \gamma_\mu v_2)  (\bar{u}_3 \gamma^\mu v_4)\,, 
\end{align}
 where the third line follows from \eqn{eq:StateSumGluon}, and the fourth line follows from $\bar{u}_1 \slashed{l} v_2=0$ and  $\bar{u}_3 \slashed{l} v_4=0$.
We see this is exactly $n(1_f,2_{\bar{f}},3_f,4_{\bar{f}})$ from \eqn{eq:TwoPairsNumerator} so our cut in \eqn{eq:CutConditionTwoPairs} is satisfied.

\begin{figure}[H]
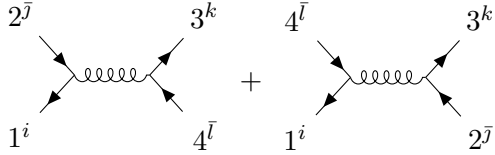


\begin{center}
$\displaystyle
 \begin{array}{c}
 \numberedfig
 \end{array}
+
 \begin{array}{c}
 \numberedfig
 \end{array}
$
\end{center}
\caption{There are two channels contributing to a four-point fermion tree-level amplitude if both pairs are the same flavor of Dirac fermions.}
\label{fig:TreeTwoSameFermions}
\end{figure}

If both pairs of fermions are of the same flavor we still have only one topology but now two channels contribute, see \fig{fig:TreeTwoSameFermions}.  We simply add two channels to the amplitude:
\begin{align}
g^{-2} \mathcal{A}(1_f^i,2_{\bar{f}}^{\bar{\jmath}},3_f^k,4_{\bar{f}}^{\bar{l}}) &= T^a_{i\bar{\jmath}} T^a_{k\bar{l}} \frac{n(1_f,2_{\bar{f}}, 3_f,4_{\bar{f}})}{s_{12}} + 
T^a_{i\bar{l}} T^a_{k\bar{\jmath}} \frac{n(1_f,4_{\bar{f}}, 3_f,2_{\bar{f}})}{s_{14}} 
\end{align}
There would generically be two ordered amplitudes in this case, the ordered amplitude of \eqn{eq:TwoPairsOrderedAmp}, and 
\be
A(1_f,4_{\bar{f}},3_f, 2_{\bar{f}})= \frac{n(1_f,4_{\bar{f}}, 3_f,2_{\bar{f}})}{s_{14}}
\ee
This is fine and holds in $D$-dimensions.

\subsection{An adjoint example of dimensional constraints.}
\label{adjointExample}
Everything has been $D$-dimensional and we have run into no constraints that require us to choose a dimension in order to satisfy the duality between color and kinematics. If it were otherwise we would see functional constraints on the nature of the spinors. This agnostic attitude towards $D$ dimensions will remain the case as long as we remain with Dirac fermions in the fundamental. If we insist on an adjoint representations for Dirac fermions, however, we will see that the duality between color and kinematics starts drawing us towards dimensional constraints on the behavior of the spinors consistent with supersymmetry. To see what a dimensional constraint would look like, it is therefore instructive to consider the same flavor case corresponding to the adjoint. Towards more familiar notation let us identify corresponding adjoint color-indices $a_1=i$, $a_2=\bar{\jmath}$, $a_3=k$, and  $a_4={\bar{l}}$.  This has
\begin{align}
 T^a_{i\bar{l}} T^a_{k\bar{\jmath}}  &\to f^{a_1 a_2 e} f^{e a_3 a_4}=c_s \, \\
 T^a_{i\bar{\jmath}} T^a_{k\bar{l}} &\to f^{a_1 a_4 e} f^{e a_3 a_2} = c_t\,.
\end{align}
The full color-dressed amplitude in this case is:
\begin{align}
  \mathcal{A}= c_s \frac{n(1_f,2_{\bar{f}}, 3_f,4_{\bar{f}})}{s}  + c_t \frac{n(1_f,4_{\bar{f}}, 3_f,2_{\bar{f}})}{t}\,.
\end{align}
Note that this representation is written in the $A(1\sigma 3)$ Kleiss-Kujif basis~\cite{DelDuca:1999rs,Kleiss:1988ne} so we can simply read off from the full amplitude the two adjoint ordered amplitudes by the coefficients of 
$c_{\bar{s}}\equiv f^{12e}f^{e43}=-c_s$ , and $c_{\bar{t}}\equiv f^{14e}f^{e23}=-c_t$, 
\begin{align}
A(1243)&=-\frac{n(1_f,2_{\bar{f}}, 3_f,4_{\bar{f}})}{s_{12}} \,, \\
A(1423)&=-\frac{n(1_f,4_{\bar{f}}, 3_f,2_{\bar{f}})}{s_{14}} .
\end{align}
To rewrite in the more familiar  $A(1\sigma4)$ basis one exploits that $c_t=c_s-c_u$ to write:
\begin{equation}
\mathcal{A}= c_s \left( \frac{n(1_f,2_{\bar{f}}, 3_f,4_{\bar{f}})}{s_{12}}  +\frac{n(1_f,4_{\bar{f}}, 3_f,2_{\bar{f}})}{s_{14}} \right) + c_u \left(-\frac{n(1_f,4_{\bar{f}}, 3_f,2_{\bar{f}})}{s_{14}} \right)
\end{equation}
with ordered amplitudes:
\begin{align}
A(1234)&=\frac{n(1_f,2_{\bar{f}}, 3_f,4_{\bar{f}})}{s_{12}} +\frac{n(1_f,4_{\bar{f}}, 3_f,2_{\bar{f}})}{s_{14}}   \\
A(1324)&=-\frac{n(1_f,4_{\bar{f}}, 3_f,2_{\bar{f}})}{s_{14}}=A(1423) 
\end{align}
Note this satisfies adjoint Kleiss-Kujif identities such as $A(1234)=-A(1243)-A(1423)$, as well as the so called Bern-Carrasco-Johansson relations~\cite{Bern:2008qj,Bern:2010ue} for adjoint amplitudes:
\begin{equation}
 \frac{A(1234)}{s_{13}}=
 \frac{A(1243)}{s_{14}}=
 \frac{A(1423)}{s_{12}}
\end{equation}
This equality follows from:
\begin{align}
n_t &= n_s - n_u\\
n(1_f,4_{\bar{f}}, 3_f,2_{\bar{f}})&=n(1_f,2_{\bar{f}}, 3_f,4_{\bar{f}}) -0 \\
 (\bar{u}_1\gamma_{\mu} v_4)(\bar{u}_3\gamma^{\mu}v_2) &= \left(-1=\frac{\text{sig}(1234)}{\text{sig}(1432)}\right)
 (\bar{u}_1\gamma_{\mu} v_2)(\bar{u}_3\gamma^{\mu}v_4)
 \end{align}
where for clarity we have made explicit the usually implicit same-flavor Fermionic signatures described in \sect{sect:swappingSameFlavorFermions}.  One may recognize the final equality as a familiar four-dimensional Fierz identity satisfied by our all outgoing convention and conservation of helicity.  E.g. for spins $(+\frac{1}{2},-\frac{1}{2},+\frac{1}{2},-\frac{1}{2})$:
\begin{align}
 (\bar{u}_{+1}\gamma_{\mu} v_{-2})(\bar{u}_{+3}\gamma^{\mu}v_{-4}) &= 2 \langle 2 4 \rangle [1 3] \\ 
 (\bar{u}_{+1}\gamma_{\mu} v_{-4})(\bar{u}_{+3}\gamma^{\mu}v_{-2}) &= 2 \langle 4 2 \rangle [1 3]  = - 2 \langle 2 4 \rangle [1 3]
\end{align}
We see the emergence of dimensional constraints from imposing the duality between color and kinematics in the adjoint. The only way color and kinematics can be satisfied for adjoint Dirac fermions is if we are in a dimension where the Fierz identity holds. 

We gain more freedom if we allow for Majorana fermions. In this case we admit an additonal channel, with potentially non-vanishing $n_u$, kklaxing the color-dual kinematic Jacobi to be:
\begin{equation}
0=  (\bar{u}_1\gamma_{\mu} v_2)(\bar{u}_3\gamma^{\mu}v_4) + (\bar{u}_1\gamma_{\mu} v_4)(\bar{u}_3\gamma^{\mu}v_2)  + (\bar{u}_1\gamma_{\mu} v_3)(\bar{u}_2\gamma^{\mu}v_4)\,.  
\end{equation}
This, and the analogous condition for pseudo-Majarona spinors, can only be satisfied in dimensions $3,4,6$ and $10$ -- those that admit supersymmetry~\cite{Chiodaroli:2013upa}.

It is gratifying to see that graphs can carry the same color-dual kinematic dressings independent of whether the fermions transform in the adjoint or the fundamental -- what changes are simply what graphs are allowed to contribute.  This speaks to a suggestive universality of minimal kinematic building blocks -- at least at tree-level.  

\subsection{Four-point  amplitudes with two external gluons and one massless fermionic pair}\label{sec:MasslessTreeQCD.3}
In the case of only one fermionic pair at tree-level four-point scattering, just as in pure gluonic scattering, we have three non-vanishing diagrams, Fig.~\ref{fig:TreeOnePairTwoGlue}. We distinguish between two different topologies.  The first, with a fermionic propagator, is used for the $s=s_{12}$ and $u=s_{13}$ channels, and the second, with a gluonic propagator, which encodes the $t=s_{14}$ channel. The duality between color and kinematics relates these topologies to each other,
\begin{equation}
  \label{eq:JacobiRelationOnePairTwoGlue}
	\begin{split}
	&c_s = T^a_{i \bar{\jmath}} T^b_{j \bar{k}}{~~~},{~~~}c_u =T^b_{i \bar{\jmath}} T^a_{j \bar{k}}{~~~},{~~~}c_t = f^{abc}T^c_{i \bar{k}}\\
	&c_s-c_u-c_t=0{~} \Leftrightarrow{~} n_s-n_u-n_t=0 
	\end{split}
\end{equation}
We therefore need only give the fermionic propagator graph a kinematic ansatz.  Here are all independent possible terms consistent with our general rules:
\begin{align}
    \text{basis} &=	\{ (k_1 \cdot \epsilon_3) \bar{u}_1 \slashed{\epsilon}_4  v_2,{~}(k_4\cdot \epsilon_3) \bar{u}_1 \slashed{\epsilon}_4  v_2,{~}(k_1 \cdot \epsilon_4 )\bar{u}_1 \slashed{\epsilon}_3  v_2,\nonumber\\
                 &\qquad {~}(k_3 \cdot \epsilon_4 )\bar{u}_1 \slashed{\epsilon}_3  v_2, 
    {~}(\epsilon_3 \cdot \epsilon_4)\bar{u}_1 \slashed{k}_3  v_2,{~}\bar{u}_1 \slashed{\epsilon}_3\slashed{k}_3\slashed{\epsilon}_4  v_2\}
\end{align} 
Our ansatz for the kinematic weight of the basis graph with a fermionic propagator is therefore given as,
\begin{equation}\label{eq:AnsatzOnePairTwoGlue}
\begin{split}
n(1_f,3_A,4_A,2_{\bar{f}}) &= a_1 (k_1 \cdot \epsilon_3) \bar{u}_1 \slashed{\epsilon}_4  v_2+a_2 ( k_4 \cdot \epsilon_3) \bar{u}_1 \slashed{\epsilon}_4  v_2+a_3 (k_1\cdot \epsilon_4) \bar{u}_1 \slashed{\epsilon}_3  v_2\\
&+a_4 (k_3\cdot\epsilon_4) \bar{u}_1 \slashed{\epsilon}_3  v_2+a_5 (\epsilon_3\cdot\epsilon_4)\bar{u}_1 \slashed{k}_3  v_2+ a_6\, \bar{u}_1 \slashed{\epsilon}_3\slashed{k}_3\slashed{\epsilon}_4  v_2 
\end{split}
\end{equation}
where the $a_i$ are free parameters to be constrained by color-dual identities and factorization.

The quickest way to fix this ansatz is by considering the $s_{13}$ cut of the ordered amplitude\footnote{These ordered amplitudes have already been given in terms of dressed cubic graphs as an example in the previous section \eqn{eq:OrderedAmplitudesOnePairTwoGlue}.} $A(1_f,3_A,4_A,2_{\bar{f}})$\,
\begin{equation}
\lim_{{s_{13} \rightarrow 0}} s_{13}A(1_f,3_A,4_A,2_{\bar{f}}) =n(1_f,3_A,4_A,2_{\bar{f}}) \mid_{s_{13}\rightarrow 0}\,.
\end{equation}
This must be equal to sewing two three-point  trees on-shell together as follows,
\begin{equation}\label{eq:CutTwoGlueOnePair}
\sum_{s\in \text{states}} A(1_f,3_A,l_s) A(-\bar{l}_{\bar{s}},4_A,2_{\bar{f}}) = \bar{u}_{1} \slashed{\epsilon}_3 (\slashed{k_1}+\slashed{k_3})\slashed{\epsilon}_{4} v_2\,.
\end{equation}
In the last equality we employ the spinor completeness relation of \eqn{eq:StateSumFermion}.  This fixes all coefficients in our ansatz yielding both
\begin{equation}\label{eq:tChannelNumeratorOnePairTwoGlue}
n\left(  \begin{array}{c}
 \numberedfig
 \end{array}
\right)
\equiv n(1_f,3_A,4_A,2_{\bar{f}})= 2 (k_1\cdot \epsilon_3) \bar{u}_1\slashed{\epsilon}_4 v_{2}+\bar{u}_1\slashed{\epsilon}_3\slashed{k}_3\slashed{\epsilon}_4 v_2
\end{equation}
and with the swap of $3\leftrightarrow 4$,
\begin{equation}\label{eq:uChannelNumeratorOnePairTwoGlue}
	n\left(  \begin{array}{c}
 \numberedfig
 \end{array}
	\right)
\equiv n(1_f,4_A,3_A,2_{\bar{f}})= 2 (k_1 \cdot\epsilon_4) \bar{u}_1\slashed{\epsilon}_3 v_2+\bar{u}_1\slashed{\epsilon}_4\slashed{k}_4\slashed{\epsilon}_3 v_2\,.
\end{equation}

We can get the numerator for the final graph, (c) of \Fig{fig:TreeOnePairTwoGlue}, from the Jacobi-like relation:
\begin{align}\label{eq:sChannelNumeratorOnePairTwoGlue}
	n(2_{\bar{f}},1_f,3_A,4_A)&=n(1_f,3_A,4_A,2_{\bar{f}})-n(1_f,4_A,3_A,2_{\bar{f}}) \nn\\
	&= 2 ((k_3\cdot\epsilon_4)\bar{u}_1\slashed{\epsilon}_3 v_2
	-(k_4 \cdot \epsilon_3)\bar{u}_1\slashed{\epsilon}_4 v_2- (\epsilon_3\cdot\epsilon_4)\bar{u}_1\slashed{k}_3 v_2)
\end{align}

Note we could have gotten the dressing for this last topology directly by constraining an ansatz given to that graph.  It is perhaps pedagogically useful to see how we arrive at this very same dressing, for this graph, by consideration of this graph's properties alone. We could e.g.~start by assigning a minimal ansatz to this numerator:
\begin{equation}\label{eq:sChannelAnsatzOnePairTwoglue}
	\begin{split}
		n(2_{\bar{f}},1_f,3_A,4_A) &= a_1 (k_1 \cdot\epsilon_3) \bar{u}_1 \slashed{\epsilon}_4  v_2+a_2 (k_4 \cdot \epsilon_3) \bar{u}_1 \slashed{\epsilon}_4  v_2+ a_3 (k_1 \cdot\epsilon_4) \bar{u}_1 \slashed{\epsilon}_3  v_2\\
		&+a_4 (k_3 \cdot \epsilon_4) \bar{u}_1 \slashed{\epsilon}_3  v_2+a_5 ( \epsilon_3 \cdot \epsilon_4) \bar{u}_1 \slashed{k}_3  v_2+ a_6 \, \bar{u}_1 \slashed{\epsilon}_3\slashed{k}_3\slashed{\epsilon}_4  v_2 
	\end{split}
\end{equation}
\\
The duality between color and kinematics requires that the numerator to be anti-symmetric under swapping two adjacent gluons:
\begin{equation*}
	\begin{split}
		n(2_{\bar{f}},1_f,4_A,3_A)=-n(2_{\bar{f}},1_f,3_A,4_A) &= (a_1-2 a_6) (k_1 \cdot\epsilon_3) \bar{u}_1 \slashed{\epsilon}_4  v_2+(a_2-2 a_6) (k_4 \cdot \epsilon_3) \bar{u}_1 \slashed{\epsilon}_4  v_2\\
		&+(a_3-2 a_6) (k_1 \cdot\epsilon_4) \bar{u}_1 \slashed{\epsilon}_3  v_2+(a_4+2a_6) ( k_3 \cdot \epsilon_4 )\bar{u}_1 \slashed{\epsilon}_3  v_2\\
		&+(a_5+2a_6) ( \epsilon_3 \cdot \epsilon_4)\bar{u}_1 \slashed{k}_3  v_2+ a_6 \bar{u}_1 \slashed{\epsilon}_3\slashed{k}_3\slashed{\epsilon}_4  v_2 
	\end{split}
\end{equation*}
\\
This imposes $a_6=0$, $a_1=-a_3$ and $a_2=-a_4$.  The next constraint comes from flipping the fermions.  Recalling the adjunct antisymmetry introduced in the color-weights and complementary phase we have:\\
\begin{align*}
n(1_f,2_{\bar{f}},3_A,4_A) &= - \left[ n(2_{\bar{f}},1_f,3_A,4_A) \equiv \Rev_{1\bar{2}} \left( n(1_f,2_{\bar{f}},3_A,4_A) \right) \right] \\
&= a_1 \left( (k_2\cdot\epsilon_3)\bar{u}_1\slashed{\epsilon}_4 v_2-(k_2\cdot\epsilon_4)\bar{u}_1\slashed{\epsilon}_3 v_2)+ a_2( (k_3\cdot\epsilon_4)\bar{u}_1\slashed{\epsilon}_3 v_2- (k_4\cdot\epsilon_3)\bar{u}_1\slashed{\epsilon}_4 v_2 \right)\\
&~~+ a_5 (\epsilon_3\cdot\epsilon_4) \bar{u}_1\slashed{k}_3 v_2
\end{align*}
\\
The above equation implies $a_1=0$. Then we impose the gauge invariance on the maximal cut targeting this graph (so $s_{34}$ must vanish):
\begin{equation*}
	n(2_{\bar{f}},1_f,3_A,4_A) \mid_{s_{34}\rightarrow 0, \epsilon_3\rightarrow k_3}=0
\end{equation*}
\\
This implies $a_2=-a_5$. So, for this graph, considering its properties alone,  we  have recovered  \eqn{eq:sChannelNumeratorOnePairTwoGlue} derived from color-dual kinematic relations.  

\subsection{Five-point  amplitudes with one gluon and two massless fermionic pairs}\label{sec:MasslessTreeQCD.4}
\begin{figure}[H]
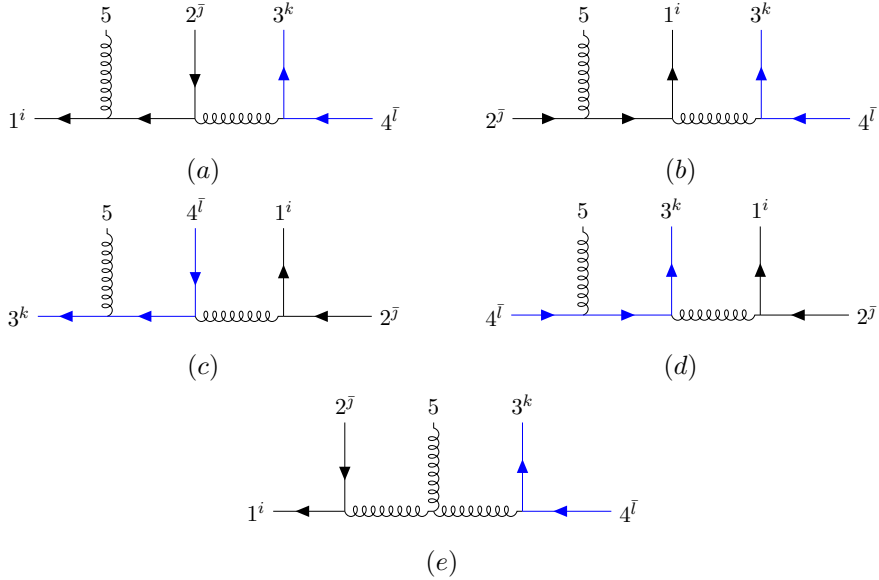

	\centering
	\begin{subfigure}{0.4\textwidth}
		\centering
	\resizebox {.9\textwidth} {!} {	
\numberedfig	
}
	\caption{\((a)\)}
	
	\end{subfigure}
	\begin{subfigure}{0.4\textwidth}
		\centering
		\resizebox {.9\textwidth} {!} {	
		\numberedfig	
		}
		\caption{\((b)\)}
		
	\end{subfigure}
	\begin{subfigure}{0.4\textwidth}
		\centering
			\resizebox {.9\textwidth} {!} {	
			\numberedfig
		}
		\caption{\((c)\)}
		
	\end{subfigure}
\begin{subfigure}{0.4\textwidth}
	\centering
	\resizebox {.9\textwidth} {!} {	
		\numberedfig
	}
	\caption{\((d)\)}
	\end{subfigure}
\begin{subfigure}{0.4\textwidth}
	\centering
	\resizebox {.9\textwidth} {!} {	
		\numberedfig
	}
	\caption{\((e)\)}
	
\end{subfigure}
	\caption{Five distinct topologies for tree-level scattering with two fermionic pairs and one gluon.}
	\label{fig:TreeTwoPairsOneGlue}
\end{figure}

In the case of five points with two distinct fermionic pairs, we have five non-vanishing graphs, Fig.~\ref{fig:TreeTwoPairsOneGlue}. The color factor for each graph is:

\begin{equation}\label{eq:ColorFactorsTwoPairsOneGlue}
\begin{split}
c_{(a)} &= T^a_{i\bar{m} }T^b_{m\bar{\jmath}} T^b_{k \bar{l}}\\
c_{(b)} &= T^a_{\bar{\jmath} m} T^b_{\bar{m}i} T^b_{k \bar{l}}= T^a_{m \bar{\jmath}} T^b_{i\bar{m}} T^b_{k \bar{l}}\\
c_{(c)} &= T^a_{k\bar{m}} T^b_{m\bar{l}} T^b_{i \bar{\jmath}}\\
c_{(d)} &= T^a_{\bar{l} m} T^b_{\bar{m}k} T^b_{i \bar{\jmath}}= T^a_{m \bar{l}} T^b_{k\bar{m}} T^b_{i \bar{\jmath}}\\
c_{(e)} &=   T^c_{i \bar{\jmath}} f^{cab}T^b_{k \bar{l}} 
\end{split}
\end{equation}

The duality between color and kinematics  relates color and kinematic weights for different graphs:

\begin{equation}\label{eq:JacobiTwoPairsOneGlue}
\begin{split}
&c_{(a)} - c_{(b)} = c_{(e)} {~~~} \Leftrightarrow 
n_{(a)} - n_{(b)} = n_{(e)}\\
&c_{(c)} - c_{(d)} = -c_{(e)} {~~~} \Leftrightarrow n_{(c)} - n_{(b)} = -n_{(e)}
\end{split}
\end{equation}
So the full color-dressed amplitude can be written:
\begin{align}
g^{-3} \mathcal{A}&= \frac{c_{(a)} 
n_{(a)}}{s_{15}s_{34}} +  \frac{c_{(b)} n_{(b)}}{s_{25}s_{34}} +  \frac{c_{(c)} n_{(c)}}{s_{35}s_{12}} +  \frac{c_{(d)} n_{(d)}}{s_{45}s_{12}} +
 \frac{c_{(e)} n_{(e)}}{s_{12}s_{34}} \\
 &= c_{(a)} \left(A_{(a)} \equiv \frac{
n_{(a)}}{s_{15}s_{34}} +  \frac{n_{(d)}}{s_{45}s_{12}}+ \frac{ n_{(e)}}{s_{12}s_{34}} \right) + \nn \\
 &~~ c_{(b)} \left(A_{(b)} \equiv \frac{ n_{(b)}}{s_{25}s_{34}}  -  \frac{n_{(d)}}{s_{45}s_{12}}- \frac{ n_{(e)}}{s_{12}s_{34}} \right ) + \nn \\
 &~~ c_{(c)} \left(A_{(c)} \equiv \frac{ n_{(c)}}{s_{35}s_{12}}  + \frac{n_{(d)}}{s_{45}s_{12}} \right )\,
\end{align}
where for the second equality we expressed the full amplitude in terms of a minimal color basis by exploiting $c_{(d)}=c_{(a)}-c_{(b)}+c_{(c)}$ and $c_{(e)}=c_{(a)}-c_{(b)}$, defining ordered (color-stripped) amplitudes of the coefficients of the remaining independent color-weights.

As we discussed in \ref{sec:ReorderingFermions}, graphs $(a)$ and $(b)$ in \Fig{fig:TreeTwoPairsOneGlue} (similarly graphs $(c)$ and $(d)$) are related to each other by reordering rules, \eqn{eq:ReorderingRules}. So, there is just one basis graph we can dress functionally. The numerators should have mass dimension three, since amplitudes in this case have dimension $[mass]^{-1}$, and we have two propagators in each diagram. Each term must have two $\bar{u}_i$ corresponding to outgoing fermions and two $ v_i$ corresponding to outgoing anti-fermions. We also need a polarization vector for the external gluon and a momentum to get the correct dimension.  Our ansatz will have $11$ terms, which can be clustered into three types:
\\
1. The general structure for this type is $(\bar{u}_i \gamma_{\mu} v_j) (\bar{u}_p \gamma^{\mu}  v_q) {k_m\cdot\epsilon_{5}}$. We have three independent terms here:
\begin{equation*}
	\begin{split}
	\{(\bar{u}_1 \gamma_{\mu} v_2) (\bar{u}_3 \gamma^{\mu}  v_4) {(k_1 \cdot\epsilon_5)},~(\bar{u}_1 \gamma_{\mu} v_2) (\bar{u}_3 \gamma^{\mu}  v_4) {(k_2\cdot\epsilon_5)}
	,~(\bar{u}_1 \gamma_{\mu} v_2) (\bar{u}_3 \gamma^{\mu}  v_4) {(k_3\cdot\epsilon_5)}\}
	\end{split}
\end{equation*}
\\
2. The general structure of this type is $(\bar{u}_i \slashed{k}_m  v_j)(\bar{u}_p \slashed{\epsilon}_5  v_q)$ and we have four different terms:
\begin{equation*}
\begin{split}
\{(\bar{u}_1 \slashed{k}_3  v_2)(\bar{u}_3 \slashed{\epsilon}_5  v_4),~(\bar{u}_3 \slashed{k}_1  v_4)(\bar{u}_1 \slashed{\epsilon}_5  v_2),~
(\bar{u}_1 \slashed{k}_4  v_2)(\bar{u}_3 \slashed{\epsilon}_5  v_4),~(\bar{u}_3 \slashed{k}_2  v_4)(\bar{u}_1 \slashed{\epsilon}_5  v_2)\}
\end{split}
\end{equation*}
\\
3. This case is a little bit more complicated than the previous cases since we have four Gamma matrices, $(\bar{u}_i \gamma_{\mu}  v_j)(\bar{u}_p \slashed{k}_m \slashed{\epsilon}_5 \gamma^{\mu}  v_q)$ There are four independent terms:
\begin{equation*}
\begin{split}
&\{(\bar{u}_1 \gamma_{\mu}  v_2)(\bar{u}_3 \slashed{k}_1 \slashed{\epsilon}_5 \gamma^{\mu}  v_4),~(\bar{u}_1 \gamma_{\mu}  v_2)(\bar{u}_3 \slashed{k}_2 \slashed{\epsilon}_5 \gamma^{\mu}  v_4),~\\
&(\bar{u}_3 \gamma_{\mu}  v_4)(\bar{u}_1 \slashed{k}_3 \slashed{\epsilon}_5 \gamma^{\mu}  v_2),~(\bar{u}_3 \gamma_{\mu}  v_4)(\bar{u}_1 \slashed{k}_4 \slashed{\epsilon}_5 \gamma^{\mu}  v_2)\}
\end{split}
\end{equation*} 
\\
First of all, we impose the following functional constraint between numerators $n_{(a)}$ and $n_{(c)}$ by flipping adjacent fermions:
\begin{equation}\label{eq:SymmetryConditionTwoPairsOneGlue}
n(1_f,5_A,2_{\bar{f}},3_f,4_{\bar{f}})  =- \left[ n(1_f,5_A,2_{\bar{f}},4_{\bar{f}},3_f)  \equiv \Rev_{3\bar{4}}  \left( n(1_f,5_A,2_{\bar{f}},3_f,4_{\bar{f}}) \right) \right]
\end{equation}
This constraint fixes 3 coefficients. To fix the remaining coefficients, we simply impose the ordered cut condition:
\begin{align}\label{eq:FirstCutConditionTwoPairsOneGlue}
\sum_{s\in\text{states}}^{} A(1_f,5_A,\bar{l}_{\bar{s}})A(-l_s,2_{\bar{f}},3_f,4_{\bar{f}}) &= \lim_{s_{15} \rightarrow 0} s_{15} A_{(a)}(1_f,5_A,2_{\bar{f}},3_f,4_{\bar{f}})\\
&=\frac{n(1_f,5_A,2_{\bar{f}},3_f,4_{\bar{f}})}{s_{34}}\mid_{s_{15}\rightarrow 0}\,.
\end{align}
With ordered lower point amplitudes from earlier in the boostrap quoted,
\begin{align}
\label{eq:FirstTreeLevelLowerAmplitudesTwoPairsOneGlue}
		A(1_f , 5_A, \bar{l}) & =- A(1_f ,\bar{l},5_A)=-\bar{u}_1\slashed{\epsilon_5} v_{l}\\
		A(-l,2_{\bar{f}},3_f,4_{\bar{f}}) &= \frac{(\bar{u}_{-l}\gamma_{\mu} v_2)(\bar{u}_3\gamma^{\mu} v_4)}{s_{34}}
\end{align}

Using the above equations, the cut condition in \eqn{eq:FirstCutConditionTwoPairsOneGlue} is equivalent to:
\begin{equation}
	-\left. \sum_{s \in \text{states}}\frac{(\bar{u}_1\slashed{\epsilon_5} v_{l,s})(\bar{u}_{-l,\bar{s}}\gamma_{\mu} v_2)(\bar{u}_3\gamma^{\mu} v_4)}{s_{34}}\right|_{l^2=0} = \left.\frac{n(1_f,5_A,2_{\bar{f}},3_f,4_{\bar{f}})}{s_{34}}\right|_{s_{15}\rightarrow 0}	
\end{equation}

We evaluate the state sum using the completeness relation, \eqn{eq:StateSumFermion}, yielding
\begin{equation}
	(\bar{u}_1\slashed{\epsilon_5}(\slashed{k}_1+\slashed{k}_5)\gamma_{\mu} v_2)(\bar{u}_3\gamma^{\mu} v_4)=n(1_f,5_A,2_{\bar{f}},3_f,4_{\bar{f}}) \mid_{s_{15}\rightarrow 0}
\end{equation}

The above constraint fixes all the remaining coefficients. So, the numerator for graphs (a) and (c) is given functionally by:
\begin{multline}\label{eq:NumeratorGraphACTwoPairsOneGlue}
n(1_f,5_A,2_{\bar{f}},3_f,4_{\bar{f}})= 
		(\bar{u}_3 \gamma_{\mu}  v_4)(\bar{u}_1 \slashed{k}_3 \slashed{\epsilon}_5 \gamma^{\mu}  v_2)+(\bar{u}_3 \gamma_{\mu}  v_4)(\bar{u}_1 \slashed{k}_4 \slashed{\epsilon}_5 \gamma^{\mu}  v_2)\\
		+2\left((\bar{u}_1 \gamma_{\mu} v_2) (\bar{u}_3 \gamma^{\mu}  v_4) {(k_1\cdot\epsilon_5)}+(\bar{u}_1 \gamma_{\mu} v_2) (\bar{u}_3 \gamma^{\mu}  v_4) {(k_2\cdot\epsilon_5)}-(\bar{u}_3 \slashed{k}_2  v_4)(\bar{u}_1 \slashed{\epsilon}_5  v_2)\right)
\end{multline}
with
\begin{align}\label{eq:NumeratorGraphATwoPairsOneGlue}
n_{(a)}&=n(1_f,5_A,2_{\bar{f}},3_f,4_{\bar{f}}) \\
\label{eq:NumeratorGraphCTwoPairsOneGlue}
n_{(c)}&=n(3_f,5_A,4_{\bar{f}},1_f,2_{\bar{f}})\,.
\end{align}
The other numerators follow from:
\begin{align}
n_{(b)}&=\Rev_{1\bar{2}}n_{(a)}\,,\\
n_{(d)}&=\Rev_{3\bar{4}}n_{(c)}\,, \\
n_{(e)}&=n_{(a)}-n_{(b)}=n_{(d)}-n_{(c)}\,.
\end{align}

It is straightforward to verify our results against Feynman rules as well as known SYM amplitudes after taking the color-weights to the adjoint and fixing helicities.
\newpage
\begin{figure}[H]
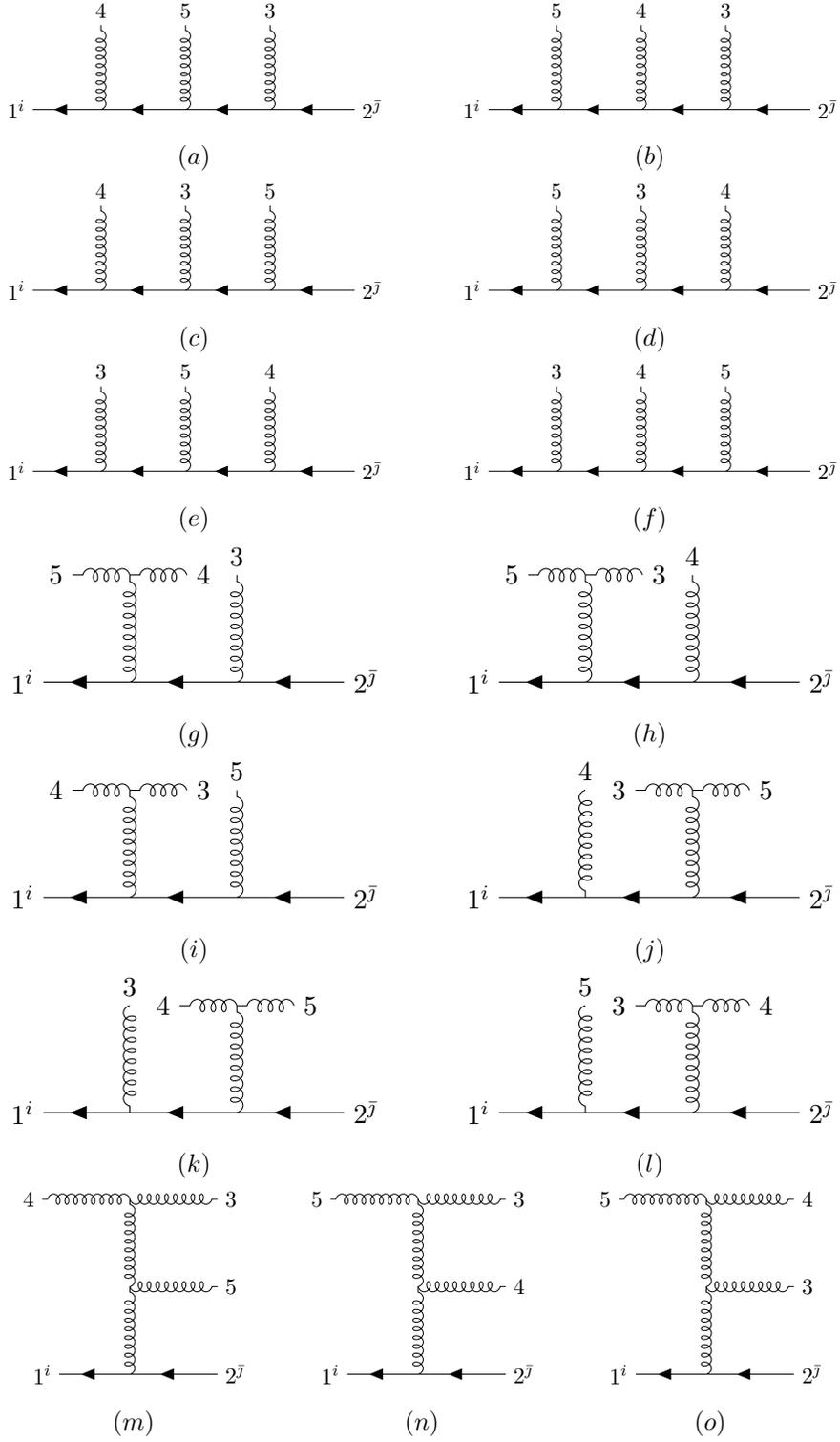

	\centering
	\begin{subfigure}{0.4\textwidth}
		\centering
		\resizebox {.9\textwidth}{!}{	
			\numberedfig
		}
		\caption{\((a)\)}
	\end{subfigure}
	\begin{subfigure}{0.4\textwidth}
	\centering
	\resizebox {.9\textwidth}{!} {	
		\numberedfig
	}
	\caption{\((b)\)}
\end{subfigure}
	\begin{subfigure}{0.4\textwidth}
	\centering
	\resizebox {.9\textwidth} {!} {	
		\numberedfig
	}
	\caption{\((c)\)}
\end{subfigure}
	\begin{subfigure}{0.4\textwidth}
	\centering
	\resizebox {.9\textwidth} {!} {	
		\numberedfig
	}
	\caption{\((d)\)}
\end{subfigure}
\begin{subfigure}{0.4\textwidth}
	\centering
	\resizebox {.9\textwidth} {!} {	
		\numberedfig
	}
	\caption{\((e)\)}
\end{subfigure}
\begin{subfigure}{0.4\textwidth}
	\centering
	\resizebox {.9\textwidth} {!} {	
		\numberedfig
	}
	\caption{\((f)\)}
\end{subfigure}
\begin{subfigure}{0.4\textwidth}
	\centering
	\resizebox {.9\textwidth} {!} {	
		\numberedfig
	}
	\caption{\((g)\)}
\end{subfigure}
\begin{subfigure}{0.4\textwidth}
	\centering
	\resizebox {.9\textwidth} {!} {	
		\numberedfig
	}
	\caption{\((h)\)}
\end{subfigure}
\begin{subfigure}{0.4\textwidth}
	\centering
	\resizebox {.9\textwidth} {!} {	
		\numberedfig
	}
	\caption{\((i)\)}
\end{subfigure}
\begin{subfigure}{0.4\textwidth}
	\centering
	\resizebox {.9\textwidth} {!} {	
		\numberedfig
	}
	\caption{\((j)\)}
\end{subfigure}
\begin{subfigure}{0.4\textwidth}
	\centering
	\resizebox {.9\textwidth} {!} {	
		\numberedfig
	}
	\caption{\((k)\)}
\end{subfigure}
\begin{subfigure}{0.4\textwidth}
	\centering
	\resizebox {.9\textwidth} {!} {	
		\numberedfig
	}
	\caption{\((l)\)}
\end{subfigure}
\begin{subfigure}{0.25\textwidth}
	\centering
	\resizebox {.9\textwidth} {!} {	
		\numberedfig
	}
	\caption{\((m)\)}
\end{subfigure}
\begin{subfigure}{0.25\textwidth}
	\centering
	\resizebox {.9\textwidth} {!} {	
		\numberedfig
	}
	\caption{\((n)\)}
\end{subfigure}
\begin{subfigure}{0.25\textwidth}
	\centering
	\resizebox {.9\textwidth} {!} {	
		\numberedfig
	}
	\caption{\((o)\)}
\end{subfigure}
	\caption{Cubic graphs for one fermionic pair and three gluons.}
	\label{fig:TreeOnePairThreeGlue}
\end{figure}

\subsection{Five-point  amplitudes with one massless fermionic pair and three gluons}\label{sec:MasslessTreeQCD.6}

Similar to the pure gluon amplitudes at five-point , we have fifteen cubic graphs as depicted \Fig{fig:TreeOnePairThreeGlue}. We just need to know the numerator for one of them. By Jacobi-like relations, reordering, and relabeling, we can get all numerators. The color factors of these diagrams are:

\begin{equation}\label{eq:ColorFactors5pointThreeGlueOnePair}
\begin{split}
&c_{(a)} = T^{b}_{i\bar{x}}T^{c}_{x\bar{y}}T^{a}_{y\bar{\jmath}}~~~c_{(b)} = T^{c}_{i\bar{x}}T^{b}_{x\bar{y}}T^{a}_{y\bar{\jmath}}~~~c_{(c)} = T^{b}_{i\bar{x}}T^{a}_{x\bar{y}}T^{c}_{y\bar{\jmath}}\\
&c_{(d)} = T^{c}_{i\bar{x}}T^{a}_{x\bar{y}}T^{b}_{y\bar{\jmath}}~~~c_{(e)} = T^{a}_{i\bar{x}}T^{c}_{x\bar{y}}T^{b}_{y\bar{\jmath}}~~~c_{(f)} = T^{a}_{i\bar{x}}T^{b}_{x\bar{y}}T^{c}_{y\bar{\jmath}}\\
&c_{(g)} = T^{l}_{i\bar{x}}f^{lcb}T^{a}_{x\bar{\jmath}}~~~c_{(h)} = T^{l}_{i\bar{x}}f^{lca}T^{b}_{x\bar{\jmath}}~~~c_{(i)} = T^{l}_{i\bar{x}}f^{lba}T^{c}_{x\bar{\jmath}}\\
&c_{(j)} = T^{b}_{i\bar{x}}f^{lac}T^{l}_{x\bar{\jmath}}~~~c_{(k)} = T^{a}_{i\bar{x}}f^{lbc}T^{l}_{x\bar{\jmath}}~~~c_{(l)} = T^{c}_{i\bar{x}}f^{lab}T^{l}_{x\bar{\jmath}}\\
&c_{(m)} = f^{lba}f^{lcx}T^{x}_{i\bar{\jmath}}~~~c_{(n)} = f^{lca}f^{lbx}T^{x}_{i\bar{\jmath}}~~~c_{(o)} = f^{lcb}f^{lax}T^{x}_{i\bar{\jmath}}\\
\end{split}
\end{equation}

The Jacobi-like relations between these graphs are:

\begin{equation}\label{eq:JacobiOnePairThreeGlue}
	\begin{split}
		&c_{(a)} -c_{(b)} = -c_{(g)}~\leftrightarrow ~
n_{(a)}-n_{(b)}=-n_{(g)}\\
		&c_{(a)} -c_{(c)} =- c_{(j)}~\leftrightarrow ~
n_{(a)}-n_{(c)}=-n_{(j)}\\
		&c_{(e)} -c_{(f)} =- c_{(k)}~\leftrightarrow ~n_{(e)}-n_{(f)}=-n_{(k)} \\
		&c_{(d)}-c_{(e)} = c_{(h)}~\leftrightarrow ~n_{(d)}-n_{(e)} =n_{(h)}\\
		&c_{(b)}-c_{(d)} =-c_{(l)}~\leftrightarrow ~n_{(b)}-n_{(d)}=-n_{(l)}\\
		&c_{(c)}-c_{(f)} = c_{(i)}~ \leftrightarrow ~n_{(c)}-n_{(f)}=n_{(i)}\\
		&c_{(b)}-c_{(c)}-c_{(d)}+c_{(f)} =- c_{(m)} ~\leftrightarrow ~ n_{(b)}-n_{(c)}-n_{(d)}+n_{(f)}=- n_{(m)}\\
		&c_{(a)}-c_{(c)}-c_{(d)}+c_{(e)} = -c_{(n)}~\leftrightarrow ~ n_{(a)}-n_{(c)}-n_{(d)}+n_{(e)} = -n_{(n)}\\
		&c_{(a)}-c_{(b)}+c_{(e)}-c_{(f)} = -c_{(o)} ~\leftrightarrow ~ n_{(a)}-n_{(b)}+n_{(e)}-n_{(f)} = -n_{(o)}
	\end{split}
\end{equation}\\
In the minimal color basis, the full color-dressed amplitude can be written as 

\begin{equation}
\label{eqn:ordAmps3A2F}
\begin{split}
(g^{-3})\mathcal{A} = &c_{(a)} \left( A_{(a)} \equiv \frac{n_{(a)}}{s_{14}s_{23}}+\frac{n_{(b)}}{s_{15}s_{23}}+\frac{n_{(c)}}{s_{14}s_{25}} +\frac{n_{(d)}}{s_{15}s_{24}}+\frac{n_{(e)}}{s_{13}s_{24}}+\frac{n_{(f)}}{s_{13}s_{25}}\right)+\\
 &c_{(j)} \left( A_{(j)} \equiv \frac{n_{(c)}}{s_{14}s_{25}}+\frac{n_{(e)}}{s_{13}s_{24}}+\frac{n_{(f)}}{s_{13}s_{25}} -\frac{n_{(h)}}{s_{35}s_{24}}+\frac{n_{(j)}}{s_{14}s_{35}}\right)\\+
 &c_{(k)} \left( A_{(k)} \equiv -\frac{n_{(b)}}{s_{15}s_{23}}-\frac{n_{(d)}}{s_{15}s_{24}}-\frac{n_{(e)}}{s_{13}s_{24}} -\frac{n_{(g)}}{s_{45}s_{23}}+\frac{n_{(k)}}{s_{13}s_{45}}\right)\\+
 &c_{(l)} \left( A_{(l)} \equiv \frac{n_{(d)}}{s_{15}s_{24}}+\frac{n_{(e)}}{s_{13}s_{24}}+\frac{n_{(f)}}{s_{13}s_{25}} -\frac{n_{(i)}}{s_{34}s_{25}}+\frac{n_{(l)}}{s_{15}s_{34}}\right)\\+
 &c_{(n)} \left( A_{(n)} \equiv -\frac{n_{(e)}}{s_{13}s_{24}}-\frac{n_{(f)}}{s_{13}s_{25}}+\frac{n_{(h)}}{s_{35}s_{24}} +\frac{n_{(i)}}{s_{34}s_{25}}+\frac{n_{(m)}}{s_{12}s_{34}}+\frac{n_{(n)}}{s_{12}s_{35}}\right)+\\
 &c_{(o)} \left( A_{(o)} \equiv \frac{n_{(b)}}{s_{15}s_{23}}+\frac{n_{(d)}}{s_{15}s_{24}}+\frac{n_{(e)}}{s_{13}s_{24}} +\frac{n_{(f)}}{s_{13}s_{25}}+\frac{n_{(g)}}{s_{45}s_{23}}-\frac{n_{(i)}}{s_{34}s_{25}}-\frac{n_{(m)}}{s_{12}s_{34}}+\frac{n_{(o)}}{s_{12}s_{45}}\right)
\end{split}
\end{equation}
This amplitude is more complicated than previous examples because we have more terms. In addition, canonical ordering of the spinor chain is more cumbersome. 

We begin by identifying the kinematics building blocks for constructing our ansatz. From dimensional analysis the mass dimension of numerators is three in $D=4$. We have one particle and one antiparticle. So, we should include one $\bar{u}$ and one $v$. To get the correct mass dimension, we need to include two momenta. In addition, each polarization vector should show up once. There are 87 independent terms for the ansatz. We categorize them into seven sets:

\begin{enumerate}
\item $\bar{u}_{1} \slashed{\epsilon_{3}}\slashed{\epsilon_{4}}\slashed{\epsilon_{5}}\slashed{k_{3}}\slashed{k_{4}} v_2$,{~}
in this case we just have one term. Any other orderings reduce to this term plus other terms we will include.
\item $\bar{u}_{1} \slashed{\epsilon_{3}}\slashed{\epsilon_{4}}\slashed{\epsilon_{5}} v_2$ $(k_{1}\cdot k_{2})$,{~}there are five terms like this in our basis.
\item $\bar{u}_{1} \slashed{\epsilon_{3}}\slashed{\epsilon_{4}}\slashed{k_{3}} v_2$ $(k_{1}\cdot\epsilon_{5})$,{~} there are eighteen independent terms similar to this in our basis.
\item $\bar{u}_{1} \slashed{\epsilon_{3}}\slashed{k_{3}}\slashed{k_{4}} v_2$ $(\epsilon_{4}\cdot\epsilon_{5})$,{~} there are three independent terms similar to this in our basis.
\item $\bar{u}_{1} \slashed{\epsilon_{3}} v_2$ ($k_{1}\cdot k_{2}$)($\epsilon_{4}\cdot\epsilon_{5}$),{~} there are fifteen independent terms similar to this in our basis.
\item $\bar{u}_{1} \slashed{\epsilon_{3}} v_2$ ($\epsilon_{4}\cdot k_{1}$)($\epsilon_{5}\cdot k_{2}$),{~} there are twenty seven independent terms similar to this in our basis.
\item$\bar{u}_{1} \slashed{k_{3}} v_2$ ($\epsilon_{3}\cdot k_{1}$)($\epsilon_{4}\cdot\epsilon_{5}$),{~} there are eighteen independent terms similar to this in our basis.
\end{enumerate}

\textbf{\textit{Imposing Constraints:}} We start by imposing cut constraints on a fermionic propagator:

\begin{equation}\label{eq:FirstCutOnePairThreeGlue}
\sum_{s\in \text{states}} A(5_A,1_f,4_A,l_s) A(-\bar{l}_{\bar{s}},2_{\bar{f}},3_A) = \lim_{s_{23} \rightarrow 0 } s_{23} A_{(a)}(5_A,1_f,4_A,2_{\bar{f}},3_A)
\end{equation}
\\
On the LHS, we have lower point ordered amplitudes which we already fixed. On the RHS, only two terms survive in this limit. So, we come up with the following constraint:
\begin{equation*}
\left.\left(\frac{
n_{(a)}}{s_{14}}+\frac{n_{(b)}}{s_{15}} \right) \right|_{s_{23}\rightarrow 0} = \frac{\bar{u}_1 \slashed{\epsilon_4}(\slashed{k_1} + \slashed{k_4})\slashed{\epsilon_5}(\slashed{k_2}+\slashed{k_3})\slashed{\epsilon_3} v_2}{s_{14}}+\frac{\bar{u}_1 \slashed{\epsilon_5}(\slashed{k_1} + \slashed{k_5})\slashed{\epsilon_4}(\slashed{k_2}+\slashed{k_3})\slashed{\epsilon_3} v_2}{s_{15}}
\end{equation*}
The above constraint leaves us with two unknown coefficients. We can use gluonic cut constraint:

\begin{equation}\label{eq:SecondCutOnePairThreeGlue}
\sum_{s \in \text{states}} A(1_f,2_{\bar{f}},l_s)A(-l_{-s},3_A,4_A,5_A)= \lim_{s_{12} \rightarrow 0 } s_{12} A_{(o)}(1_f,2_{\bar{f}},3_A,4_A,5_A)
\end{equation}
\\
This relation fixes another coefficient and we are left with just one unfixed coefficient remaining. This remaining coefficient, we call $a_0$ does not show up in any gauge-invariant observables and so represents total generalized gauge freedom. \eqn{eq:BasisNumeratorOnePairThreeGlue} shows the  basis numerator from which all the others follow from linear relations:

\begin{equation}\label{eq:BasisNumeratorOnePairThreeGlue}
\begin{split}
n_{a}&=n(1_f,4_A,5_A,3_A,2_{\bar{f}}){~}= {~} \bar{u}_1 \slashed{\epsilon_3} \slashed{\epsilon_4} \slashed{\epsilon_5} \slashed{k_3} \slashed{k_4}  v_2-2\bar{u}_1 \slashed{\epsilon_3} \slashed{\epsilon_4} \slashed{\epsilon_5}  v_2 k_{1} \cdot k_{2}-2\bar{u}_1 \slashed{\epsilon_3} \slashed{\epsilon_4} \slashed{\epsilon_5}  v_2 k_{1} \cdot k_{3}-\\
&\frac{4}{3}\bar{u}_1 \slashed{\epsilon_3} \slashed{\epsilon_4} \slashed{\epsilon_5}  v_2 k_{1} \cdot k_{4}-2\bar{u}_1 \slashed{\epsilon_3} \slashed{\epsilon_4} \slashed{k_3}  v_2 (k_{1}\cdot\epsilon_{4})+2\bar{u}_1 \slashed{\epsilon_3} \slashed{\epsilon_4} \slashed{k_3}  v_2 (k_{1}\cdot\epsilon_{5})-\\
&\frac{4}{3}\bar{u}_1 \slashed{\epsilon_3} \slashed{\epsilon_4} \slashed{\epsilon_5}  v_2 k_{2} \cdot k_{3}-2\bar{u}_1 \slashed{\epsilon_3} \slashed{\epsilon_4} \slashed{\epsilon_5}  v_2 k_{2} \cdot k_{4}+2\bar{u}_1 \slashed{\epsilon_3} \slashed{\epsilon_4} \slashed{k_4}  v_2 (k_{2}\cdot\epsilon_{3})-\\
&4\bar{u}_1\slashed{\epsilon_5} v_2 (k_{1}\cdot\epsilon_{4}) (k_{2}\cdot\epsilon_{3}) +4\bar{u}_1\slashed{\epsilon_4} v_2 (k_{1}\cdot\epsilon_{5}) (k_{2}\cdot\epsilon_{3})+2\bar{u}_1 \slashed{\epsilon_3} \slashed{\epsilon_4} \slashed{k_3}  v_2 (k_{2}\cdot\epsilon_{5})+\\
&4\bar{u}_1\slashed{\epsilon_4} v_2 (k_{2}\cdot\epsilon_{3}) (k_{2}\cdot\epsilon_{5})+2\bar{u}_1 \slashed{\epsilon_3} \slashed{\epsilon_4} \slashed{k_3}  v_2 (k_{3}\cdot\epsilon_{5})+4\bar{u}_1\slashed{\epsilon_4} v_2 (k_{2}\cdot\epsilon_{3}) (k_{3}\cdot\epsilon_{5})-\\
&2\bar{u}_1 \slashed{\epsilon_4} \slashed{\epsilon_5} \slashed{k_3}  v_2 (k_{4}\cdot\epsilon_{3})+2\bar{u}_1 \slashed{\epsilon_5} \slashed{k_3} \slashed{k_4}  v_2 (\epsilon_{3}\cdot\epsilon_{4})+4\bar{u}_1\slashed{\epsilon_5} v_2 k_{1} \cdot k_{2} (\epsilon_{3}\cdot\epsilon_{4})+\\
&4\bar{u}_1\slashed{\epsilon_5} v_2 k_{1} \cdot k_{3} (\epsilon_{3}\cdot\epsilon_{4})-4\bar{u}_1\slashed{k_3} v_2 (k_{1}\cdot\epsilon_{5}) (\epsilon_{3}\cdot\epsilon_{4})+4\bar{u}_1\slashed{\epsilon_5} v_2 k_{2} \cdot k_{4} (\epsilon_{3}\cdot\epsilon_{4})-\\
&4\bar{u}_1\slashed{k_3} v_2 (k_{2}\cdot\epsilon_{5}) (\epsilon_{3}\cdot\epsilon_{4})-4\bar{u}_1\slashed{k_3} v_2 (k_{3}\cdot\epsilon_{5}) (\epsilon_{3}\cdot\epsilon_{4})-2\bar{u}_1 \slashed{\epsilon_4} \slashed{k_3} \slashed{k_4}  v_2 (\epsilon_{3}\cdot\epsilon_{5})-\\
&4\bar{u}_1\slashed{\epsilon_4} v_2 k_{1} \cdot k_{2} (\epsilon_{3}\cdot\epsilon_{5})-4\bar{u}_1\slashed{\epsilon_4} v_2 k_{1} \cdot k_{3} (\epsilon_{3}\cdot\epsilon_{5})+4\bar{u}_1\slashed{k_3} v_2 (k_{1}\cdot\epsilon_{4}) (\epsilon_{3}\cdot\epsilon_{5})-\\
&\frac{10}{3}\bar{u}_1\slashed{\epsilon_4} v_2 k_{2} \cdot k_{3}( \epsilon_{3}\cdot\epsilon_{5})-4\bar{u}_1\slashed{\epsilon_4} v_2 k_{2} \cdot k_{4} (\epsilon_{3}\cdot\epsilon_{5})-\frac{2}{3}\bar{u}_1\slashed{\epsilon_3} v_2 k_{1} \cdot k_{4} (\epsilon_{4}\cdot\epsilon_{5})+\\
&(\frac{4}{3}-\frac{a_0}{2})\bar{u}_1\slashed{\epsilon_5} v_2 k_{1} \cdot k_{4} (\epsilon_{3}\cdot\epsilon_{4})+(\frac{4}{3}-\frac{a_0}{2})\bar{u}_1\slashed{\epsilon_5} v_2 k_{2} \cdot k_{3} (\epsilon_{3}\cdot\epsilon_{4})-\\
&(\frac{4}{3}-\frac{a_0}{2})\bar{u}_1\slashed{\epsilon_4} v_2 k_{1} \cdot k_{4} (\epsilon_{3}\cdot\epsilon_{5})+(\frac{4}{3}+\frac{a_0}{2})\bar{u}_1\slashed{\epsilon_3} v_2 k_{2} \cdot k_{3} (\epsilon_{4}\cdot\epsilon_{5})
\end{split}
\end{equation}

Where $a_0$ is an unfixed parameter representing generalized gauge freedom, which cancels out in any physical observables like ordered or color-dressed amplitudes. One can check that the above numerator is not same as the numerator that we find from Feynman rules in Feynman gauge, but the requisite ordered amplitudes are the same.  

\subsection{Massless six-point  amplitudes with three fermionic pairs} \label{sec:MasslessTreeQCD.5}
\begin{figure}[t]
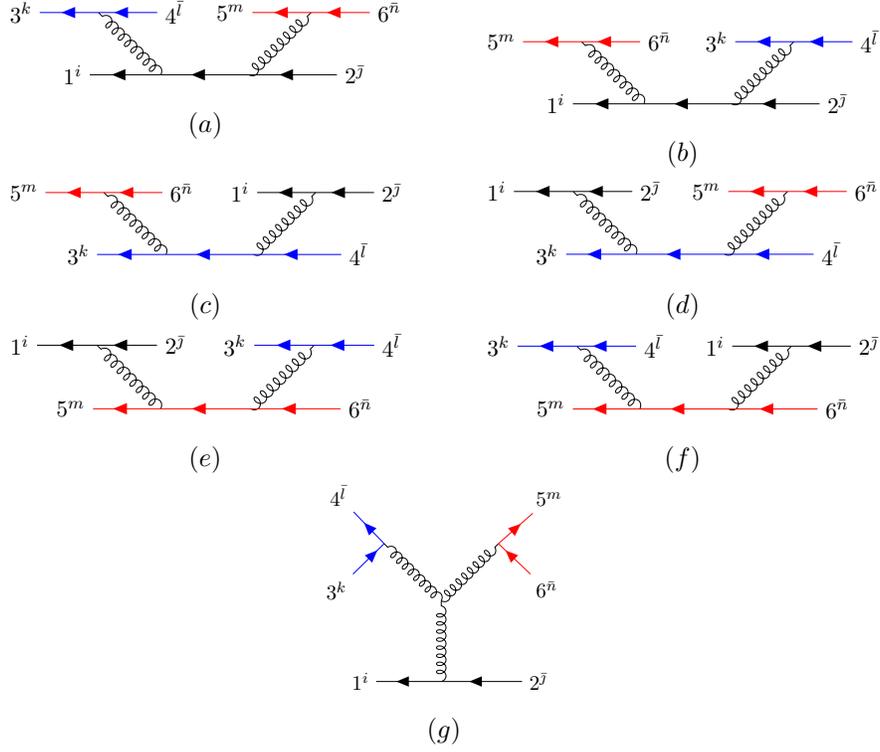

\centering
	\begin{subfigure}{0.4\textwidth}
		\centering
		\resizebox {.9\textwidth} {!} {	
			\numberedfig
		}
		\caption{\((a)\)}\
		\label{fig:TreeThreePairs.a}
	\end{subfigure}
		\begin{subfigure}{0.4\textwidth}
		\centering
		\resizebox {.9\textwidth} {!} {	
			\numberedfig
		}
		\caption{\((b)\)}
		\label{fig:TreeThreePairs.b}
	\end{subfigure}
\begin{subfigure}{0.4\textwidth}
	\centering
	\resizebox {.9\textwidth} {!} {	
		\numberedfig
	}
	\caption{\((c)\)}
	\label{fig:TreeThreePairs.c}
\end{subfigure}
	\begin{subfigure}{0.4\textwidth}
		\centering
		\resizebox {.9\textwidth} {!} {	
			\numberedfig
		}
		\caption{\((d)\)}
		\label{fig:TreeThreePairs.d}
	\end{subfigure}
\begin{subfigure}{0.4\textwidth}
	\centering
	\resizebox {.9\textwidth} {!} {	
		\numberedfig
	}
	\caption{\((e)\)}
	\label{fig:TreeThreePairs.e}
\end{subfigure}
\begin{subfigure}{0.4\textwidth}
	\centering
	\resizebox {.9\textwidth} {!} {	
		\numberedfig
	}
	\caption{\((f)\)}
	\label{fig:TreeThreePairs.f}
\end{subfigure}
\begin{subfigure}{0.25\textwidth}
	\centering
	\resizebox {.9\textwidth} {!} {	
		\numberedfig
	}
	\caption{\((g)\)}
	\label{fig:TreeThreePairs.g}
\end{subfigure}
	\caption{Cubic graphs for the six-point  amplitude with three fermionic pairs.}
	\label{fig:TreeThreePairs}
\end{figure}

This case is simpler than the previous one because we just have external fermions. The mass dimension of amplitude must be $-2$ in $D=4$, and because we have three propagators, our numerators have mass dimension four. We need three $\bar{u}_i$ and three $ v_i$ corresponding to fermions and antifermions, respectively. Like the five-point  case, we include one momentum to get the correct mass dimension. We have seven non-vanishing diagrams, \Fig{fig:TreeThreePairs}. The color factor for each graph is:

\begin{equation}\label{eq:ColorFactorThreePairs}
\begin{split}
c_{(a)} &= T^{a}_{i\bar{x}}T^{a}_{k\bar{l}}T^{b}_{x\bar{\jmath}}T^{b}_{m\bar{n}}\\
c_{(b)} &= T^{a}_{i\bar{x}}T^{a}_{m\bar{n}}T^{b}_{x\bar{\jmath}}T^{b}_{k\bar{l}}\\
c_{(c)} &= T^{a}_{k\bar{x}}T^{a}_{m\bar{n}}T^{b}_{x\bar{l}}T^{b}_{i\bar{\jmath}}\\
c_{(d)} &= T^{a}_{k\bar{x}}T^{a}_{i\bar{\jmath}}T^{b}_{x\bar{l}}T^{b}_{m\bar{n}}\\
c_{(e)} &= T^{a}_{m\bar{x}}T^{a}_{i\bar{\jmath}}T^{b}_{x\bar{n}}T^{b}_{k\bar{l}}\\
c_{(f)} &= T^{a}_{m\bar{x}}T^{a}_{k\bar{l}}T^{b}_{x\bar{n}}T^{b}_{i\bar{\jmath}}\\
c_{(g)} &= f^{abc}T^{a}_{i\bar{\jmath}}T^{b}_{k\bar{l}}T^{c}_{m\bar{n}}
\end{split}
\end{equation}

Color-kinematics duality relates different topologies:

\begin{equation}\label{eq:JacobiThreePairs}
	\begin{split}
		&c_{(a)} -c_{(b)} = -c_{(g)}~\leftrightarrow ~
n_{(a)}-n_{(b)}=-n_{(g)}\\
		&c_{(c)} -c_{(d)} = -c_{(g)}~\leftrightarrow ~n_{(c)}-n_{(d)}=-n_{(g)}\\
		&c_{(e)} -c_{(f)} = -c_{(g)}~\leftrightarrow ~n_{(e)}-n_{(f)}=-n_{(g)} 
	\end{split}
\end{equation}

The full color-dressed amplitude can be written:
\begin{align}
g^{-4} \mathcal{A}&= \frac{c_{(a)} 
n_{(a)}}{s_{134}s_{56}s_{34}} +  \frac{c_{(b)} n_{(b)}}{s_{156}s_{56}s_{34}} + \frac{c_{(c)} n_{(c)}}{s_{356}s_{56}s_{12}} +  \frac{c_{(d)} n_{(d)}}{s_{123}s_{56}s_{12}} +
 \frac{c_{(e)} n_{(e)}}{s_{125}s_{12}s_{34}}+\\
 & \frac{c_{(f)} n_{(f)}}{s_{345}s_{12}s_{34}}+ \frac{c_{(g)} n_{(g)}}{s_{12}s_{34}s_{56}} \\
 &= c_{(a)} \left(A_{(a)} \equiv \frac{
n_{(a)}}{s_{134}s_{56}s_{34}} +  \frac{n_{(b)}}{s_{156}s_{56}s_{34}}\right) +\\
 &~~ c_{(c)}\left(A_{(c)} \equiv \frac{ n_{(c)}}{s_{356}s_{56}s_{12}} +  \frac{n_{(d)}}{s_{123}s_{56}s_{12}}\right) + \\
 &~~  c_{(f)}\left(A_{(f)} \equiv \frac{ n_{(e)}}{s_{125}s_{34}s_{12}} +  \frac{n_{(f)}}{s_{345}s_{34}s_{12}}\right) + \\
 &~~  c_{(g)}\left(A_{(g)} \equiv \frac{ n_{(b)}}{s_{156}s_{34}s_{56}} +  \frac{n_{(d)}}{s_{123}s_{56}s_{12}}- \frac{ n_{(e)}}{s_{125}s_{34}s_{12}} +  \frac{n_{(g)}}{s_{34}s_{56}s_{12}}\right)\,\,
\end{align}
where in the second equality we wrote down terms in the  minimal color basis.

 The structure of the first six graphs are the same. If we can fix the numerator for one graph, by relabeling we can get the others. We have two types of terms in our ansatz in this case:

1. The general structure of these terms is $(\bar{u}_i \gamma_{\mu}  v_j)(\bar{u}_p \gamma^{\mu}  v_q)(\bar{u}_m \slashed{k}_l  v_n)$. There are nine independent terms:
\begin{equation*}
	\begin{split}
		&\{(\bar{u}_1 \gamma_{\mu}  v_2)(\bar{u}_3 \gamma^{\mu}  v_4)(\bar{u}_5 \slashed{k}_2  v_6),~(\bar{u}_1 \gamma_{\mu}  v_2)(\bar{u}_3 \gamma^{\mu}  v_4)(\bar{u}_5 \slashed{k}_3  v_6),~(\bar{u}_1 \gamma_{\mu}  v_2)(\bar{u}_3 \gamma^{\mu}  v_4)(\bar{u}_5 \slashed{k}_4  v_6),~\\
		&(\bar{u}_1 \gamma_{\mu}  v_2)(\bar{u}_5 \gamma^{\mu}  v_6)(\bar{u}_3 \slashed{k}_2  v_4),~(\bar{u}_1 \gamma_{\mu}  v_2)(\bar{u}_5 \gamma^{\mu}  v_6)(\bar{u}_3 \slashed{k}_5  v_4),~(\bar{u}_1 \gamma_{\mu}  v_2)(\bar{u}_5 \gamma^{\mu}  v_6)(\bar{u}_3 \slashed{k}_6  v_4),~\\
		&(\bar{u}_3 \gamma_{\mu}  v_4)(\bar{u}_5 \gamma^{\mu}  v_6)(\bar{u}_1 \slashed{k}_3  v_2),~(\bar{u}_3 \gamma_{\mu}  v_4)(\bar{u}_5 \gamma^{\mu}  v_6)(\bar{u}_1 \slashed{k}_4  v_2),~(\bar{u}_3 \gamma_{\mu}  v_4)(\bar{u}_5 \gamma^{\mu}  v_6)(\bar{u}_1 \slashed{k}_5  v_2)\}
	\end{split}
\end{equation*}

2. This case is similar to the first one just with two more Gamma matrices, \\
 $(\bar{u}_i \gamma_{\mu}  v_j)(\bar{u}_p \gamma_{\rho}  v_q)(\bar{u}_m \gamma^\mu \slashed{k}_l \gamma^\rho  v_n)$. There are nine independent terms:
\begin{align}
\big\{(\bar{u}_1 \gamma_{\mu}  v_2)(\bar{u}_3 \gamma_{\rho}  v_4)(\bar{u}_5 \gamma^\mu \slashed{k}_2 \gamma^\rho  v_6),&
~(\bar{u}_1 \gamma_{\mu}  v_2)(\bar{u}_3 \gamma_{\rho}  v_4)(\bar{u}_5 \gamma^\mu \slashed{k}_3 \gamma^\rho  v_6),\nn\\
~(\bar{u}_1 \gamma_{\mu}  v_2)(\bar{u}_3 \gamma_{\rho}  v_4)(\bar{u}_5 \gamma^\mu \slashed{k}_4 \gamma^\rho  v_6),&
~(\bar{u}_1 \gamma_{\mu}  v_2)(\bar{u}_5 \gamma_{\rho}  v_6)(\bar{u}_3 \gamma^\mu \slashed{k}_2 \gamma^\rho  v_4),\nn\\
~(\bar{u}_1 \gamma_{\mu}  v_2)(\bar{u}_5 \gamma_{\rho}  v_6)(\bar{u}_3 \gamma^\mu \slashed{k}_5 \gamma^\rho  v_4),&
~(\bar{u}_1 \gamma_{\mu}  v_2)(\bar{u}_5 \gamma_{\rho}  v_6)(\bar{u}_3 \gamma^\mu \slashed{k}_6 \gamma^\rho  v_4),\\
~(\bar{u}_3 \gamma_{\mu}  v_4)(\bar{u}_5 \gamma_{\rho}  v_6)(\bar{u}_1 \gamma^\mu \slashed{k}_3 \gamma^\rho  v_2),&
~(\bar{u}_3 \gamma_{\mu}  v_4)(\bar{u}_5 \gamma_{\rho}  v_6)(\bar{u}_1 \gamma^\mu \slashed{k}_4 \gamma^\rho  v_2),\nn\\
~(\bar{u}_3 \gamma_{\mu}  v_4)(\bar{u}_5 \gamma_{\rho}  v_6)(\bar{u}_1 \gamma^\mu \slashed{k}_5 \gamma^\rho  v_2) \big\}\nn&
\end{align}

The quickest approach to fixing the ansatz is simply by applying cut constraints to one of the smaller ordered amplitudes:
\begin{equation}\label{eq:CutThreePairs}
	\begin{split}
		\sum_{s\in\text{states}}^{} A(1_f,\bar{l}_{\bar{s}},3_f,4_{\bar{f}})A({-l_s},2_{\bar{f}},5_f,6_{\bar{f}}) &= \lim_{s_{134} \rightarrow 0} s_{134} A_{(a)}(1_f,3_f,4_{\bar{f}},5_f,6_{\bar{f}},2_{\bar{f}})\\
		&=\frac{n(1_f,3_f,4_{\bar{f}},5_f,6_{\bar{f}},2_{\bar{f}})}{s_{34}s_{56}}\mid_{s_{134} \rightarrow 0}
	\end{split}
\end{equation}

Using \eqn{eq:TwoPairsOrderedAmp} we can write down the LHS as:
\begin{equation}
\left. \sum_{s \in \text{states}} \frac{(\bar{u}_1\gamma_{\rho} v_{l,\bar{s}})(\bar{u}_3\gamma^{\rho} v_4)(\bar{u}_{-l,s}\gamma_{\mu} v_2)(\bar{u}_5\gamma^{\mu} v_6)}{s_{34}s_{56}}\right|_{l^2\to0}=\left.\frac{n(1_f,3_f,4_{\bar{f}},5_f,6_{\bar{f}},2_{\bar{f}})}{s_{34}s_{56}}\right|_{s_{134} \to 0}
\end{equation}
Using the state sum relation for spinors, we will come up with a simple equation:
\begin{equation}\label{eq:BasisNumeratorThreePairs}
\begin{split}
n(1_f,3_f,4_{\bar{f}},5_f,6_{\bar{f}},2_{\bar{f}})&=(\bar{u}_1 \gamma_{\mu}(\slashed{k}_3+\slashed{k}_4)\gamma_{\rho}  v_2)(\bar{u}_3 \gamma^\mu  v_4)(\bar{u}_5 \gamma^\rho  v_6)\\
&+2(\bar{u}_1 \gamma_{\rho}  v_2)(\bar{u}_3 \slashed{k}_1  v_4)(\bar{u}_5 \gamma^\rho  v_6)
\end{split}
\end{equation}
By relabeling indices and applying color-kinematic duality, we can get other diagrams. This result is consistent with~\cite{Johansson:2015oia}.

\section{Tree level QCD amplitudes, massive cases}\label{sec:MassiveTreeQCD}

For massive fermions, we have the same constraining relations for three-point  and four-point  with two fermionic pairs and symbolically the same ansatz building blocks. 
Our expressions change only by the meaning of the spinors -- they now satisfy the solution to the massive Dirac equations. As a result the amplitudes are otherwise trivially the same. The first difference in ansatz building blocks between the expressions between massless and massive cases is at four-point  with one fermionic pair and two gluons. The first time we will see actually distinct amplitudes at tree-level is at five-points.

\subsection{Four-point  amplitudes with two gluons and one massive fermionic pair}\label{sub:MassiveTreeQCD.1}
We build up our ansatz precisely as in the massless cases described in the preceding sections, except now we have access to a few more building blocks,
\begin{multline}
\{k_1 \cdot\epsilon_3 \bar{u}_1 \slashed{\epsilon}_4  v_2,{~}k_4\cdot\epsilon_3 \bar{u}_1 \slashed{\epsilon}_4  v_2,{~}k_1 \cdot\epsilon_4 \bar{u}_1 \slashed{\epsilon}_3  v_2,{~}k_3\cdot\epsilon_4 \bar{u}_1 \slashed{\epsilon}_3  v_2,{~}\epsilon_3\cdot\epsilon_4\bar{u}_1 \slashed{k}_3  v_2,{~}\bar{u}_1 \slashed{\epsilon}_3\slashed{\epsilon}_4 \slashed{k}_3 v_2,\\
m\, \bar{u}\slashed{\epsilon_3}\slashed{\epsilon_4} v_2, m\,  \bar{u}_1  v_2 \epsilon_3 . \epsilon_4  \}\,.
\end{multline}
Imposing Jacobi-like relations, \eqn{eq:JacobiRelationOnePairTwoGlue}, reordering rules for two adjacent fermions in the third diagram in \Fig{fig:TreeOnePairTwoGlue}, and gauge invariance, fixes all coefficients up to an overall factor which can be constrained by factorization. Here then is the first numerator in \Fig{fig:TreeOnePairTwoGlue} and one can get the others from this:
\begin{equation}\label{eq:BasisNumeratorOnePairTwoGlueMassive}
\begin{split}
n(1_f,3_A,4_A,2_{\bar{f}}) = 
 2(k_1 \cdot\epsilon_3)\bar{u}_1\slashed{\epsilon}_4 v_2+\bar{u}_1\slashed{\epsilon}_3\slashed{k}_3\slashed{\epsilon}_4 v_2
\end{split}
\end{equation}\\
We see the above numerator is exactly the same as the massless case, \eqn{eq:tChannelNumeratorOnePairTwoGlue}. So, in this case we can also use same result for both massive and massless cases. 

\subsection{Five-point  amplitudes with one gluon and two massive fermionic pairs}\label{sec:MassiveTreeQCD.2}
To build five-point  amplitudes with two massive fermionic pairs, we must consider four additional types of terms in our ansatz, which would be zero in the massless limit.
Again in this case we will land on the same symbolic expression as the massless case.  The four classes of additional terms can be described as: 
\begin{enumerate}
    \item Three terms similar to $(\bar{u}_1 v_2)( \bar{u}_3 v_4)( k_{a}\cdot\epsilon_5)$. (Why only three? Recall that $k_4=-k_1-k_2-k_3-k_5$ so $k_4 \cdot \epsilon_5$ is spanned by having $k_{1,2,3} \cdot \epsilon_5$ already in our ansatz.)

\item Three terms similar to $(\bar{u}_1\gamma^{\alpha}\gamma^{\beta} v_2)( \bar{u}_3 \gamma_{\alpha}\gamma_{\beta} v_4)( k_{a}\cdot\epsilon_5)$.

\item Four terms similar to  $(\bar{u}_1 v_2)( \bar{u}_3 \slashed{k_1}\slashed{\epsilon_5} v_4)$. 

\item Four terms of the form $m_1 (\bar{u}_1 v_2)( \bar{u}_3 \slashed{\epsilon_5} v_4)$.
\end{enumerate}
The quickest way to fix the ansatz is from a single ordered cut. Similar to the massless case, we have:
\begin{align}\label{eq:FirstCutConditionTwoPairsOneGlueMassive}
\sum_{s\in\text{states}}^{} A(1_f,5_A,\bar{l}_{\bar{s}})A(-l_s,2_{\bar{f}},3_f,4_{\bar{f}}) &= \lim_{s_{15} \rightarrow m^{2}_1} s_{15} A_{(a)}(1_f,5_A,2_{\bar{f}},3_f,4_{\bar{f}})\\
&=\frac{n(1_f,5_A,2_{\bar{f}},3_f,4_{\bar{f}})}{s_{34}}\mid_{s_{15}\rightarrow m^{2}_1}\,.
\end{align}
This cut fixes all coefficients in our ansatz and we have:
\begin{align}\label{eq:MassiveBasisNumeratorTwoPairsOneGlue}
n(1_f,5_A,2_{\bar{f}},3_f,4_{\bar{f}})&=2(\bar{u}_1 \gamma_{\mu} v_2) (\bar{u}_3 \gamma^{\mu}  v_4) {(k_1 \cdot\epsilon_5)}+\nn\\
		&2(\bar{u}_1 \gamma_{\mu} v_2) (\bar{u}_3 \gamma^{\mu}  v_4) {(k_2\cdot\epsilon_5)}-2(\bar{u}_3 \slashed{k}_2  v_4)(\bar{u}_1 \slashed{\epsilon}_5  v_2)+\\
		&(\bar{u}_3 \gamma_{\mu}  v_4)(\bar{u}_1 \slashed{k}_3 \slashed{\epsilon}_5 \gamma^{\mu}  v_2)+(\bar{u}_3 \gamma_{\mu}  v_4)(\bar{u}_1 \slashed{k}_4 \slashed{\epsilon}_5 \gamma^{\mu}  v_2)\nn
\end{align}
This is exactly \eqn{eq:NumeratorGraphATwoPairsOneGlue} which we derived for the massless case from a general ansatz. So, again, we can use the above functional form both for massive and massless cases.

\subsection{Five-point  amplitudes with three gluons and one massive fermionic pair}\label{sec:MassiveTreeQCD.3}
The general ansatz in this case is similar to the massless case, we just need to add terms proportional to $m$ or $m^2$.  The new terms proportional to $m^2$ are given,
\begin{equation}
  \text{basis}_{m^2}=  \{ m^2 \bar{u}_1 \slashed{\epsilon}_3  \slashed{\epsilon}_4  \slashed{\epsilon}_5  v_2, m^2 \bar{u}_1 \slashed{\epsilon}_3   v_2 (\epsilon_4\cdot\epsilon_5), m^2 \bar{u}_1 \slashed{\epsilon}_4   v_2 (\epsilon_3\cdot\epsilon_5), m^2 \bar{u}_1 \slashed{\epsilon}_5   v_2 (\epsilon_3\cdot\epsilon_4)\}
\end{equation}
Terms proportional to a single power of $m$ begin to proliferate, but we can categorize them as follows:
\begin{enumerate}
\item $m\bar{u}_1 \slashed{\epsilon}_3  \slashed{\epsilon}_4  \slashed{\epsilon}_5 \slashed{k}_3  v_2$ , and $m\bar{u}_1 \slashed{\epsilon}_3  \slashed{\epsilon}_4  \slashed{\epsilon}_5 \slashed{k}_4  v_2$.
\item $m\bar{u}_1 \slashed{\epsilon}_3 \slashed{\epsilon}_4  v_2 (k_1 \cdot\epsilon_5) $, among nine independent terms like this.
\item $m\bar{u}_1 \slashed{\epsilon}_3 \slashed{k}_3  v_2 (\epsilon_4\cdot\epsilon_5) $,  among six independent terms like this.
\item $m\bar{u}_1  v_2 (k_1 \cdot \epsilon_3) ( \epsilon_4\cdot\epsilon_5 )$, among nine independent terms like this.
\end{enumerate}
Imposing the duality between color and kinematics reduces the number of basis graphs to one. Graphically the ordered amplitudes can all be arranged as per \eqn{eqn:ordAmps3A2F}. All remaining freedom can either be fixed via unitarity cuts of ordered amplitudes or perhaps most simply by imposing gauge invariance on a single ordered amplitude:
\begin{equation}\label{eq:GImassiveThreeGlueOnePair}
A_{(a)}(1_f,2_{\bar{f}},3_A,4_A,5_A) \mid_{\epsilon_3 \rightarrow k_3} = 0
\end{equation}
Enforcing this constraint and matching with the massless result in the $m\rightarrow 0$ limit, we achieve:
\begin{equation}\label{eq:BasisMassiveOnePairThreeGlue}
\begin{split}
    n_{(a)}&=n(1_f,4_A,5_A,3_A,2_{\bar{f}}){~}= {~} \bar{u}_1 \slashed{\epsilon_3} \slashed{\epsilon_4} \slashed{\epsilon_5} \slashed{k_3} \slashed{k_4}  v_2-2\bar{u}_1 \slashed{\epsilon_3} \slashed{\epsilon_4} \slashed{\epsilon_5}  v_2 k_{1} \cdot k_{2}-2\bar{u}_1 \slashed{\epsilon_3} \slashed{\epsilon_4} \slashed{\epsilon_5}  v_2 k_{1} \cdot k_{3}-\\
&\frac{4}{3}\bar{u}_1 \slashed{\epsilon_3} \slashed{\epsilon_4} \slashed{\epsilon_5}  v_2 k_{1} \cdot k_{4}-2\bar{u}_1 \slashed{\epsilon_3} \slashed{\epsilon_4} \slashed{k_3}  v_2 (k_{1}\cdot\epsilon_{4})+2\bar{u}_1 \slashed{\epsilon_3} \slashed{\epsilon_4} \slashed{k_3}  v_2 (k_{1}\cdot\epsilon_{5})-\\
&\frac{4}{3}\bar{u}_1 \slashed{\epsilon_3} \slashed{\epsilon_4} \slashed{\epsilon_5}  v_2 k_{2} \cdot k_{3}-2\bar{u}_1 \slashed{\epsilon_3} \slashed{\epsilon_4} \slashed{\epsilon_5}  v_2 k_{2} \cdot k_{4}+2\bar{u}_1 \slashed{\epsilon_3} \slashed{\epsilon_4} \slashed{k_4}  v_2 (k_{2}\cdot\epsilon_{3})-\\
&4\bar{u}_1\slashed{\epsilon_5} v_2 (k_{1}\cdot\epsilon_{4}) (k_{2}\cdot\epsilon_{3}) +4\bar{u}_1\slashed{\epsilon_4} v_2 (k_{1}\cdot\epsilon_{5}) (k_{2}\cdot\epsilon_{3})+2\bar{u}_1 \slashed{\epsilon_3} \slashed{\epsilon_4} \slashed{k_3}  v_2 (k_{2}\cdot\epsilon_{5})+\\
&4\bar{u}_1\slashed{\epsilon_4} v_2 (k_{2}\cdot\epsilon_{3}) (k_{2}\cdot\epsilon_{5})+2\bar{u}_1 \slashed{\epsilon_3} \slashed{\epsilon_4} \slashed{k_3}  v_2 (k_{3}\cdot\epsilon_{5})+4\bar{u}_1\slashed{\epsilon_4} v_2 (k_{2}\cdot\epsilon_{3}) (k_{3}\cdot\epsilon_{5})-\\
&2\bar{u}_1 \slashed{\epsilon_4} \slashed{\epsilon_5} \slashed{k_3}  v_2 (k_{4}\cdot\epsilon_{3})
+2\bar{u}_1 \slashed{\epsilon_5} \slashed{k_3} \slashed{k_4}  v_2 (\epsilon_{3}\cdot\epsilon_{4})+4\bar{u}_1\slashed{\epsilon_5} v_2 k_{1} \cdot k_{2} (\epsilon_{3}\cdot\epsilon_{4})+\\
&4\bar{u}_1\slashed{\epsilon_5} v_2 k_{1} \cdot k_{3} (\epsilon_{3}\cdot\epsilon_{4})-4\bar{u}_1\slashed{k_3} v_2 (k_{1}\cdot\epsilon_{5}) (\epsilon_{3}\cdot\epsilon_{4})+4\bar{u}_1\slashed{\epsilon_5} v_2 k_{2} \cdot k_{4} (\epsilon_{3}\cdot\epsilon_{4})-\\
&4\bar{u}_1\slashed{k_3} v_2 (k_{2}\cdot\epsilon_{5}) (\epsilon_{3}\cdot\epsilon_{4})-4\bar{u}_1\slashed{k_3} v_2 (k_{3}\cdot\epsilon_{5}) (\epsilon_{3}\cdot\epsilon_{4})-2\bar{u}_1 \slashed{\epsilon_4} \slashed{k_3} \slashed{k_4}  v_2 (\epsilon_{3}\cdot\epsilon_{5})-\\
&4\bar{u}_1\slashed{\epsilon_4} v_2 k_{1} \cdot k_{2} (\epsilon_{3}\cdot\epsilon_{5})-4\bar{u}_1\slashed{\epsilon_4} v_2 k_{1} \cdot k_{3} (\epsilon_{3}\cdot\epsilon_{5})+4\bar{u}_1\slashed{k_3} v_2 (k_{1}\cdot\epsilon_{4}) (\epsilon_{3}\cdot\epsilon_{5})-\\
&\frac{10}{3}\bar{u}_1\slashed{\epsilon_4} v_2 k_{2} \cdot k_{3} (\epsilon_{3}\cdot\epsilon_{5})-4\bar{u}_1\slashed{\epsilon_4} v_2 k_{2} \cdot k_{4} (\epsilon_{3}\cdot\epsilon_{5})-\frac{2}{3}\bar{u}_1\slashed{\epsilon_3} v_2 k_{1} \cdot k_{4} (\epsilon_{4}\cdot\epsilon_{5})+\\
&(\frac{4}{3}-\frac{a_0}{2})\bar{u}_1\slashed{\epsilon_5} v_2 k_{1} \cdot k_{4} (\epsilon_{3}\cdot\epsilon_{4})+(\frac{4}{3}-\frac{a_0}{2})\bar{u}_1\slashed{\epsilon_5} v_2 k_{2} \cdot k_{3} (\epsilon_{3}\cdot\epsilon_{4})-\\
&(\frac{4}{3}-\frac{a_0}{2})\bar{u}_1\slashed{\epsilon_4} v_2 k_{1} \cdot k_{4} (\epsilon_{3}\cdot\epsilon_{5})+(\frac{4}{3}+\frac{a_0}{2})\bar{u}_1\slashed{\epsilon_3} v_2 k_{2} \cdot k_{3} (\epsilon_{4}\cdot\epsilon_{5})-\\
&2m^2 \bar{u}_1 \slashed{\epsilon}_3  \slashed{\epsilon}_4  \slashed{\epsilon}_5  v_2+4m^2\bar{u}_1 \slashed{\epsilon}_5   v_2 (\epsilon_3\cdot\epsilon_4)-4m^2\bar{u}_1 \slashed{\epsilon}_4   v_2 (\epsilon_3\cdot\epsilon_5)
\end{split}
\end{equation}
Where $a_0$ represents generalized gauge freedom which does not contribute to any on-shell quantity.

\subsection{Massive six-point  amplitudes with three fermionic pairs}\label{sec:MassiveTreeQCD.4}
Like the previous example, we need to add terms to our ansatz which are zero in the massless limit. There are nine categories:
\begin{enumerate}
\item $m_1(\bar{u}_1 v_2)(\bar{u}_3 v_4) (\bar{u}_5 v_6)$, there are three terms like this because of the three fermion masses.

\item $(\bar{u}_1 v_2)(\bar{u}_3 v_4) (\bar{u}_5\slashed{k_2} v_6)$, there are nine terms like this.

\item $m_1(\bar{u}_1\gamma^{\mu} v_2)(\bar{u}_3 \gamma_{\mu}v_4) (\bar{u}_5 v_6)$, there are nine terms like this.

\item $(\bar{u}_1\gamma^{\mu} v_2)(\bar{u}_3 v_4) (\bar{u}_5\slashed{k_2}\gamma_{\mu} v_6)$, there are eighteen terms like this.

\item $m_1(\bar{u}_1\gamma^{\mu} \gamma^{\nu} v_2)(\bar{u}_3 \gamma_{\mu}\gamma_{\nu} v_4) (\bar{u}_5v_6)$, there are nine terms like this.

\item $m_1(\bar{u}_1\gamma^{\mu} \gamma^{\nu} v_2)(\bar{u}_3 \gamma_{\mu} v_4) (\bar{u}_5\gamma_{\nu}v_6)$, there are nine terms like this.

\item $(\bar{u}_1\gamma^{\mu} \gamma^{\nu} v_2)(\bar{u}_3 \gamma_{\mu}\gamma_{\nu} v_4) (\bar{u}_5\slashed{k_2} v_6)$, there are nine terms like this.

\item $(\bar{u}_1\gamma^{\mu} \gamma^{\nu} v_2)(\bar{u}_3 \gamma_{\mu} v_4) (\bar{u}_5\slashed{k_2} \gamma_{\nu} v_6)$, there are eighteen terms like this.

\item $(\bar{u}_1\gamma^{\mu}\slashed{k_3} \gamma^{\nu} v_2)(\bar{u}_3 \gamma_{\mu}\gamma_{\nu} v_4) (\bar{u}_5 v_6)$, there are eighteen terms like this.
    
\end{enumerate}

Imposing cut constraints on the massive fermionic propagator as per,
\begin{equation}\label{eq:CutThreePairsMassive}
	\begin{split}
		\sum_{s\in\text{states}}^{} A(1_f,\bar{l}_{\bar{s}},3_f,4_{\bar{f}})A({-l_s},2_{\bar{f}},5_f,6_{\bar{f}}) &= \lim_{s_{134} \rightarrow m^{2}_1} s_{134} A_{(a)}(1_f,3_f,4_{\bar{f}},5_f,6_{\bar{f}},2_{\bar{f}})\\
		&=\frac{n(1_f,3_f,4_{\bar{f}},5_f,6_{\bar{f}},2_{\bar{f}})}{s_{34}s_{56}}\mid_{s_{134} \rightarrow m^{2}_1} \,,
	\end{split}
\end{equation}
fixes all the coefficients in our ansatz:
\begin{equation}\label{eq:MassiveBasisNumeratorThreePairs}
\begin{split}
n(1_f,3_f,4_{\bar{f}},5_f,6_{\bar{f}},2_{\bar{f}})&=(\bar{u}_1 \gamma_{\mu}(\slashed{k}_3+\slashed{k}_4)\gamma_{\rho}  v_2)(\bar{u}_3 \gamma^\mu  v_4)(\bar{u}_5 \gamma^\rho  v_6)\\
&+2(\bar{u}_1 \gamma_{\rho}  v_2)(\bar{u}_3 \slashed{k}_1  v_4)(\bar{u}_5 \gamma^\rho  v_6)
\end{split}
\end{equation}\\
 We see that in this case, like the previous examples for massive amplitudes, we are allowed to use the same ansatz for both massive and massless amplitudes, with no explicit mass dependence.

\section{One loop QCD amplitudes}\label{sec:one-loopQCD}
In this section we turn to one-loop four-point calculations.  We will build the integrand using similar techniques as tree level, then verify by comparing to known results.  Namely we  begin by writing down a general ansatz for numerators at the integrand level, now including dependence on $l$, the off-shell loop momentum. When closed loop states involve massive particles we will allow for both the on-shell mass of the particle and the off-shell $l^2$ to contribute to the integrand of the ansatz. Imposing the duality between color and kinematics at the integrand level as per \eqn{eq:genericLoopQCDAmplitudes}, which will allow us to reduce the number of independent kinematic numerators -- necessitating a far smaller ansatz than if we had dressed each graph independently. Using unitarity cuts and symmetries of diagrams, we will see that we can fix  the coefficients in our ansatz up to  pure generalized gauge terms that will cancel out in any physical observable. Verification will proceed via integral reduction and comparison to equivalent calculations preformed via Feynman rules.  For ease of book-keeping we consider ourselves in even dimensions when applying any spinor-trace identities, but will otherwise leave dimension generic.  In this work we focus on the traditionally cut-constructible aspects of the integrand (excluding tadpoles and bubbles on external legs). As discussed in ~\cite{Bern:1995db}, such  contributions---while subtle to access from a unitarity perspective---seperate and can be constrained by consideration of known UV and IR behavior.
 Of course gauge invariance for external vectors in an ordered sub-amplitude is required and checked in the process of verification of our calculation.

\subsection{ One-loop massive four-point  amplitudes with two different fermionic pairs}\label{sub:1.one-loopQCD}
\captionsetup[subfigure]{labelformat=empty}
\begin{figure}[t]
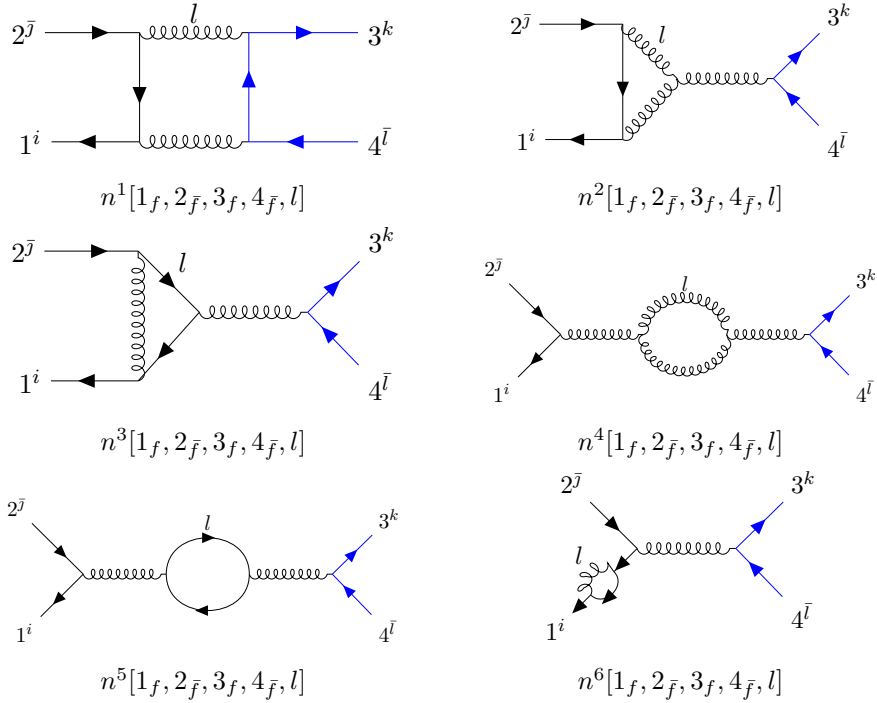

	\centering
	\begin{subfigure}{0.4\textwidth}
		\centering
	\resizebox {.9\textwidth} {!} {	
	\numberedfig
}
	\caption{$n^{1}[1_f,2_{\bar{f}},3_f,4_{\bar{f}},l]$}
	\end{subfigure}
	\begin{subfigure}{0.4\textwidth}
		\centering
	\resizebox {.8\textwidth} {!} {	
	\numberedfig
}
\caption{$n^{2}[1_f,2_{\bar{f}},3_f,4_{\bar{f}},l]$}
	\end{subfigure}
	\begin{subfigure}{0.4\textwidth}
		\centering
	\resizebox {.9\textwidth} {!} {	
	\numberedfig
}
\caption{$n^{3}[1_f,2_{\bar{f}},3_f,4_{\bar{f}},l]$}
	\end{subfigure}
	\begin{subfigure}{0.4\textwidth}
		\centering
	\resizebox {.9\textwidth} {!} {	
	\numberedfig
}
\caption{$n^{4}[1_f,2_{\bar{f}},3_f,4_{\bar{f}},l]$}
	\end{subfigure}
	\begin{subfigure}{0.4\textwidth}
		\centering
	\resizebox {.9\textwidth} {!} {	
	\numberedfig
}
\caption{$n^{5}[1_f,2_{\bar{f}},3_f,4_{\bar{f}},l]$}
\end{subfigure}
	\begin{subfigure}{0.4\textwidth}
		\centering
	\resizebox {.65\textwidth} {!} {	
	\numberedfig
}
\caption{$n^{6}[1_f,2_{\bar{f}},3_f,4_{\bar{f}},l]$}
\end{subfigure}

\caption{Different topologies for two fermionic pairs at one-loop. The momentum flow of loop leg $l$ is taken to the right in each graph independent of the orientation of the spinor arrow.}
	\label{fig:one-loopTwoFermionicPairs}
\end{figure}

Let us now consider the one-loop correction to two-to-two fermion scattering between distinct flavor quarks with mass squared $k_1^2=k_2^2=m_1^2$ and $k_3^2=k_4^2=m_2^2$ respectively.  We here consider the case explicitly for $N_f=2$, as additional flavors of fermions would only contribute an overall-factor to the massive fermion bubble. With two distinct fermionic pairs we can have six different topologies as drawn in \Fig{fig:one-loopTwoFermionicPairs}. Indeed there are further distinct graphs from swapping flavors for graphs 2,3,5, and 6, but their contributions are straightforwardly related via relabeling. If there were no relations between kinematic weights we would need to write down six different ansatze.  Imposing the duality between color and kinematics functionally imposes constraints which allow us to reduce to a smaller basis:
\begin{align}
\label{eq:JacobiLoopTwoPairsn2Sol}
n^{2}[1_f,2_{\bar{f}},3_f,4_{\bar{f}},l] &=n^{1}[1_f,2_{\bar{f}},3_f,4_{\bar{f}},l] -\Rev_{3\bar{4}} \left( n^{1}[1_f,2_{\bar{f}},3_f,4_{\bar{f}},l] \right)\,,\\
\label{eq:JacobiLoopTwoPairsn4Sol}
n^{4}[1_f,2_{\bar{f}},3_f,4_{\bar{f}},l] &=n^{2}[1_f,2_{\bar{f}},3_f,4_{\bar{f}},l] +n^{2}[1_f,2_{\bar{f}},3_f,4_{\bar{f}},-l-1_{f}-2_{\bar{f}}]\,,\\
\label{eq:JacobiLoopTwoPairsn6Sol}
n^{6}[1_f,2_{\bar{f}},3_f,4_{\bar{f}},l] &=n^{2}[1_f,2_{\bar{f}},3_f,4_{\bar{f}},-l-1_{f}-2_{\bar{f}}]+n^{3}[1_f,2_{\bar{f}},3_f,4_{\bar{f}},l+2_{\bar{f}}]\,.
\end{align}

As per the discussion around \eqn{eq:ReorderingRules}, the operation $\Rev_{ij}$ applies the appropriate fermionic relabeling on fermionic pair $\{i,j\}$. There are three independent equations, which we can use to specify $n^1$, $n^3$ and $n^5$ as basis numerators.  As per the discussion around \eqn{eq:ReorderingRules}, the operation $\Rev_{i\jmath}$ applies the appropriate fermionic relabeling for the fermionic pair $\{i,\jmath\}$. 
There are four independent equations, which we can use to specify $n^1$ and $n^3$ as basis numerators.

\subsubsection{General ansatz }\label{sub:1.1 one-loopQCD}
The mass dimension of the ansatz at this level is four. With four spinors, we require two additional momenta to get the correct dimension. Since some graphs have four fermionic vertices and two fermionic propagators, we can have six Gamma matrices at most. We characterize the different available terms in the ansatz in Table \ref{tab:CategoriesAnsatzTwoPairs}.

\begin{center}
\begin{table}
\begin{tabular}{ |c|c|c|c| } 
\hline
Categories & Number of Gamma matrices & Number of independent terms \\
\hline
\(\displaystyle \bar{u}_1  v_2 \bar{u}_3  v_4 k_1 . k_2 \) &  0 & $9$  \\ 
\hline
\(\displaystyle \bar{u}_1  v_2 \bar{u}_3 \slashed{l}  v_4 m_1 \) &  1 & $8$  \\ 
\hline
\(\displaystyle \bar{u}_1 \slashed{k_3}  v_2 \bar{u}_3 \slashed{k_1}  v_4 \)  &  2 & $4$  \\ 
\hline
\(\displaystyle \bar{u}_1 \gamma^\mu  v_2 \bar{u}_3 \gamma_\mu  v_4 k_1 . k_2 \)  &  2 & $9$  \\ 
\hline
\(\displaystyle \bar{u}_1   v_2 \bar{u}_3 \slashed{k_1} \slashed{l}  v_4 \) &  2 & $2$  \\ 
\hline
\(\displaystyle \bar{u}_1 \gamma^\mu  v_2 \bar{u}_3 \slashed{k_1} \gamma_\mu  v_4 m_1 \) &  3 & $8$  \\ 
\hline
\(\displaystyle \bar{u}_1 \gamma^\mu  v_2 \bar{u}_3 \slashed{k_1} \slashed{l} \gamma_\mu  v_4 \)  &  4 & $2$  \\ 
\hline
\(\displaystyle \bar{u}_1 \slashed{l} \gamma^\mu  v_2 \bar{u}_3 \slashed{k_1} \gamma_\mu  v_4 \)  &  4 & $4$  \\ 
\hline
\(\displaystyle \bar{u}_1  \gamma^\mu \gamma^\nu  v_2 \bar{u}_3  \gamma_\mu \gamma_\nu  v_4 k_1 . k_2 \)  & 4 & $9$  \\ 
\hline
\(\displaystyle \bar{u}_1 \gamma^\mu \gamma^\nu  v_2 \bar{u}_3 \slashed{k_1} \gamma_\mu \gamma_\nu  v_4 m_1 \) &  5 & $8$  \\ 
\hline
\(\displaystyle \bar{u}_1 \slashed{k_3}  \gamma^\mu \gamma^\nu  v_2 \bar{u}_3 \slashed{k_1}  \gamma_\mu \gamma_\nu  v_4  \)  &  6 & $4$  \\ 
\hline
\(\displaystyle \bar{u}_1  \gamma^\mu \gamma^\nu \gamma^\delta   v_2 \bar{u}_3  \gamma_\mu \gamma_\nu \gamma_\delta  v_4 (k_1 \cdot k_2)  \)  &  6 & $9$  \\ 
\hline
\end{tabular}
\caption{Different categories in the general ansatz for one-loop amplitudes with two distinct massive fermionic pairs }
\label{tab:CategoriesAnsatzTwoPairs}
\end{table}
\end{center}

The total number of independent terms are $76$ and because we have three graphs in our basis, we will start with $228$ unknown coefficients to be constrained by cuts and color-dual relations.

\subsubsection{Bootstrapping the general ansatz}\label{sub:1.2 one-loopQCD}

\begin{figure}[H]
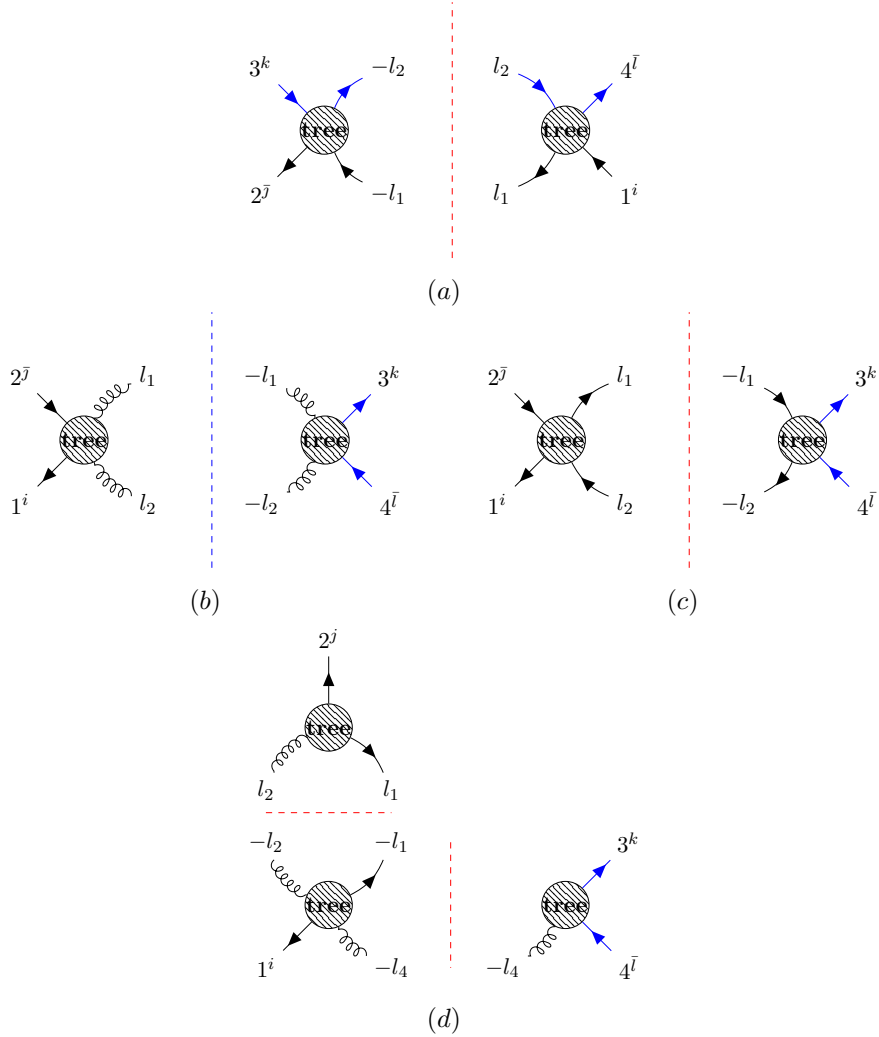

	\centering
	\begin{subfigure}[h]{0.4\textwidth}
		\centering
	\resizebox {.9\textwidth} {!} {	
	\numberedfig
}
	\caption{\((a)\)}
	\end{subfigure}
	
		\begin{subfigure}[h]{0.4\textwidth}
		\centering
	\resizebox {.9\textwidth} {!} {	
	\numberedfig
}
	\caption{\((b)\)}
	\end{subfigure}
		\begin{subfigure}[h]{0.4\textwidth}
		\centering
	\resizebox {.9\textwidth} {!} {	
	\numberedfig
}
	\caption{\((c)\)}
	\end{subfigure}
	\begin{subfigure}[h]{0.4\textwidth}
		\centering
	\resizebox {.9\textwidth} {!} {	
	\numberedfig
}
	\caption{\((d)\)}
	\end{subfigure}
		\caption{Generalized unitarity cuts for four-point  one-loop amplitudes with two different massive fermionic pairs. (a) Two different internal fermions, $l^{2}_{1} = m^2_1$,$ l^{2}_{2} =m^2_3$ (b) Two gluons, $l^{2}_{1} = l^{2}_{2} =0$ (c) Same internal fermions, $l^{2}_{1} = l^{2}_{2} =m^2_1$ (d) Two gluons, $l_{2}^{2}=l_{4}^{2}=0$ and one internal fermion $l_{1}^{2}=m_{1}^{2}$}
	\label{fig:GeneralizedUnitarityCutsOneLoopTwoPairs}
\end{figure}

 We start by imposing generalized unitarity cuts. Topologically there are four distinct ordered cuts as drawn in see \Fig{fig:GeneralizedUnitarityCutsOneLoopTwoPairs}.
The constraining data comes from the four-point  ordered amplitudes which we calculated in previous sections.

The simplest cut is the cut with two distinct cut fermions, cut (a) of \Fig{fig:GeneralizedUnitarityCutsOneLoopTwoPairs}, with only the graph topology (1) from \fig{fig:one-loopTwoFermionicPairs} contributing:
\begin{multline}\label{eq:CutTwoDifferentMassiveFermions}
 \sum_{s_i \in \text{states}} A^{\text{tree}}[1^{m_1},\bar{l}^{m_1}_{1,\bar{s}_1},l^{m_2}_{2,s_2},\bar{4}^{m_2}] A^{\text{tree}}[-l^{m_1}_{1,s_1},\bar{2}^{m_1},3^{m_2},-\bar{l}^{m_2}_{2,\bar{s}_2}] =
\frac{n^1[1_f,2_{\bar{f}},3_f,4_{\bar{f}},l_1-k_2] }{(l_1 + k_1)^2 (k_2-l_1)^2}\,.
\end{multline}
The gluonic cut, \Fig{fig:GeneralizedUnitarityCutsOneLoopTwoPairs}(b), has four graphs contributing,
\begin{multline}\label{eq:eq:GluonicCutMassive}
\sum_{s_i \in \text{states}}  A^{\text{tree}}[1^{m_1}, \bar{2}^{m_1}, l^{s_1}_{1},l^{s_2}_{2}] A^{\text{tree}}[-l^{\overline{s_2}}_{2},-l^{\overline{s_1}}_{1},3^{m_2},\bar{4}^{m_2}] =\frac{n^2[1,\bar{2},3,\bar{4},l_1]}{((l_1 + k_2)^2-m^{2}_{1}) (k_1+ k_2)^2}+\\
\frac{n^1[1,\bar{2},3,\bar{4},l_1]}{((l_1 + k_2)^2-m^{2}_{1}) ((l_1- k_3)^2-m^{2}_{2})}+ \frac{n^2[3,\bar{4},1,\bar{2},l_1+k_1+k_2]}{((l_1 - k_3)^2-m^{2}_{2}) (k_1+ k_2)^2}+\frac{n^4[1,\bar{2},3,\bar{4},l_1]}{((k_1+ k_2)^2)^2}\,,
\end{multline}
but we can use the Jacobi relations between kinematic numerators to write them all in terms of the kinematic weight of $n^1$.  These two cuts fix all coefficients in our ansatz for $n^1$.

To constrain the terms in numerator 3, first we impose the cut depicted in \Fig{fig:GeneralizedUnitarityCutsOneLoopTwoPairs}(c):
\begin{multline}\label{eq:CutTwoSameMassiveFermionsNumerator3}
\sum_{s \in \text{states}}  A^{\text{tree}}[1^{m_1}_f,2^{m_1}_{\bar{f}},\bar{l}^{m_1}_{1, s_1},\bar{l}^{m_1}_{2,s_2}]A^{\text{tree}}[-\bar{l}^{m_1}_{2,\bar{s_2}},-\bar{l}^{m_1}_{1,\bar{s_1}},3^{m_2}_f,4^{m_2}_{\bar{f}}] = 
\frac{n^3[1_f,2_{\bar{f}},3_f,4_{\bar{f}},l_1]}{(l_1+ k_2)^2 (k_1+ k_2)^2}
\end{multline}
This cut fixes $68$ coefficients.  Additionally, from \eqn{eq:RelationReorderedGraphs}, we have:
\begin{equation}\label{eq:AdjacentFermionsReorderingConstraintN3}
n^3[1_f,2_{\bar{f}},3_f,4_{\bar{f}},l] =  -\left[n^3[1_f,2_{\bar{f}},4_{\bar{f}},3_f,l]\equiv \Rev_{3\bar{4}}\left(n^3[1_f,2_{\bar{f}},3_f,4_{\bar{f}},l] \right)\right]\,.
\end{equation}
From flipping and relabeling momenta of $n^3[1_f,2_{\bar{f}},3_f,4_{\bar{f}},l]$ we also have,
\begin{equation}\label{eq:MirroringTwoMassivePair}
n^3[1_f,2_{\bar{f}},3_f,4_{\bar{f}},l] = \left[n^3[2_{\bar{f}},1_f,4_{\bar{f}},3_f,-l-1_{f}-2_{\bar{f}}]\equiv \Rev_{1\bar{2}} \left(n^3[1_f,2_{\bar{f}},3_f,4_{\bar{f}},-l-1_{f}-2_{\bar{f}}] \right)\right]\,.
\end{equation}
These considerations leave us with two unknown coefficients. To fix them, we impose the three-particle cut depicted in  \Fig{fig:GeneralizedUnitarityCutsOneLoopTwoPairs}(d):

\begin{multline}\label{eq:ThreeParticleCut}
\sum_{s \in \text{states}}  A^{\text{tree}}[1^{m_1}_f,-l_{2}^{s_1},-l_{4}^{s_2},-\bar{l}^{m_1}_{1,s_3}]A^{\text{tree}}[l^{s_1}_{2},2_{\bar{f}},l^{m_1}_{1,s_3}]A^{\text{tree}}[l_{4}^{s_2},3_f,4_{\bar{f}}] = 
\frac{n^3[1_f,2_{\bar{f}},3_f,4_{\bar{f}},l_1]}{(l_1+ k_1+k_2)^2 -m_{1}^2}\\
+\frac{n^2[1_f,2_{\bar{f}},3_f,4_{\bar{f}},-l_1-k_2]}{(-l_1+ k_1)^2 -m_{1}^2}\,,
\end{multline}
To constrain the terms in numerator 5, we impose the cut in \Fig{fig:GeneralizedUnitarityCutsOneLoopTwoPairs}(c):

\begin{multline}\label{eq:CutTwoSameMassiveFermionsNumerator5}
\sum_{s \in \text{states}}  A^{\text{tree}}[\bar{l}^{m_1}_{2,s_2},1^{m_1}_f,2^{m_1}_{\bar{f}},\bar{l}^{m_1}_{1, s_1}]A^{\text{tree}}[-\bar{l}^{m_1}_{2,\bar{s_2}},-\bar{l}^{m_1}_{1,\bar{s_1}},3^{m_2}_f,4^{m_2}_{\bar{f}}] = 
\frac{n^5[1_f,2_{\bar{f}},3_f,4_{\bar{f}},l_1]}{((k_1+ k_2)^2)^2}
\end{multline}
This cut fixes $68$ coefficients. Similar to the \eqn{eq:AdjacentFermionsReorderingConstraintN3}, we have:
\begin{equation}\label{eq:AdjacentFermionsReorderingConstraintN5}
n^5[1_f,2_{\bar{f}},3_f,4_{\bar{f}},l] = -\left[n^5[1_f,2_{\bar{f}},4_{\bar{f}},3_f,l]\equiv \Rev_{3\bar{4}}\left(n^5[1_f,2_{\bar{f}},3_f,4_{\bar{f}},l] \right)\right]\,.
\end{equation}
The above constraint leaves us with four unknown coefficients. To fix them, note that in $n^5[1_f,2_{\bar{f}},3_f,4_{\bar{f}},l]$ we consider the massive fermionic loop that has same flavor as pair $\{1_f,2_{\bar{f}}\}$. It is possible to have a fermionic loop with same flavor as pair $\{3_f,4_{\bar{f}}\}$. To distinguish them we show the first one with $n^5[1_f,2_{\bar{f}},3_f,4_{\bar{f}},l,m_1] $ and the second one with $n^5[1_f,2_{\bar{f}},3_f,4_{\bar{f}},l,m_2]$. We can define a new numerator $N^5[1_f,2_{\bar{f}},3_f,4_{\bar{f}},l]=n^5[1_f,2_{\bar{f}},3_f,4_{\bar{f}},l,m_1]+n^5[1_f,2_{\bar{f}},3_f,4_{\bar{f}},l,m_2]$. Symmetry of the diagram implies:
\begin{equation}\label{eq:ConstraintBubble}
N^5[1_f,2_{\bar{f}},3_f,4_{\bar{f}},l] = N^5[3_f,4_{\bar{f}},1_f,2_{\bar{f}},l+1_{f}+2_{\bar{f}}]\,,
\end{equation}
which fixes the remaining freedom in $n^5$.

The explicit expression for our basis numerators $n^1$, $n^3$ and $n^5$ are given,
\begin{align}\label{eq:Numerator1FourFermions}
n^{1}[1_f,2_{\bar{f}},3_f,4_{\bar{f}},l] =& -8\bar{u}_{1} \slashed{k_3} v_{2} \bar{u}_{3} \slashed{l} v_{4} +(2D-4)\bar{u}_{1} \slashed{l} v_{2} \bar{u}_{3} \slashed{l} v_{4} +2\bar{u}_{1}\slashed{k_3}\slashed{l}\gamma_{\alpha} v_{2} \bar{u}_{3}\gamma^{\alpha}v_{4}-\nn\\
&2\bar{u}_{1}\gamma^{\alpha} v_{2} \bar{u}_{3}\slashed{k_1} \slashed{l}\gamma_{\alpha}v_{4}-\bar{u}_{1}\slashed{l}\gamma^{\alpha}\gamma^{\beta}v_{2}\bar{u}_{3}\slashed{l}\gamma_{\alpha}\gamma_{\beta}v_{4}+\nn\\
&4\bar{u}_{1}\gamma^{\alpha}v_{2}\bar{u}_{3}\gamma_{\alpha}v_{4}(m^{2}_{1} + k_{1} \cdot( k_{2} + k_{3} +  l)+ (k_{2} \cdot l)+ l^2)+\nn\\
&m_{2}\bar{u}_{1}\slashed{l}v_{2} \bar{u}_{3}v_{4}\,,\\
\label{eq:Numerator3FourFermions}
n^{3}[1_f,2_{\bar{f}},3_f,4_{\bar{f}},l] =&(12-2D)\bar{u}_{1} \slashed{l} v_{2} \bar{u}_{3} \slashed{k_1} v_{4}+(2D-4)\bar{u}_{1} \slashed{l} v_{2} \bar{u}_{3} \slashed{l} v_{4}\nn\\
&+(2D-8)m_{1}\bar{u}_{1} v_{2} \bar{u}_{3} \slashed{k_1} v_{4}-(2D)m_{1}\bar{u}_{1} v_{2} \bar{u}_{3} \slashed{l} v_{4}\nn\\
&+(6-D)m_{1}^2 u_{1} \gamma^{\alpha} v_{2} u_{3} \gamma^{\alpha} v_{4} + 4 (k_{1} \cdot l)u_{1} \gamma^{\alpha} v_{2} u_{3} \gamma^{\alpha} v_{4} \nn \\
&+(8-2D)(k_{2} \cdot l)u_{1} \gamma^{\alpha} v_{2} u_{3} \gamma^{\alpha} v_{4} + (2-D)(l^2)u_{1} \gamma^{\alpha} v_{2} u_{3} \gamma^{\alpha} v_{4}\,,\\
\label{eq:Numerator5FourFermions}
n^{5}[1_f,2_{\bar{f}},3_f,4_{\bar{f}},l] =&2^{D/2}\left(2\bar{u}_{1} \slashed{l} v_{2} \bar{u}_{3} \slashed{l} v_{4}+m_{1}^2 u_{1} \gamma^{\alpha} v_{2} u_{3} \gamma^{\alpha} v_{4} \right.\nn\\
&- (k_{1} \cdot l)u_{1} \gamma^{\alpha} v_{2} u_{3} \gamma^{\alpha} v_{4}- (k_{2} \cdot l)u_{1} \gamma^{\alpha} v_{2} u_{3} \gamma^{\alpha} v_{4} \nn \\
&\left. - l^2 u_{1} \gamma^{\alpha} v_{2} u_{3} \gamma^{\alpha} v_{4}\right)\,. 
\end{align}
It is straightforward to use momentum conservation to show that these are exactly what one would expect from Feynman rules.
This is not entirely surprising as, after all, there are no contact term contributions demanding generalized freedom for these diagrams, but we could have been required to distribute Jacobi-like zeros in the form of contacts that would cancel between graphs for symmetry purposes. It is notable that the basis graphs under color-kinematics duality can be dressed with Feynman rules and that that Jacobi-like relations automatically propagate their information to the full amplitude. The expression for all descendent numerators are available in the machine readable Mathematica form in ancillary files associated with the arxiv version of this paper. 

We can write down the full color-dressed amplitude as:
\begin{equation}\label{eq:FullColorDressed1Loop4fermions}
\begin{split}
(g^{-4})\mathcal{A}^{\text{one-loop}}(1_f,2_{\bar{f}},3_f,4_{\bar{f}}) =& \int \frac{d^{D} l}{(2\pi)^D}(\frac{c^{1}[1_f,2_{\bar{f}},3_f,4_{\bar{f}}]n^{1}[1_f,2_{\bar{f}},3_f,4_{\bar{f}}]}{l^{2}((l-k_2)^2-m^{2}_{1})((l-k_1-k_2)^2)((l+k_3)^2-m^{2}_2)}+\\
&\frac{c^{1}[1_f,2_{\bar{f}},4_{\bar{f}},3_f]\Rev_{3\bar{4}}(n^{1}[1_f,2_{\bar{f}},3_f,4_{\bar{f}}])}{l^{2}((l-k_2)^2-m^{2}_{1})((l-k_1-k_2)^2)((l+k_4)^2-m^{2}_2)}+\\
&\frac{c^{2}[1_f,2_{\bar{f}},3_f,4_{\bar{f}}]n^{2}[1_f,2_{\bar{f}},3_f,4_{\bar{f}}]}{l^{2}((l-k_2)^2-m^{2}_{1})((l-k_1-k_2)^2)(k_3+k_4)^2}+\\
&\frac{c^{2}[3_f,4_{\bar{f}},1_f,2_{\bar{f}}]n^{2}[3_f,4_{\bar{f}},1_f,2_{\bar{f}}]}{l^{2}((l-k_4)^2-m^{2}_{2})((l-k_3-k_4)^2)(k_1+k_2)^2}+\\
&\frac{c^{3}[1_f,2_{\bar{f}},3_f,4_{\bar{f}}]n^{3}[1_f,2_{\bar{f}},3_f,4_{\bar{f}}]}{(l^2-m^{2}_1)(l-k_2)^2((l-k_1-k_2)^2-m^{2}_1)(k_3+k_4)^2}+\\
&\frac{c^{3}[3_f,4_{\bar{f}},1_f,2_{\bar{f}}]n^{3}[3_f,4_{\bar{f}},1_f,2_{\bar{f}}]}{(l^2-m^{2}_2)(l-k_4)^2((l-k_3-k_4)^2-m^{2}_2)(k_1+k_2)^2}+\\
&\frac{1}{2}\frac{c^{4}[1_f,2_{\bar{f}},3_f,4_{\bar{f}}]n^{4}[1_f,2_{\bar{f}},3_f,4_{\bar{f}}]}{l^{2}(l-k_1-k_2)^{2}((k_1+k_2)^2)^2}\\
&-c^{5}[1_f,2_{\bar{f}},3_f,4_{\bar{f}}] \left(\frac{n^{5}[1_f,2_{\bar{f}},3_f,4_{\bar{f}},m_1]}{(l^{2}-m^{2}_1)((l-k_1-k_2)^{2}-m^{2}_1)((k_1+k_2)^2)^2} \right.\\
&\left. +\frac{n^{5}[1_f,2_{\bar{f}},3_f,4_{\bar{f}},m_2]}{(l^{2}-m^{2}_2)((l-k_1-k_2)^{2}-m^{2}_2)((k_1+k_2)^2)^2}\right)\,. 
\end{split}
\end{equation}
The symmetry factor of $\frac{1}{2}$ comes from the gluon bubble and the fermion loop contributes a minus sign. The explicit expression for the color factors are:

\begin{equation}\label{eq:ExplicitColorFactor1LoopFourFermion}
\begin{split}
&c^{1}[1_f,2_{\bar{f}},3_f,4_{\bar{f}}]=T^{a}_{i\bar{x}}T^{b}_{x\bar{\jmath}}T^{a}_{s\bar{l}}T^{b}_{k\bar{s}}\\
&c^{1}[1_f,2_{\bar{f}},4_{\bar{f}},3_f]=T^{a}_{i\bar{x}}T^{b}_{x\bar{\jmath}}T^{b}_{s\bar{l}}T^{a}_{k\bar{s}}\\
&c^{2}[1_f,2_{\bar{f}},3_f,4_{\bar{f}}]=-T^{a}_{i\bar{x}}T^{b}_{x\bar{\jmath}}f^{abc}T^{c}_{k\bar{l}}\\
&c^{2}[3_f,4_{\bar{f}},1_f,2_{\bar{f}}]=-T^{a}_{k\bar{x}}T^{b}_{x\bar{l}}f^{abc}T^{c}_{i\bar{\jmath}}\\
&c^{3}[1_f,2_{\bar{f}},3_f,4_{\bar{f}}]=T^{a}_{i\bar{x}}T^{a}_{s\bar{\jmath}}T^{b}_{x\bar{s}}T^{b}_{k\bar{l}}\\
&c^{3}[3_f,4_{\bar{f}},1_f,2_{\bar{f}}]=T^{a}_{k\bar{x}}T^{a}_{s\bar{l}}T^{b}_{x\bar{s}}T^{b}_{i\bar{\jmath}}\\
&c^{4}[1_f,2_{\bar{f}},3_f,4_{\bar{f}}]=T^{a}_{i\bar{\jmath}}f^{abc}f^{bdc}T^{d}_{k\bar{l}}\\
&c^{5}[1_f,2_{\bar{f}},3_f,4_{\bar{f}}]=T^{a}_{i\bar{\jmath}}T^{a}_{s\bar{t}}T^{b}_{t\bar{s}}T^{b}_{k\bar{l}}\\
\end{split}
\end{equation}

From the Jacobi-like identities, we have the following relations between color factors:

\begin{equation}\label{eq:JacobiColorFactor1LoopFourFermion}
\begin{split}
&c^{1}[1_f,2_{\bar{f}},4_{\bar{f}},3_f] = c^{1}[1_f,2_{\bar{f}},3_f,4_{\bar{f}}]-\frac{1}{2}c^{4}[1_f,2_{\bar{f}},3_f,4_{\bar{f}}] \\
&c^{2}[1_f,2_{\bar{f}},3_f,4_{\bar{f}}] = \frac{1}{2}c^{4}[1_f,2_{\bar{f}},3_f,4_{\bar{f}}] \\
&c^{2}[3_f,4_{\bar{f}},1_f,2_{\bar{f}}] = \frac{1}{2}c^{4}[1_f,2_{\bar{f}},3_f,4_{\bar{f}}] \\
\end{split}
\end{equation}
We see that we have four independent color factors. In this color basis, we can write down the total amplitude in \eqn{eq:FullColorDressed1Loop4fermions} as:

\begin{multline}\label{eq:FullColorDressed1Loop4fermionsInColorBasis}
(g^{-4})\mathcal{A}^{\text{one-loop}}(1_f,2_{\bar{f}},3_f,4_{\bar{f}}) = \int \frac{d^{D} l}{(2\pi)^D} \Biggl[\\
c^{1}[1_f,2_{\bar{f}},3_f,4_{\bar{f}}] 
\left(\frac{n^{1}[1_f,2_{\bar{f}},3_f,4_{\bar{f}}]}{l^{2}((l-k_2)^2-m^{2}_{1})((l-k_1-k_2)^2)((l+k_3)^2-m^{2}_2)} + \right. \\
\left. \frac{\Rev_{3\bar{4}}(n^{1}[1_f,2_{\bar{f}},3_f,4_{\bar{f}}])}{l^{2}((l-k_2)^2-m^{2}_{1})((l-k_1-k_2)^2)((l+k_4)^2-m^{2}_2)} \right)+ \\
 \frac{1}{2} c^{4}[1_f,2_{\bar{f}},3_f,4_{\bar{f}}] \left(\frac{n^{4}[1_f,2_{\bar{f}},3_f,4_{\bar{f}}]}{l^{2}(l-k_1-k_2)^{2}((k_1+k_2)^2)^2}-\right.\\
 \frac{\Rev_{3\bar{4}}(n^{1}[1_f,2_{\bar{f}},3_f,4_{\bar{f}}])}{l^{2}((l-k_2)^2-m^{2}_{1})((l-k_1-k_2)^2)((l+k_4)^2-m^{2}_2)}+\\
 \frac{n^{2}[1_f,2_{\bar{f}},3_f,4_{\bar{f}}]}{l^{2}((l-k_2)^2-m^{2}_{1})((l-k_1-k_2)^2)(k_3+k_4)^2}+\\
\left. \frac{n^{2}[3_f,4_{\bar{f}},1_f,2_{\bar{f}}]}{l^{2}((l-k_4)^2-m^{2}_{2})((l-k_3-k_4)^2)(k_1+k_2)^2} \right)+\\
c^{3}[1_f,2_{\bar{f}},3_f,4_{\bar{f}}] \left(\frac{n^{3}[1_f,2_{\bar{f}},3_f,4_{\bar{f}}]}{(l^2-m^{2}_1)(l-k_2)^2((l-k_1-k_2)^2-m^{2}_1)(k_3+k_4)^2}\right)+\\
c^{3}[3_f,4_{\bar{f}},1_f,2_{\bar{f}}] \left(\frac{n^{3}[3_f,4_{\bar{f}},1_f,2_{\bar{f}}]}{(l^2-m^{2}_2)(l-k_4)^2((l-k_3-k_4)^2-m^{2}_2)(k_1+k_2)^2}\right)- \\
 c^{5}[1_f,2_{\bar{f}},3_f,4_{\bar{f}}]\left(\frac{n^{5}[1_f,2_{\bar{f}},3_f,4_{\bar{f}},m_1]}{(l^{2}-m^{2}_1)((l-k_1-k_2)^{2}-m^{2}_1)((k_1+k_2)^2)^2}\right.\\
\left. +\frac{n^{5}[1_f,2_{\bar{f}},3_f,4_{\bar{f}},m_2]}{(l^{2}-m^{2}_2)((l-k_1-k_2)^{2}-m^{2}_2)((k_1+k_2)^2)^2}\right)\Biggr]\,.
\end{multline}
The kinematic coefficients associated with each color factor in this minimal basis corresponds to a gauge invariant quantity which can be called a color ordered amplitude. We have verified these color-ordered amplitudes using FeynCalc~\cite{MERTIG1991345,Shtabovenko:2020gxv,Shtabovenko:2016sxi}

 \subsection{Four-point  amplitudes with four external gluons with a massive fermionic loop}\label{sec:one-loopQCDFourExternalGlueOneFermionicLoop}
 \begin{figure}[H]
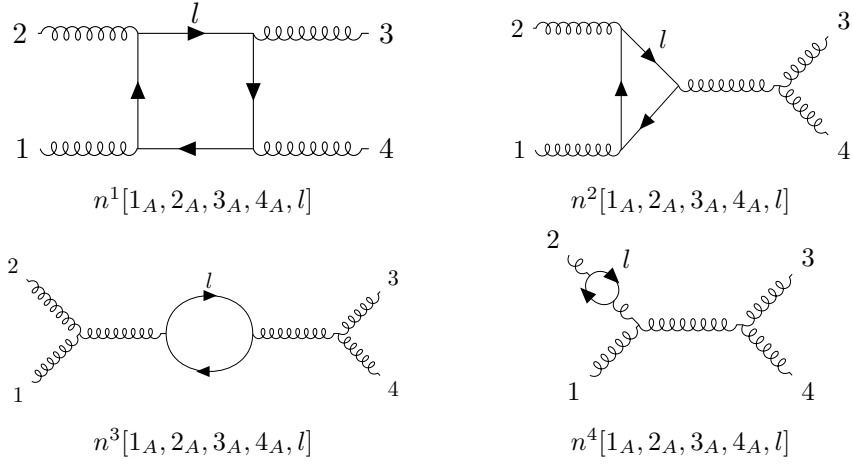

	\centering
	\begin{subfigure}{0.4\textwidth}
		\centering
	\resizebox {.9\textwidth} {!} {	
	\numberedfig
}
	\caption{$n^{1}[1_A,2_A,3_A,4_A,l]$}
	\end{subfigure}
	\begin{subfigure}{0.4\textwidth}
		\centering
	\resizebox {.8\textwidth} {!} {	
	\numberedfig
}
\caption{$n^{2}[1_A,2_A,3_A,4_A,l]$}
	\end{subfigure}
		\begin{subfigure}{0.4\textwidth}
		\centering
	\resizebox {.9\textwidth} {!} {	
	\numberedfig
}
\caption{$n^{3}[1_A,2_A,3_A,4_A,l]$}
\end{subfigure}
	\begin{subfigure}{0.4\textwidth}
		\centering
	\resizebox {.65\textwidth} {!} {	
	\numberedfig
}
\caption{$n^{4}[1_A,2_A,3_A,4_A,l]$}
\end{subfigure}

\caption{Different topologies for four gluons  with one massive fermionic loop, excluding tadpole diagrams.  The direction of the loop momentum as labeled is to be taken to the right in each graph independent of the orientation of the spinor arrow. }  
	\label{fig:DifferentTopologiesFourPointOneFermionicLoop}
\end{figure}

 Now we consider the case with external gluons and a massive fermionic loop of mass $m$.  We do not bother writing the contribution of internal gluon diagrams as these are known to be color-dual and indeed have been constructed in the literature\cite{Bern:2013yya}.
We need concern ourselves with four different topologies, excluding any tadpole diagrams, \Fig{fig:DifferentTopologiesFourPointOneFermionicLoop}.

The numerator of these topologies are related by the color-kinematics duality:
\begin{equation}\label{eq:JacobiOneFermionicLoop}
\begin{split}
&n^{2}[1_A,2_A,3_A,4_A,l] = n^{1}[1_A,2_A,3_A,4_A,l]-n^{1}[1_A,2_A,4_A,3_A,l]\\
&n^{3}[1_A,2_A,3_A,4_A,l] = n^{2}[1_A,2_A,3_A,4_A,l]-n^{2}[2_A,1_A,4_A,3_A,l]\\
&n^{4}[1_A,2_A,3_A,4_A,l] = n^{2}[1_A,2_A,3_A,4_A,l]+n^{2}[1_A,2_A,4_A,3_A,-l-2]
\end{split}
\end{equation}
We just have one basis graph and we choose the box diagram, $n^1$, as our basis kinematic numerator. There are three different classes of Lorentz invariants in our general ansatz:\\

$1$. $(\epsilon_{i}\cdot\epsilon_{j})(\epsilon_{k}\cdot\epsilon_{l}) (k_{m} \cdot k_{n})(k_{p} \cdot k_{q}$), there are $84$ terms like this.\\

$2$. $(\epsilon_{i}\cdot\epsilon_{j})(k_{k}\cdot\epsilon_{l})(k_{m}\cdot\epsilon_{n})( k_{p} \cdot k_{q}$), there are $378$ terms like this.\\

$3$. $(k_{i}\cdot\epsilon_{j})(k_{k}\cdot\epsilon_{l})(k_{m}\cdot\epsilon_{n})(k_{p}\cdot\epsilon_{q})$, there are $81$ terms like this.\\

So, there are 543 independent terms in our ansatz. 

\subsubsection{Constraints}\label{sec:4Glue1FermionicLoop}

\begin{figure}[H]
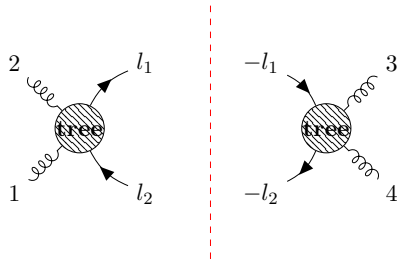

	\centering
			\begin{subfigure}[h]{0.4\textwidth}
		\centering
	\resizebox {.9\textwidth} {!} {	
	\numberedfig
}
	\caption{\(\)}
	\end{subfigure}
		\caption{Generalized unitarity cuts for four-point  amplitudes with four gluons with a massive fermionic loop, $l^{2}_{1} = l^{2}_{2} = m^2$. }
	\label{fig:GeneralizedUnitarityCutOneMassiveFermionicLoop}
\end{figure}

As drawn in \Fig{fig:GeneralizedUnitarityCutOneMassiveFermionicLoop}, we have only one type of cut that we need to consider, 
\begin{equation}\label{eq:OneFermionicLoopCut}
\begin{split}
&\sum_{s_i \in \text{states}} A^{\text{tree}}[1,2,l^{s_1},\bar{l}_2{}^{\overline{s_2}}] A^{\text{tree}}[-l^{s_2}_2,-\bar{l}_1{}^{\overline{s_1}},3,4]=\\
&\frac{n^1[1,2,3,4,l_1]}{((l_1+ k_2)^2-m^2) ((l_1- k_3)^2-m^2)}+\frac{n^2[1,2,3,4,l_1]}{((l_1+ k_2)^2-m^2) (k_1+ k_2)^2}+\\
&\frac{n^2[3,4,1,2,l_1+k_1+k_2]}{((l_1- k_3)^2-m^2) (k_1+ k_2)^2}+\frac{n^3[1,2,3,4,l_1]}{((k_1+ k_2)^2)^2}
\end{split}
\end{equation}
This cut condition fixes 364 coefficients. We apply additional constraints from rotating and mirroring graphs:
\begin{equation}\label{eq:OneFermionicLoopRotating}
\begin{split}
n^{1}[1_A,2_A,3_A,4_A,l]&= n^{1}[4_A,1_A,2_A,3_A,l+k_2]\\
n^{3}[1_A,2_A,3_A,4_A,l]&= n^{3}[3_A,4_A,1_A,2_A,l-k_1-k_2]\\
n^{2}[1_A,2_A,3_A,4_A,l]&= -n^{2}[2_A,1_A,3_A,4_A,-l-k_1-k_2]\\
n^{1}[1_A,2_A,3_A,4_A,l]&= n^{1}[2_A,1_A,4_A,3_A,-l]
\end{split}
\end{equation}
These fix a further 169 coefficients, leaving us with 10 remaining coefficients parameterizing a type of generalized gauge freedom.  These cancel in any gauge-invariant quantity upon integration. 

\subsubsection{Reduction to scalar integral basis}\label{sec:IntegralReductionOneFermionicLoop}
To compare with previous results in the literature it is conventient to express our results in terms of scalar integral basis. It has been shown that in $D$ dimensions, one-loop amplitudes can be expressed in terms of basis integrals\footnote{See, e.g.~refs.~\cite{VANNEERVEN1984241,Bern:1995db,Britto:2010xq,Ellis:2011cr} and references therein.},
\begin{equation}\label{eq:OneLoopIntegralReduction}
A^{one-loop} = \sum_{i} C^{i}_{D}  I^{i}_{D} + \sum_{j} C^{j}_{D-1}  I^{j}_{D-1}  + ... + \sum_{k} C^{k}_{2}  I^{k}_{2}  +  \mbox{rational terms}\,
\end{equation}
where $C^{i}_{D} $ are functions of external kinematics and  $I^{i}_{D}$ are a scalar integral basis. Rational terms are not four-dimensionally cut-constructible, but clever exploitation of dimensional regularization allow their contributions to be determined~\cite{Bern:1995db}. 

There are four independent basis integrals at one-loop which are known as tadpole, bubble, triangle and box integrals, respectively,
\begin{subequations}\label{eq:ExplicitIntegralBasis}
\begin{equation}\label{eq:TadPole}
I_{1}(m^{2}_{1}) = \frac{\mu^{4-D}}{i \pi^{\frac{D}{2}}}  \int \frac{d^{D}l}{d_1}\,,
\end{equation}
\begin{equation}\label{eq:Bubble}
I_{2}(r^2_{10};m^{2}_{1},m^{2}_{2}) = \frac{\mu^{4-D}}{i \pi^{\frac{D}{2}}}  \int \frac{d^{D}l}{d_1 d_2}\,,
\end{equation}
\begin{equation}\label{eq:Triangle}
I_{3}(r^2_{10},r^2_{12},r^2_{20};m^{2}_{1},m^{2}_{2},m^{3}_{3}) = \frac{\mu^{4-D}}{i \pi^{\frac{D}{2}}}  \int \frac{d^{D}l}{d_1 d_2 d_3}\,,
\end{equation}
\begin{equation}\label{eq:Box}
I_{4}(r^2_{10},r^2_{12},,r^2_{23},,r^2_{30},r^2_{20},r^2_{13};m^{2}_{1},m^{2}_{2},m^{3}_{3},m^{4}_{4}) = \frac{\mu^{4-D}}{i \pi^{\frac{D}{2}}}  \int \frac{d^{D}l}{d_1 d_2 d_3 d_4}\,,
\end{equation}
\end{subequations}
where $\mu$ is a renormalization factor, $d_k = (l+q_{k-1})^2-m^2_k + i\epsilon$, $q_n = \sum^{n}_i k_i$, $q_0=0$, and $r_{ij}= (q_i - q_j)^2$.

We will discuss how to go from our representation to the coefficients of basis integrals via Passarino-Veltman reduction.  As a simple example, consider a vector integral like:
\begin{equation*}
\int \frac{d^{D} l}{(2\pi)^D} \frac{l^\mu}{l^2 (l-k_1)^2 (l-k_1-k_2)^2 (l+k_4)^2}
\end{equation*}
The only vectors we are allowed are external momenta, three of which are independent, so we can write down the above integral as a linear combinations of coefficients of the three vectors,
\begin{equation*}
\int \frac{d^{D} l}{(2\pi)^D} \frac{l^\mu}{l^2 (l-k_1)^2 (l-k_1-k_2)^2 (l+k_4)^2} = c_1 k^{\mu}_1 + c_2 k^{\mu}_2 + c_3 k^{\mu}_3\,,
\end{equation*}
where $c_{(i)}$ are scalar integrals. By dotting both side of the above equation, we arrive at three equations which can be solved to find the $c_{(i)}$. 

In our numerators, we have tensor integrals up to rank four and the reduction is more complicated but proceeds apace. There exists many convenient software packages like {FeynCalc}~\cite{MERTIG1991345,Shtabovenko:2020gxv,Shtabovenko:2016sxi} which can be used to automate such reduction. After this reduction procedure, one can read off the coefficients $C^{i}_D$ for \eqn{eq:OneLoopIntegralReduction}. We include these in an ancillary file associated with the arXiv version of this paper.  

\subsubsection{Ordered amplitudes}\label{sec:OrderedAmplitudesFermionicLoop} 
To verify our construction, we want to write down ordered amplitudes at one-loop. For four external gluons with a massive fundamental fermion in the loop, we can write down the full amplitude~\cite{Bern:1995db}:
\begin{equation}\label{eq:GenericAmplitudeOneFermionicLoopTraceBase}
M^{one-loop}(1,2,3,4) = g^4 \mu^{2\epsilon}_R \sum_{\sigma} tr(T^{a_{\sigma (1)}}T^{a_{\sigma (2)}}T^{a_{\sigma (3)}}T^{a_{\sigma (4)}})A(\sigma (1),\sigma (2),\sigma (3),\sigma (4))
\end{equation}
Where $g$ is a coupling constant, $\mu_R$ is a renormalization scale, $T^a$ are fundamental representation color matrices (normalized as $\text{tr}(T^a T^b)=\delta^{ab}$), and $A(\ldots)$ represents the various ordered amplitudes. The sum is over all non-cyclic permutations of the indices $\sigma (n)$. With this decomposition, just Feynman diagrams with a fixed cyclic ordering of external legs contribute to ordered amplitudes. For example, $A(1_A,2_A,3_A,4_A)$ would be (excluding tadpole and snail diagrams):
\begin{equation}\label{eq:Explicitone-loopOrderedAmplitude}
\begin{split}
A^{one-loop}(1_A,2_A,3_A,4_A) &=\int \frac{d^{D}l}{(2\pi)^D} \left(\frac{n^{3}[4_A,1_A,2_A,3_A,l]}{(l^2-m^2) ((l-k_1-k_4)^2-m^2)((k_2+k_3)^2)^2}+ \right.\\
&\frac{n^{2}[1_A,2_A,3_A,4_A,l]}{(l^2-m^2) ((l-k_2)^2-m^2) ((l-k_1-k_2)^2-m^2) (k_1+ k_2)^2}+\\
&\frac{n^{2}[3_A,4_A,1_A,2_A,l]}{(l^2-m^2) ((l-k_4)^2-m^2) ((l-k_1-k_4)^2-m^2) (k_1+ k_2)^2}+\\
&\frac{n^{3}[1_A,2_A,3_A,4_A,l]}{(l^2-m^2) ((l-k_1-k_2)^2-m^2)((k_1+ k_2)^2)^2}+\\
&\frac{n^{2}[4_A,1_A,2_A,3_A,l]}{(l^2-m^2) ((l-k_1)^2-m^2) ((l-k_1-k_4)^2-m^2) (k_2+k_3)^2}+\\
&\frac{n^{2}[2_A,3_A,4_A,1_A,l]}{(l^2-m^2) ((l-k_3)^2-m^2) ((l-k_2-k_3)^2-m^2) (k_2+k_3)^2}+\\
&\left.   \frac{n^{1}[1_A,2_A,3_A,4_A,l]}{(l^2-m^2) ((l-k_2)^2-m^2) ((l-k_1-k_2)^2-m^2) ((l+k_3)^2-m^2)} \right)
\end{split}
\end{equation}
Where $n^{i}$ are the numerators for diagrams in \Fig{fig:DifferentTopologiesFourPointOneFermionicLoop} we calculated in \sect{sec:4Glue1FermionicLoop}. After Passarino-Veltman reduction, we recover the coefficients needed for \eqn{eq:OneLoopIntegralReduction}.  An important check at this stage is to verify the gauge invariance of amplitude, after reduction to a minimal color and integral basis, which we have preformed.

We additionally verified our results numerically against the results provided in~\cite{Bern:1995db} for various helicity configurations taking the external gluons to four dimensions. For instance, for all plus gluons, all coefficients of basis integrals vanish except for the box integral and in spinor-helicity language it is:
\begin{equation}\label{eq:OneFermionicLoopBoxCoefficient}
C_{\Box} =\frac{2i}{16\pi^2} \frac{[12][34]}{\langle 12 \rangle \langle 34 \rangle} m^4.
\end{equation}
This is a very sharp check that we have correctly accounted for $D$-dimensional data crossing the cuts as such rational terms are famously inaccessible via four-dimensional cut construction~\cite{Bern:1995db}.

There is a nice additional check of this approach, which is to consider the analogous calculation with a massive scalar running around the loop.  We leave the details of the analogous scalar calculation to Appendix~\ref{sec:FourExternalGlueOneScalarLoop}, but quote the results for all gluons taken with external plus-helicity, 
\begin{equation}\label{eq:OneLoopBoxOneMassiveScalar}
C^{\text{scalar loop}}_{\Box} =\frac{i}{16\pi^2} \frac{[12][34]}{\langle 12 \rangle \langle 34 \rangle} m^4
\end{equation}
We have recovered the well known  factor of one-half~\cite{Bern:1995db} relative to the fermionic case of \eqn{eq:OneFermionicLoopBoxCoefficient}. 

\section{Summary and Conclusion}\label{sec:conclusion}
In this paper, we have established that color-kinematics can be a tool for actual quantum chromodynamics at the integrand level. Unlike related previous double-copy work which fixes the coefficients of the one-loop integral basis in either the gauge theory or double-copy gravity theory via unitarity\footnote{Admittedly by far the most efficient approach to one-loop calculation.}, e.g.~Refs.~\cite{Primo:2016omk,Bern:2020buy}, here we find color-dual gauge theory integrands at four-points involving massive Dirac fermions in the fundamental.  The integrands satisfy $D$-dimensional cuts as well as the duality between color and kinematics. As we show, unitarity cuts and imposing the duality between color and kinematics are sufficient constraints to fix the gauge-invariant amplitudes. Specifically, we start with the three-point  graph and produce tree level amplitudes up through six-points, as well as the four-point  one-loop amplitudes with external fermions as well as  four external gluons with a massive fermion and scalar loop particle. We verify all amplitudes against known results from Feynman rules.  While individual graphs may be dressed differently than from Feynman gauge, indeed we only employ cubic graphs, all gauge invariant observables are equal.

We understand the UV behavior of QCD so it is sufficient~\cite{Bern:1995db} to control the IR by hand according to the renormalization scheme we are interested in.  We can therefore be content, especially at one-loop, to generate the integrands relevant to $d$-dimensional cut-construction as we do here -- requiring only that they satisfy the requisite physical cuts, and defer contributions that can be delicate for unitarity such as tadpoles and bubbles on external legs, often called snails, to other consideration.  However we can aspire to something more -- color-kinematics relates the weights of graphs.  It has the potential to transport physical information consistently from graphs that are trivial to access via unitarity methods to graphs that are not.  This is by no means trivial.  The book-keeping required to handle tadpoles and snails in concordance with color-kinematics is still being developed, e.g.~Refs.~{\cite{Bern:2012uf,Bern:2013yya,Edison:2022smn,Li:2022tir,Edison:2022jln}. We expect this to be an important avenue in future investigation. 

    It is intriguing\footnote{Despite similar statements being recognized for gauge invariant ordered kinematic amplitudes~\cite{Ochirov:2019mtf}, or indeed the universality of off-shell kinematic components of Feynman rules, the key point here is the dressing of graphs to satisfy algebraic properties can often mean adding zero to an amplitude by introducing contact contributions that cancel between graphs.  A priori one could imagine different functional forms for the dressing of graphs that satisfy different algebraic properties. } to note that the kinematic weights of individual color-dual graphs were independent of the nature of the color-dressing being in the fundamental or adjoint.  This echos similar results in the massive scalar case of ref.~\cite{Carrasco:2020ywq}, In many cases we were able to verify our tree-level results not only with Feynman rules for the entire amplitude but also by considering the kinematic weights of the same graphs contributing to supersymmetric theories (and thus having fermions dressed in the adjoint).  We expect this universality of kinematic color-dual weights to be an important future thread to explore.  One should be cautious about such a universality at loop-level -- given the explicit dimensional dependence of fermion traces, and the fact that adjoint conditions can relate graphs with closed fermion loops to graphs without.  The latter relation does not occur for Dirac fermions in the fundamental, but adjoint color-dual relations would relate $n^5$ of the one-loop four-fermion integrand to $n^3$.  Ultimately we expect this to simply fix the dimension where adjoint fermionic theories can be color-dual (the same dimensions that admit supersymmetry) in the manner of our example at four-point tree-level in \sect{adjointExample}.

A feature of the fermionic developments of this paper is that it dovetails incredibly well with a color-dual compositional approach~\cite{Carrasco:2019yyn,Low:2019wuv,Low:2020ubn,Carrasco:2021ptp} to building  amplitudes associated with higher-derivative operators in both gauge and gravity theories, of relevance both for effective field theory searches for beyond the standard model physics, and formal ultraviolet completion in gravity theories, which we look forward to exploring in forthcoming work. 

The relevance of these types of results in the double-copy to considering at least toy examples of Kerr-Schild black-hole interactions in the classical limit does not escape us -- encoding both  massive spin-1/2 scattering when double-copied with the results of \cite{Carrasco:2020ywq} as well as massive spin-1 scattering when double-copied with itself. Furthermore the five-point  one-loop calculation with massive fermions and emitted gluon is well within reach and will allow for a description of the one-loop correction gravitational radiative processes involving spinning matter which is a natural target as is of course higher-loops with associated corresponding progress in the $\kappa$, or Post-Minkowskian, expansion for higher-spin. Data already exists at one-loop in the conservative sector for fixed order results at arbitrarily oriented spin at four-points from direct cut-construction of one-loop amplitudes~\cite{Bern:2020buy} applying double-copy only at tree-level. An intriguing open question is if composition can be used to efficiently go after all order-spin scattering at loop level using color-dual massive loop-graph numerators, such as we present here, as building-blocks.

\section{Acknowledgment}
We would like to thank Alex Edison, Henrik Johansson, James Mangan, Nicolas Pavao, Radu Roiban, Bogdan Stoica, and Suna Zekioglu for helpful discussions, related collaboration, and encouragement along various stages of this project, including detailed feedback and helpful comments on earlier drafts of this paper. This work was supported by the DOE under contract DE-SC0015910 and by the Alfred P. Sloan Foundation. A.S. acknowledges the Northwestern University Amplitudes and Insight group, the Department of Physics and Astronomy, and Weinberg College for support.

\appendix

\section{External glue with a massive scalar loop}

\label{sec:FourExternalGlueOneScalarLoop}
This case is very similar to the fermionic example of \sect{sec:4Glue1FermionicLoop}. Here we simply allow a massive scalar to run in the loop instead of the quark, but otherwise  we have the same topologies and color-dual relations between their kinematic weights, \Fig{fig:TopologiesOneScalarLoop}.
\begin{figure}[H]
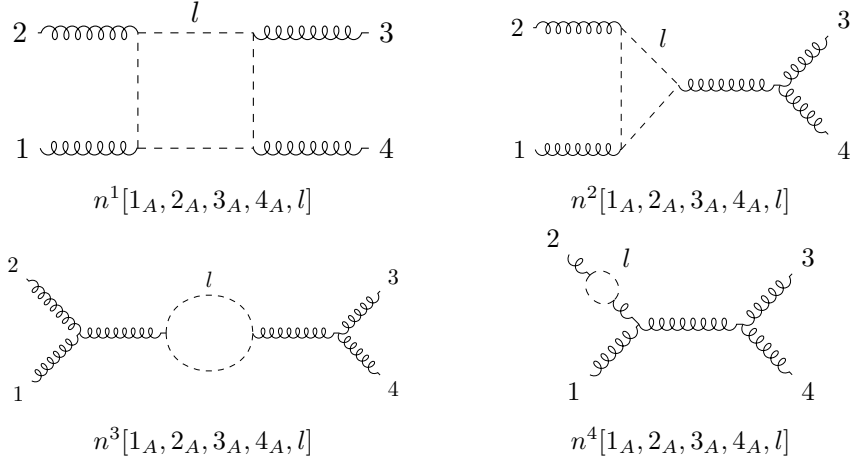

	\centering
	\begin{subfigure}{0.4\textwidth}
		\centering
	\resizebox {.9\textwidth} {!} {	
	\numberedfig
}
	\caption{$n^{1}[1_A,2_A,3_A,4_A,l]$}
	\end{subfigure}
	\begin{subfigure}{0.4\textwidth}
		\centering
	\resizebox {.8\textwidth} {!} {	
	\numberedfig
}
\caption{$n^{2}[1_A,2_A,3_A,4_A,l]$}
	\end{subfigure}
		\begin{subfigure}{0.4\textwidth}
		\centering
	\resizebox {.9\textwidth} {!} {	
	\numberedfig
}
\caption{$n^{3}[1_A,2_A,3_A,4_A,l]$}
\end{subfigure}
	\begin{subfigure}{0.4\textwidth}
		\centering
	\resizebox {.65\textwidth} {!} {	
	\numberedfig
}
\caption{$n^{4}[1_A,2_A,3_A,4_A,l]$}
\end{subfigure}

\caption{Different topologies for four gluons  with one massive scalar loop, The direction of the loop momentum as labeled is to be taken to the right in each graph. Note equivalent graphs contribute to two-to-two scattering for graphs 2,3, 5, and 6 hold with flavors swapped. Taking the numerators as functional allows us to dress those graphs via relabeling.}
	\label{fig:TopologiesOneScalarLoop}
\end{figure}
In our unitarity cut, we sew two four-point tree-level amplitudes. Each tree amplitude is between two scalars and two gluons. Using a bootstrap imposing the duality between color and kinematics and  the consistency of factorization, one can find find numerators and then ordered amplitudes ~\cite{Carrasco:2020ywq}. For example, the numerator for \Fig{fig:TreeTwoGlueTwoScalar}:
\begin{figure}[H]
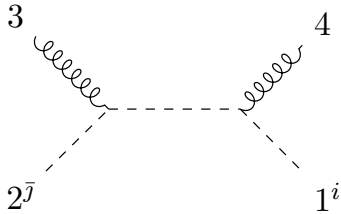

	\centering
\begin{subfigure}{0.5\textwidth}
		\centering
	\resizebox {.65\textwidth} {!} {	
	\numberedfig
}
\caption{}
\end{subfigure}

\caption{Basis Feynman diagram for two gluons and two scalar}
\label{fig:TreeTwoGlueTwoScalar}
\end{figure}

\begin{equation*}
n(1_S,2_S,3_A,4_A) = 4(k_1 . \epsilon_4)(k_2 . \epsilon_3)-2(k_1 . k_1)(\epsilon_3 . \epsilon_4)-2(k_1 . k_2)(\epsilon_3 . \epsilon_4)-2(k_1 . k_3)(\epsilon_3 . \epsilon_4)
\end{equation*}

\captionsetup[subfigure]{labelformat=empty}
\begin{figure}[H]
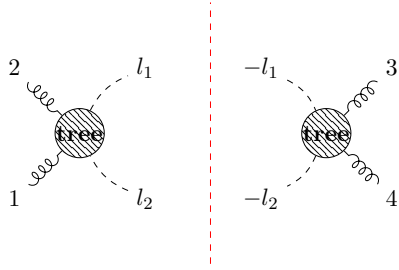

	\centering
			\begin{subfigure}[h]{0.4\textwidth}
		\centering
	\resizebox {.9\textwidth} {!} {	
	\numberedfig
}
	\caption{\(\)}
	\end{subfigure}
		\caption{Two-particle unitarity cut of the one loop correction to two-to-two gluon scattering with a massive scalar loop, $l^{2}_{1} = l^{2}_{2} = m^2$. }
	\label{fig:UnitarityCutOneScalarLoop}
\end{figure}
Now, we are equipped to do the two particle cut in the scalar case, \Fig{fig:UnitarityCutOneScalarLoop}. Repeating the same procedures as for the fermionic loop, we can find ordered amplitudes which are gauge-independent. For example, like the fermionic case, for all plus gluons, all coefficients of basis integrals will vanish except the box integral. As expressed in 
\eqn{eq:OneLoopBoxOneMassiveScalar}, we find half the result of the fermionic case, matching the well known result~\cite{Bern:1995db}.

\bibliographystyle{JHEP}
\bibliography{references}
\end{document}